\documentclass[%
 reprint,
superscriptaddress,
preprintnumbers,
nofootinbib,
 bibnotes,
 amsmath,amssymb,
 aps,
prb,
]{revtex4-2}
\usepackage[utf8]{inputenc}
\usepackage[T1]{fontenc}
\usepackage{graphicx}
\usepackage{dcolumn}
\usepackage{bm}
\usepackage{hyperref}
\usepackage{amsmath}	
\usepackage{amssymb}	
\usepackage{slashed}
\usepackage[normalem]{ulem}
\usepackage[dvipsnames]{xcolor}
\usepackage{xspace}
\usepackage{multirow}
\usepackage{booktabs}
\usepackage{comment}
\usepackage{lineno}
\RequirePackage{ifxetex}
\def\apj{ApJ}%
\def\apjl{ApJ}%
\def\apjs{ApJS}%
\def\ao{Appl.~Opt.}%
\def\aap{A\&A}%
\def\mnras{MNRAS}%
\def\prd{Phys.~Rev.~D}%
\def\prl{Phys.~Rev.~Lett.}%
\def\pasp{PASP}%
\def\pasj{PASJ}%
\def\jcap{JCAP}%
\def\physrep{Phys.~Rep.}%
\newcommand{\redmagic}{\textsc{redMaGiC}}
\newcommand{\maglim}{\textsc{MagLim}}
\newcommand{\metacal}{\textsc{Metacalibration}}

\newcommand{\sqdeg}{{\rm deg}^{2}}
\newcommand{\planck}{{\it Planck}}
\newcommand{\nn}{$\langle \delta_{\rm g} \delta_{\rm g} \rangle$}
\newcommand{\ngc}{$\langle \delta_{\rm g} \gamma_{\rm t} \rangle$}
\newcommand{\nk}{$\langle \delta_{\rm g} \kappa_{\rm CMB}\rangle $}
\newcommand{\gk}{$\langle \gamma_{\rm t}\kappa_{\rm CMB}\rangle$}
\newcommand{\fivetwo}{$5$$\times$$2{\rm pt}$}
\newcommand{\threetwo}{$3$$\times$$2{\rm pt}$}

\newcommand{\sixtwo}{$6$$\times$$2{\rm pt}$}
\newcommand{\nkgk}{$\langle \delta_{\rm g} \kappa_{\rm CMB}\rangle + \langle \gamma_{\rm t}\kappa_{\rm CMB}\rangle$}

\definecolor{ircol}{HTML}{B72654}

\newcommand{\results}[1]{{\color{black} #1}}

\renewcommand{\thempfootnote}{\arabic{footnote}}

\usepackage[utf8]{inputenc}
\usepackage{newunicodechar,graphicx}
\DeclareRobustCommand{\okina}{%
  \raisebox{\dimexpr\fontcharht\font`A-\height}{%
    \scalebox{0.8}{`}%
  }%
}
\newunicodechar{ʻ}{\okina}
\newcommand*\hawaii{Hawai\okina{}i}

\begin{document}

\title[Joint analysis of DES Year 3 data and CMB lensing from SPT and Planck I: Modeling]{Joint analysis of DES Year 3 data and CMB lensing from SPT and Planck I: \\Construction of CMB Lensing Maps and Modeling Choices}


\author{Y.~Omori}
\affiliation{Department of Astronomy and Astrophysics, University of Chicago, Chicago, IL 60637, USA}
\affiliation{Kavli Institute for Cosmological Physics, University of Chicago, Chicago, IL 60637, USA}
\affiliation{Department of Physics, Stanford University, 382 Via Pueblo Mall, Stanford, CA 94305, USA}
\affiliation{Kavli Institute for Particle Astrophysics \& Cosmology, P. O. Box 2450, Stanford University, Stanford, CA 94305, USA}
\author{E.~J.~Baxter}
\affiliation{Institute for Astronomy, University of \hawaii, 2680 Woodlawn Drive, Honolulu, HI 96822, USA}
\author{C.~Chang}
\affiliation{Department of Astronomy and Astrophysics, University of Chicago, Chicago, IL 60637, USA}
\affiliation{Kavli Institute for Cosmological Physics, University of Chicago, Chicago, IL 60637, USA}
\author{O.~Friedrich}
\affiliation{Kavli Institute for Cosmology, University of Cambridge, Madingley Road, Cambridge CB3 0HA, UK}
\author{A.~Alarcon}
\affiliation{Argonne National Laboratory, 9700 South Cass Avenue, Lemont, IL 60439, USA}
\author{O.~Alves}
\affiliation{Department of Physics, University of Michigan, Ann Arbor, MI 48109, USA}
\affiliation{Laborat\'orio Interinstitucional de e-Astronomia - LIneA, Rua Gal. Jos\'e Cristino 77, Rio de Janeiro, RJ - 20921-400, Brazil}
\author{A.~Amon}
\affiliation{Kavli Institute for Particle Astrophysics \& Cosmology, P. O. Box 2450, Stanford University, Stanford, CA 94305, USA}
\author{F.~Andrade-Oliveira}
\affiliation{Department of Physics, University of Michigan, Ann Arbor, MI 48109, USA}
\author{K.~Bechtol}
\affiliation{Physics Department, 2320 Chamberlin Hall, University of Wisconsin-Madison, 1150 University Avenue Madison, WI  53706-1390}
\author{M.~R.~Becker}
\affiliation{Argonne National Laboratory, 9700 South Cass Avenue, Lemont, IL 60439, USA}
% ...

\collaboration{DES \& SPT Collaborations}

\date{\today}

\begin{abstract}
Joint analyses of cross-correlations between measurements of galaxy positions, galaxy lensing, and lensing of the cosmic microwave background (CMB) offer powerful constraints on the large-scale structure of the Universe.  In a forthcoming analysis, we will present cosmological constraints from the analysis of such cross-correlations measured using Year 3 data from the Dark Energy Survey (DES), and CMB data from the South Pole Telescope (SPT) and {\it Planck}.  Here we present two key ingredients of this analysis: (1) an improved CMB lensing map in the SPT-SZ survey footprint, and (2) the analysis methodology that will be used to extract cosmological information from the cross-correlation measurements.  Relative to previous lensing maps made from the same CMB observations, we have implemented techniques to remove contamination from the thermal Sunyaev Zel'dovich effect, enabling the extraction of cosmological information from smaller angular scales of the cross-correlation measurements than in previous analyses with DES Year 1 data.  We describe our model for the cross-correlations between these maps and DES data, and validate our modeling choices to demonstrate the robustness of our analysis. We then forecast the expected cosmological constraints from the galaxy survey-CMB lensing auto and cross-correlations.  We find that 
the galaxy-CMB lensing and galaxy shear-CMB lensing correlations will on their own provide a constraint on $S_8=\sigma_8 \sqrt{\Omega_{\rm m}/0.3}$ at the few percent
level, providing a powerful consistency check for the DES-only constraints.  
We  explore scenarios where external priors on shear calibration are removed, finding that the joint analysis of CMB lensing cross-correlations can provide constraints on the shear calibration amplitude at the 5 to 10\% level.
\end{abstract}

\preprint{DES-2021-0647}
\preprint{FERMILAB-PUB-22-194-PPD}

\maketitle

\section{Introduction} \label{sec:intro}

Cross-correlations of galaxy surveys with overlapping measurements of cosmic microwave background (CMB) lensing offer a powerful way to probe the large-scale structure (LSS) of the Universe. Galaxy imaging surveys use measurements of the positions of galaxies and of the gravitational shearing of galaxy images to trace the LSS. For current imaging surveys \citep{des2005, deJong2013, Aihara2018}, these measurements typically become less sensitive at $z \gtrsim 1$, as galaxies become more difficult to detect and characterize at higher redshifts. Gravitational lensing of the CMB probes the LSS across a broad range of redshift, and is most sensitive to structures at $z\sim2$. Cross-correlations of galaxy surveys with CMB lensing can exploit this sensitivity to achieve tighter constraints on the high-redshift Universe than with galaxy surveys alone \cite[e.g.][]{Giannantonio16,omori18a,omori18b,namikawa2019,Krolewski2020,darwish2020,robertson2021,white2022}. CMB lensing also offers a probe of LSS that shares (almost) no sources of systematic error with measurements from galaxy surveys. For instance, unlike galaxies used to measure gravitational lensing, the redshift of the CMB is precisely known.  CMB lensing is also not impacted by effects such as intrinsic alignments.  Consequently, cross-correlations of galaxy and CMB lensing are expected to offer especially robust probes of LSS \cite[e.g.][]{Baxter:2016,Schaan:2017}.  This is an exciting prospect since control of systematic uncertainties in LSS surveys has become increasingly important as statistical uncertainties have continued to decrease.

The Dark Energy Survey \cite[DES,][]{des2005} and the South Pole Telescope \cite[SPT,][]{carlstrom2011} provide state-of-the-art galaxy and CMB data sets, respectively, that overlap across a large area on the sky, and are therefore very well suited to cross-correlation analyses.  DES has recently completed a six year survey of roughly 5,000 $\sqdeg$, with cosmological constraints from the first three years (Y3) of data presented in \cite{y3-3x2ptkp}.  The SPT-SZ survey was completed in 2011, and provides roughly 2,500 $\sqdeg$ of high-sensitivity and high-angular resolution CMB data that overlaps with DES observations.  At the same time, {\it Planck} provides maps of CMB lensing that overlap with the full 5,000 $\sqdeg$ DES survey region, albeit with higher noise and lower angular resolution than SPT-SZ \cite{planck18-8}.   

Several recent analyses have used cross-correlations between earlier DES data and SPT-SZ measurements of CMB lensing to constrain cosmology \cite[e.g.][]{Giannantonio16, omori18a, omori18b, 5x2y1}. In particular, \cite{5x2y1} presented a joint analysis of cross-correlations between first year (Y1) data from DES and CMB lensing measurements from SPT-SZ and {\it Planck}, using these correlations to constrain cosmological parameters, and to test for consistency between the galaxy survey and CMB lensing measurements. In that work, we analyzed six two-point functions between the galaxy density, galaxy lensing, and CMB lensing fields; we refer to this combination as \sixtwo{}.  When leaving out the CMB lensing auto-correlation, we refer to the remaining combination of probes as \fivetwo{}; the combination of two-point functions between galaxy density and galaxy lensing is referred to as \threetwo{}.  A challenge for the \fivetwo{} analysis presented in \cite{5x2y1} was  contamination of the CMB lensing maps by the thermal Sunyaev-Zel'dovich (tSZ) effect. This contamination prevented us from using the two-point function measurements at small scales, resulting in a significant reduction in signal-to-noise ratio: 19.9 to 9.9 and 10.8 to 6.8 for the galaxy-CMB lensing and shear-CMB lensing correlations respectively \cite{omori18a,omori18b}.

In this work, we present an updated CMB lensing map as well as the modeling framework and analysis choices that will be applied to the forthcoming analysis of cross-correlations between Year 3  data from DES and CMB lensing maps from SPT-SZ and {\it Planck}. The CMB lensing map presented here is constructed in a way that removes contamination from the tSZ, enabling a much larger fraction of the measured signal (and in particular the information at small angular scales) to be used to constrain cosmology. We apply several tests to the new CMB lensing maps to show that they are free from significant biases. 

The modeling framework that we present is similar to that developed in \cite{baxter2019}, but incorporates several improvements. These include new models for intrinsic alignments, the impact of lensing magnification of the galaxy sample,  modeling of nonlinear galaxy bias, and the use of lensing ratios.  We additionally describe the estimation of a covariance matrix for the cross-correlation measurements, and perform detailed validation of this estimate.  Finally, we determine a set of analysis choices, that when applied to simulated data designed to replicate the real DES, SPT-SZ and {\it Planck} data, yield robust and unbiased constraints on cosmological models. The methodology developed here will be applied to data in a companion paper.

The highest signal-to-noise measurement of the CMB lensing power spectrum to date is from the full-sky {\it Planck} mission \citep{planck18-8}.  Therefore, as in \cite{5x2y1}, we plan to present joint constraints that combine the {\it Planck} lensing power spectrum measurements with the \fivetwo{} measurements presented here.  As we demonstrate below, since {\it Planck} covers the full sky and since the CMB lensing power spectrum is primarily sensitive to higher redshifts than the \fivetwo{} combination, covariance between the two is negligible.  We therefore consider the CMB lensing auto-spectrum as an external probe, and focus the methodological developments in this paper entirely on \fivetwo{}.

The paper is organized as follows. In Section \ref{sec:mapmaking}, we present the methodology used to construct the CMB lensing map from SPT and {\it Planck} data, as well as tests of these maps.  We quantify the noise level in the maps, a key ingredient for determining the covariance of the cross-correlation measurements.  In Section~\ref{sec:modeling} we present our models for the correlations between these maps and DES galaxies and shears. In Section~\ref{sec:model_fitting} we describe our procedure for fitting the theoretical models to the two-point measurements, including our modeling and validation for the covariance matrix. In Section~\ref{sec:analysischoices}, we describe our procedure for selecting parts of the full data vector (i.e. the correlation measurements) for which we are sufficiently certain of the accuracy of our model that we can use the measurements to constrain cosmological parameters. We present forecasts for cosmological constraints in Section~\ref{sec:results}. We conclude in Section \ref{sec:discussion}.

\section{tSZ-free CMB lensing map}
\label{sec:mapmaking}

We begin by describing the data and methodology used to generate a CMB lensing map from SPT-SZ and {\it Planck} data that is not biased by contamination from the tSZ effect.

\subsection{Data} \label{sec:data}
\subsubsection{SPT-SZ temperature map}

The SPT is a millimeter/sub-millimeter telescope with a 10~m aperture that is located at the National Science Foundation Amundsen-Scott South Pole station in Antarctica.  
The SPT data used in this analysis is the same as used in \cite{story13,omori2017,chown2018}, namely data from the 2500 deg$^2$ SPT-SZ survey, which was conducted between 2008 and 2011.  While the SPT-SZ camera had three frequency channels, we primarily focus on the 150\ GHz data since its noise level ($\sim\!18\ \mu {\rm K}$-${\rm arcmin}$) is lower than that of the 90 and 220 GHz data (40 and 70 $\mu {\rm K}$-${\rm arcmin}$, respectively) \cite{bleem2015}. We start with the same data products as in \cite{omori2017} and reprocess the data to optimize for cross-correlation analyses.  In particular, we reduce the number of masked regions\footnote{ In \cite{omori2017}, clusters detected with S/N greater than 5 in \cite{bleem2015} were masked. In this study, we only mask clusters detected above S/N 10 in the temperature map before performing the lensing reconstruction.} around clusters before performing the lensing reconstruction procedure, since the tSZ-nulling method will eliminate the tSZ bias. The nulling procedure is described in Section \ref{sec:QE}.

\subsubsection{Planck data}

The {\it Planck} satellite was launched in 2009 by the European Space agency, with the goal of making clean maps of the CMB by observing the sky at nine frequencies ranging from 30 to 857 GHz \cite{tauber2010,planck2011a}).  We rely on two different temperature maps from {\it Planck}:
\begin{itemize}
\item {\bf Planck 143 GHz temperature map}. By combining the {\it Planck} data and SPT-SZ data over the same footprint, we can improve signal-to-noise by recovering the modes that are removed in the SPT-SZ data due to filtering.  
To this end, we use the $\planck$ 143 GHz full mission temperature map from the 2018 data release \cite{planck18-3}.\footnote{The maps are publicly available from the Planck Legacy Archive: \url{https://pla.esac.esa.int}.} Additionally, we use the 300 Full Focal Plane (FFP10)
full mission noise realizations for the purposes of computing the {\it Planck} noise power.  We describe the process of combining the SPT-SZ 150 GHz and {\it Planck} 143 GHz temperature data to improve signal-to-noise in Section \ref{sec:combining}. 
\item {\bf Planck SMICA tSZ-nulled (SMICAnoSZ) temperature map}. Our reconstruction of the CMB lensing field from the CMB temperature data relies on the quadratic estimator \cite{okamoto2003}, which estimates the lensing field using two (differently filtered) temperature maps, or ``legs.'' In \cite{omori2017}, the minimum-variance combination of SPT 150 GHz and {\it Planck} 143 GHz was used for both legs.

In this study, we replace one of the legs with a lower-resolution and higher-noise, but tSZ-cleaned temperature map generated from {\it Planck} data. Specifically, we use {\it Planck} maps generated with the Spectral Matching Independent Component Analysis (\textsc{SMICA})  algorithm \cite{delabrouille2003,cardoso2008}.  SMICA takes the linear combinations of all three LFI and six HFI {\it Planck} frequency channels from 30 to 857 GHz \cite{planck18-4} to produce the minimum-variance map of the CMB. The tSZ-free variant of this map, SMICAnoSZ, exploits the known frequency dependence of the tSZ signal to remove the tSZ signal, in exchange for a slight increase in the noise and potential bias from the cosmic infrared background (CIB).\footnote{A similar result has been obtained by \cite{bobin2016} using their LGMCA algorithm based on the blind source separation technique.} Similar approaches have been used to make tSZ-nulled CMB maps in other studies \cite{madhavacheril2020, bleem2021}.
This temperature map is also the input for the SMICAnoSZ variant of the lensing map released by the \planck{} collaboration. 
\end{itemize}
 
\subsection{CMB simulations}
\label{sec:sims}

Simulations of the CMB data are necessary to compute quantities such as the response function, mean-field bias, and noise bias terms that are used to produce normalized and debiased CMB lensing maps and CMB lensing auto-spectra \cite{planck13-17,planck15-15,omori2017,planck18-8}.  We begin by generating unlensed CMB realizations at the {\it Planck} 2018 best-fit cosmology \cite{planck18-6} with $N_{\rm side}=8192$, and also Gaussian realizations of the lensing potential, which we use to deflect the unlensed CMB maps using the \textsc{LensPix} package\cite{lenspix}.

We also simulate contributions to the sky from secondary (i.e. non-CMB) sources of emission.  We split these contributions into Gaussian and Poisson components. For the Gaussian component, we largely follow the simulation pipeline that was used in \cite{omori2017}: we take the best-fit model power spectrum of thermal SZ, kinematic SZ, cosmic infrared background (CIB), and radio sources from \cite{george2015} and generate Gaussian realizations from those power spectra.\footnote{As noted in \cite{baxter2019}, these simulations using Gaussian realizations are not sufficient to asses biases coming from high-order correlations, however they are sufficient to estimate the noise-levels and calculating  quantities such as the lensing response function.}
For the Poisson term, we place detected point sources with their measured fluxes at their observed locations.

We generate 150 full-sky realizations of lensed CMB and Gaussian secondary realizations, and extract two patches at the opposite hemispheres. After extracting two SPT-SZ-sized patches from each realization---for a total of 300 simulations of the SPT-SZ survey---we add clusters detected above $5\sigma$ in \cite{bleem2015} and point sources with fluxes between 6.4 and 50 mJy in 150 GHz \cite{everett2020} and place them at their observed locations. This ensures that these sources are at the same locations in all of the realizations, which is important for  computing the mean-field bias after reconstructing the lensing map.  

From the sum of the simulated lensed CMB and foreground maps, we generate mock SPT-SZ and {\it Planck} maps. For SPT-SZ, we pass the extracted maps through a mock-observing pipeline.  As described in \cite{omori2017} and \cite{chown2018}, we compare the outputs of the 300 realizations from the mock observations with the input maps to compute the filter transfer function. We then add noise realizations obtained using the half-difference technique, where half of the observations are multiplied with a minus sign, such that when the sum of all the observations are taken, the sky signal is nulled and noise is left. For {\it Planck} 143 GHz mocks, we simply convolve the input sky maps with the 143 GHz channel beam,\footnote{\texttt{HFI\_RIMO\_R3.00.FITS} available from the Planck Legacy Archive.} and add the noise realizations from the FFP10 simulations.

Generating simulated maps corresponding to the SMICAnoSZ maps is somewhat more involved because these use data from nine frequency channels.  Generating foreground models across these bands would require detailed knowledge of the foreground emission. We take a simplified approach, using the mock 143 GHz channel map with modified amplitudes for the tSZ and CIB components (the other two components, radio sources and kSZ, are subdominant). The tSZ component is simply removed since it is not present in the SMICAnoSZ maps. To modify the amplitude of the CIB component, we first generate maps of the CIB at all of the frequency channels used to construct the SMICAnoSZ map by scaling the Gaussian CIB realizations at 150 GHz, using the scaling relation based on the CIB map amplitudes in \cite{sehgal2010} at low frequencies and maps at \cite{lenz2019}  at higher frequencies.  The CIB maps generated this way are then passed through the SMICAnoSZ weights,\footnote{The weights are publicly available as part of the SMICA weight propagation code at the $Planck$ legacy archive.} to generate a mock SMICAnoSZ CIB map.  The mock CIB map used in the analysis is finally generated by multiplying the Gaussian 150 GHz CIB map by the multipole-dependent ratio of power spectra of the mock SMICAnoSZ CIB map and the Gaussian 150 GHz CIB map.

\subsection{Combining  SPT-SZ and Planck data}\label{sec:combining}

In order to capture modes in the SPT-SZ temperature map that are lost due to filtering and to improve the signal-to-noise of the CMB observations,  we combine the SPT-SZ 150 GHz and {\it Planck} 143 GHz maps using inverse variance weighting.  {\it Planck} data are used to fill in the spherical harmonic modes $\ell<500$ as well as modes with $m<250$. Modes where both SPT-SZ and {\it Planck} are noise dominated ($\ell>1600$ and $m<250$) are filtered out.

Starting with the 300 noise realizations, we compute the average 2D noise power spectrum $\langle |N_{\ell m}|^{2}\rangle$, where $N_{\ell m}$ are the coefficients of the spherical harmonic decomposition of the noise map.  The SPT-SZ 150 GHz and {\it Planck} 143 GHz maps are then combined (we denote the combined map with the superscript $x$) using the same inverse noise weighted combining technique\footnote{We increase the number of simulations from 200 to 300 realizations in the present study. The number is limited by the number of FFP10 noise realizations available.} as used in \cite{crawford2016,omori2017,chown2018}:
\begin{equation}
T_{\ell m}^{x}=\frac{w_{\ell m}^{\rm SPT}}{w_{\ell m}^{\rm SPT}+w_{\ell m}^{Planck}}\frac{T_{\ell m}^{\rm SPT}}{\mathcal{T}^{\rm SPT}_{\ell m}}+\frac{w_{\ell m}^{Planck}}{w_{\ell m}^{\rm SPT}+w_{\ell m}^{Planck}}\frac{T_{\ell m}^{Planck}}{b^{\it Planck}_{\ell}},
\end{equation}
where $T_{\ell m}$ are the temperature spherical harmonic coefficients and $w_{\ell m}$ are the weights per mode, which are taken to be $w_{\ell m}=1/\langle |N_{\ell m}|^{2}\rangle$. $\mathcal{T}_{\ell m}^{\rm SPT},b_{\ell}^{Planck}$ are the SPT-SZ transfer function (a combination of the beam and filter transfer function) and the {\it Planck} beam, respectively. Once the high-resolution SPT-SZ+{\it Planck} maps are produced, point sources detected by SPT-SZ with flux $F$ in the range  $6.4<F<200$ mJy ($6.4<F<50$ mJy for simulations) are inpainted using the Gaussian constrained inpainting method \cite{hoffman91,benoitlevy2013,omori2017} out to 3 and 5 arcminutes for sources below and above 50 mJy respectively . 

We similarly compute the combined noise power using:
\begin{equation}
N_{\ell m}^{x}=\frac{w_{\ell m}^{\rm SPT}}{w_{\ell m}^{\rm SPT}+w_{\ell m}^{Planck}}\frac{N_{\ell m}^{\rm SPT}}{\mathcal{T}^{\rm SPT}_{\ell m}}+\frac{w_{\ell m}^{Planck}}{w_{\ell m}^{\rm SPT}+w_{\ell m}^{Planck}}\frac{N_{\ell m}^{Planck}}{b^{\it Planck}_{\ell}}.
\end{equation}

\subsection{Construction of an unbiased CMB lensing map from SPT and Planck data}
\label{sec:maps}

\subsubsection{Bias from the thermal Sunyaev-Zel'dovich effect}

The tSZ effect induces a frequency-dependent signal into CMB temperature maps that is correlated with the large-scale structure.  As shown in \cite{baxter2019,madhavacheril18}, this signal can propagate through the standard quadratic estimator used to estimate CMB lensing, resulting in a bias to correlations between CMB lensing maps and galaxies or galaxy lensing.  In principle, since the frequency-dependence of the tSZ is known, one could combine multi-frequency CMB observations in a way that nulls the contribution from tSZ, but preserves the underlying CMB signal.  However, for the noise levels of SPT-SZ data, carrying out this procedure results in a tSZ-cleaned map that has significantly higher noise than the original tSZ-biased maps.  Since the noise level in the reconstructed lensing map is proportional to the temperature noise level squared, this results in a significant degradation in the signal-to-noise of the CMB lensing cross-correlations. 

Several approaches have been proposed in the literature to remove foreground biases in CMB lensing with minimal noise penalty, ranging from using a polarization-only lensing reconstruction \cite{osbourne2014}, to using a lensing reconstruction estimator based on shear instead of convergence \cite{schaan18}.  The approach that we adopt in this work is based on using a modified quadratic estimator \cite{madhavacheril18,darwish2020} with two maps, only one of which has been tSZ-cleaned.  In effect, by only cleaning one of the maps, the tSZ bias can be removed from the final lensing map, without the high noise penalty incurred from cleaning both maps entering the quadratic estimator.  Here we implement the same methodology as \cite{darwish2020}, but without flat-sky approximations. 

\begin{figure}
\begin{center}
\includegraphics[width=0.9\linewidth]{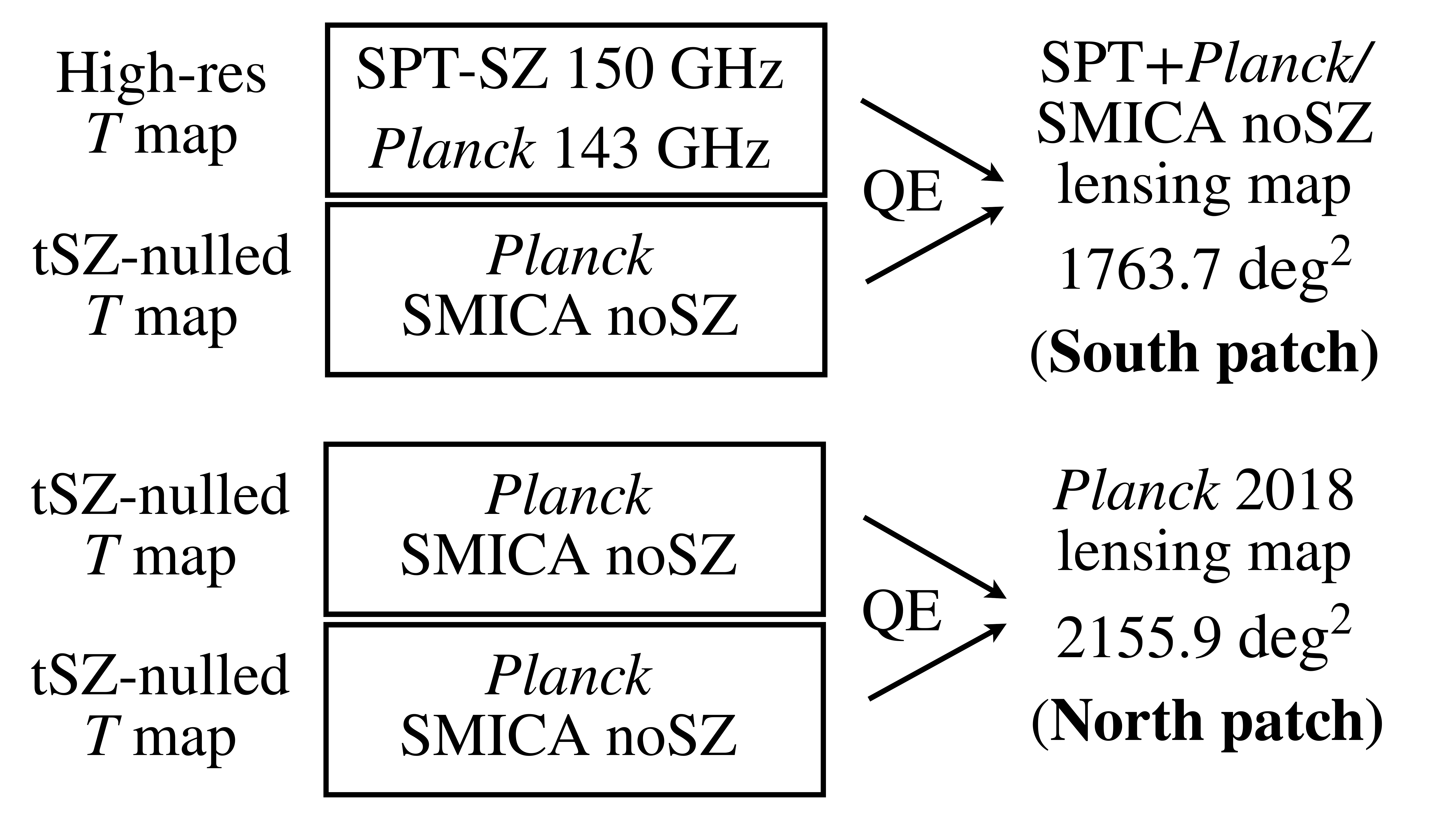}\\[0.25cm]
\includegraphics[width=1.0\linewidth,bb=120 10 500 380]{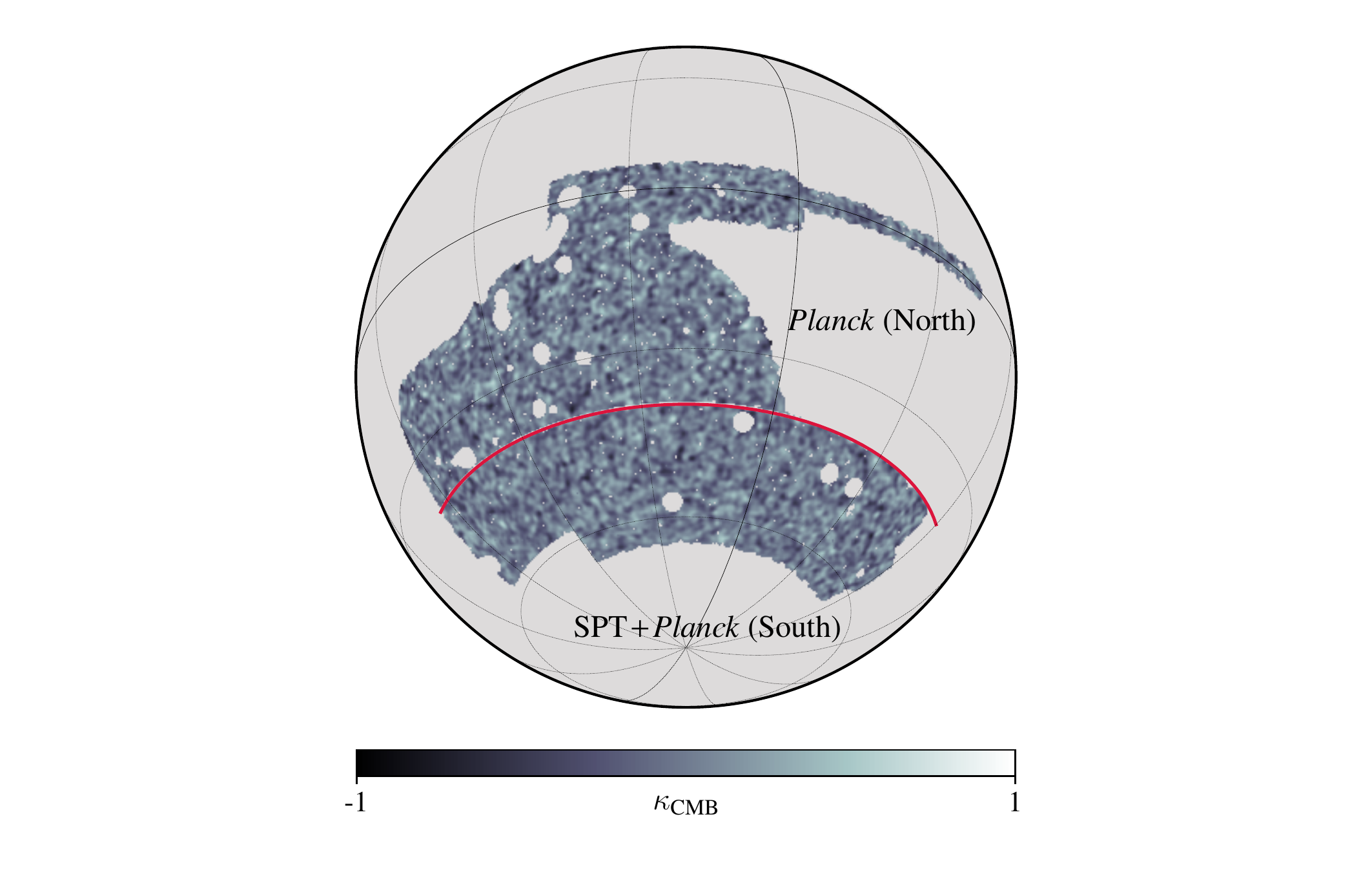}
\caption{{\bf Upper:} Diagram illustrating the input temperature maps used to construct the two different lensing maps utilized in this analysis. The operation ``QE'' (quadratic estimator) is the lensing reconstruction step described in Section \ref{sec:QE}. {\bf Lower:} Illustration of the sky coverage and lensing maps for the North ({\it Planck}) and South (SPT+{\it Planck}) patches.  The red line indicates the cut in declination (dec=$-40^{\circ}$) that divides the two regions. The union of the DES mask used in the DES Y3 analysis and the {\it Planck} lensing map mask is applied.
}
\label{fig:QE_flowchart}
\end{center}
\end{figure}

\begin{figure*}
\begin{center}
\includegraphics[width=0.8\linewidth]{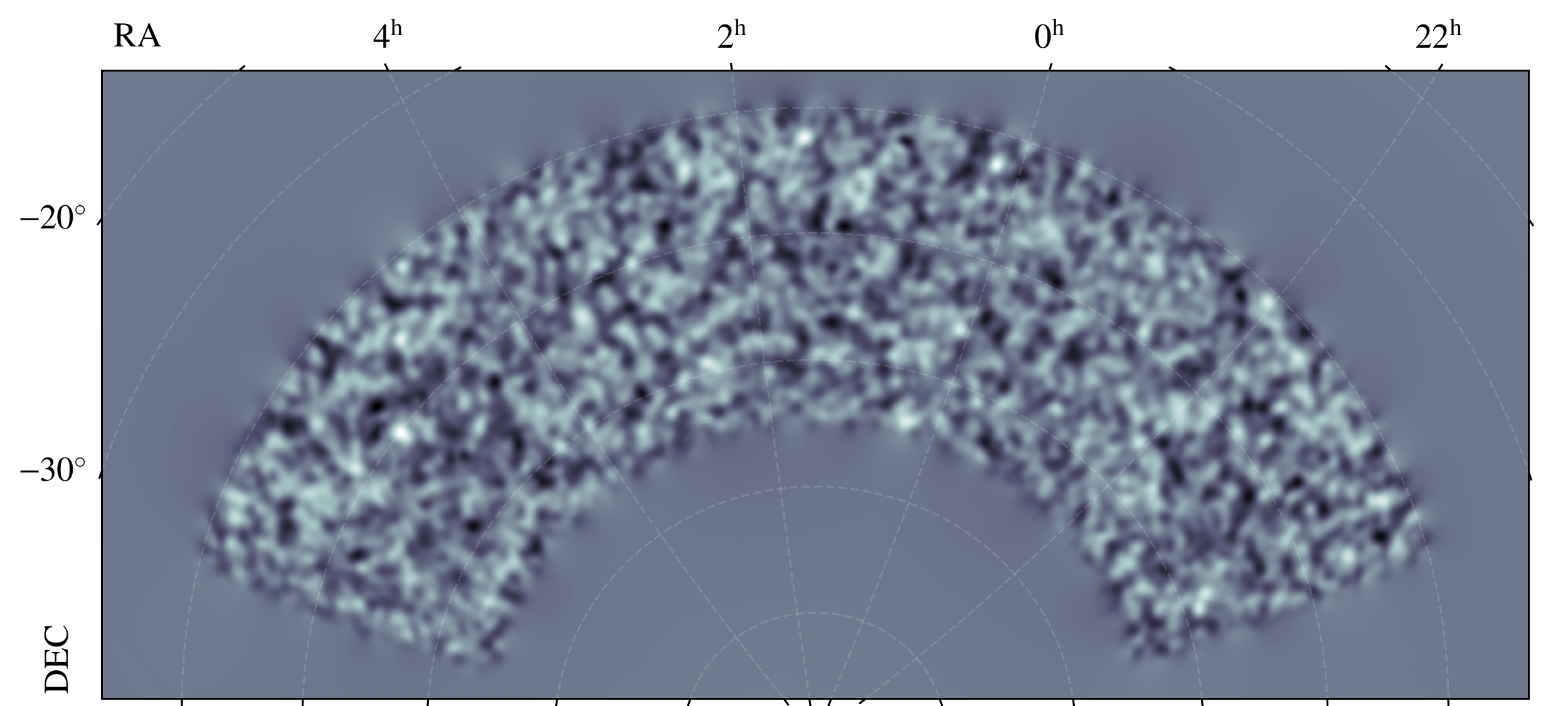}
\caption{CMB convergence map generated using the tSZ nulling method described in the text. The map has been smoothed with a Gaussian beam with ${\rm FWHM}=60'$ for visualization purposes.}
\label{fig:kcmb_map_ZEA}
\end{center}
\end{figure*}

\subsubsection{tSZ-cleaned lensing reconstruction}\label{sec:QE}

Prior to running the lensing reconstruction procedure, we filter the temperature maps with the filter $\mathcal{F}_{\ell m}=(C_{\ell}^{TT}+\langle |N_{\ell m}|^{2}\rangle )^{-1}$, such that $\bar{T}_{\ell m}=\mathcal{F}_{\ell m}T_{\ell m}=T_{\ell m}/(C_{\ell}^{TT}+\langle |N_{\ell m}|^{2} \rangle )$ for modes in the range $100<\ell<4000$ and zero otherwise \cite{planck13-17,planck15-15,omori2017}.  Note here that we use the 1D power spectrum for the signal component $C_{\ell}^{TT}$, but use a 2D filtering noise spectrum $\langle |N_{\ell m}|^{2} \rangle$ to account for possible anisotropies in the  noise. We then use the quadratic estimator:
\begin{multline}\label{eq:QE}
\bar{\phi}_{\ell m}=\frac{(-1)^{M}}{2}\sum_{\ell_{1} m_{1}\ell_{2} m_{2}}\begin{pmatrix}
\ell_{1} & \ell_{2} & L \\
-m_{1} & -m_{2} & M \\
\end{pmatrix} \times \\
W_{\ell_{1}\ell_{2}L}^{\phi}\bar{T}^{x}_{\ell_{1}m_{1}}\bar{T}^{\rm SMICAnosz}_{\ell_{2}m_{2}},
\end{multline}
where the term in brackets is the Wigner-$3j$ symbol, and $W_{\ell_{1}\ell_{2}L}^{\phi}$ is the weight function defined as
\begin{align}
&W_{\ell_{1}\ell_{2}L}^{\phi} \nonumber\\ 
&=-\sqrt{\frac{(2\ell_{1}+1)(2\ell_{2}+1)(2L+1)}{4\pi}}\sqrt{L(L+1)\ell_{1}(\ell_{1}+1)}\nonumber\\
&\times C_{\ell_{1}}^{TT}\left(\frac{1+(-1)^{\ell_{1}+\ell_{2}+L}}{2}\right) 
\begin{pmatrix}
\ell_{1} & \ell_{2} & L\\
1 & 0 & -1
\end{pmatrix} +(\ell_{1}\leftrightarrow \ell_{2}),
\end{align}
where the last term implies an identical term with $\ell_{1}$ and $\ell_{2}$ flipped.
Equation~\ref{eq:QE} requires two temperature maps (i.e. the ``legs").  Here we use the high resolution SPT-SZ+{\it Planck} temperature map $\bar{T}_{\ell m }^{x}$ and foreground cleaned temperature map $\bar{T}_{\ell m }^{\rm SMICAnoSZ}$ (see Figure \ref{fig:QE_flowchart}). The CMB lensing maps of \cite{omori2017} could be effectively  recovered\footnote{This will not be a perfect recovery since analysis choices have been changed slightly including the difference in simulations and masking choices.} by replacing the $\bar{T}_{\ell m}^{\rm SMICAnoSZ}$ with $\bar{T}_{\ell m }^{x}$.  If instead we were to use the tSZ-free maps for both legs of the estimator (i.e. using $\bar{T}_{\ell m}^{\rm SMICAnoSZ}$ for both), the resulting lensing map would also be tSZ-free, but would have higher noise owing to the higher noise levels of the $\bar{T}_{\ell m}^{\rm SMICAnoSZ}$ maps.  \cite{madhavacheril18} and \cite{darwish2020} have shown that the effect of the tSZ bias can be reduced with a small penalty in signal-to-noise ratio using this technique. 

We convert the lensing potential map to lensing convergence, $\kappa$, after subtracting the mean field $\bar{\phi}^{\rm MF}_{LM}$ and applying the lensing response function $R^{\phi}_{L}$:
\begin{equation}
\hat{\kappa}_{LM}=\frac{L(L+1)}{2}(\mathcal{R}^{\phi}_{L})^{-1}(\bar{\phi}_{LM}-\bar{\phi}_{LM}^{\rm MF} ).
\end{equation}

\begin{figure}
\begin{center}
\includegraphics[width=1.0\linewidth]{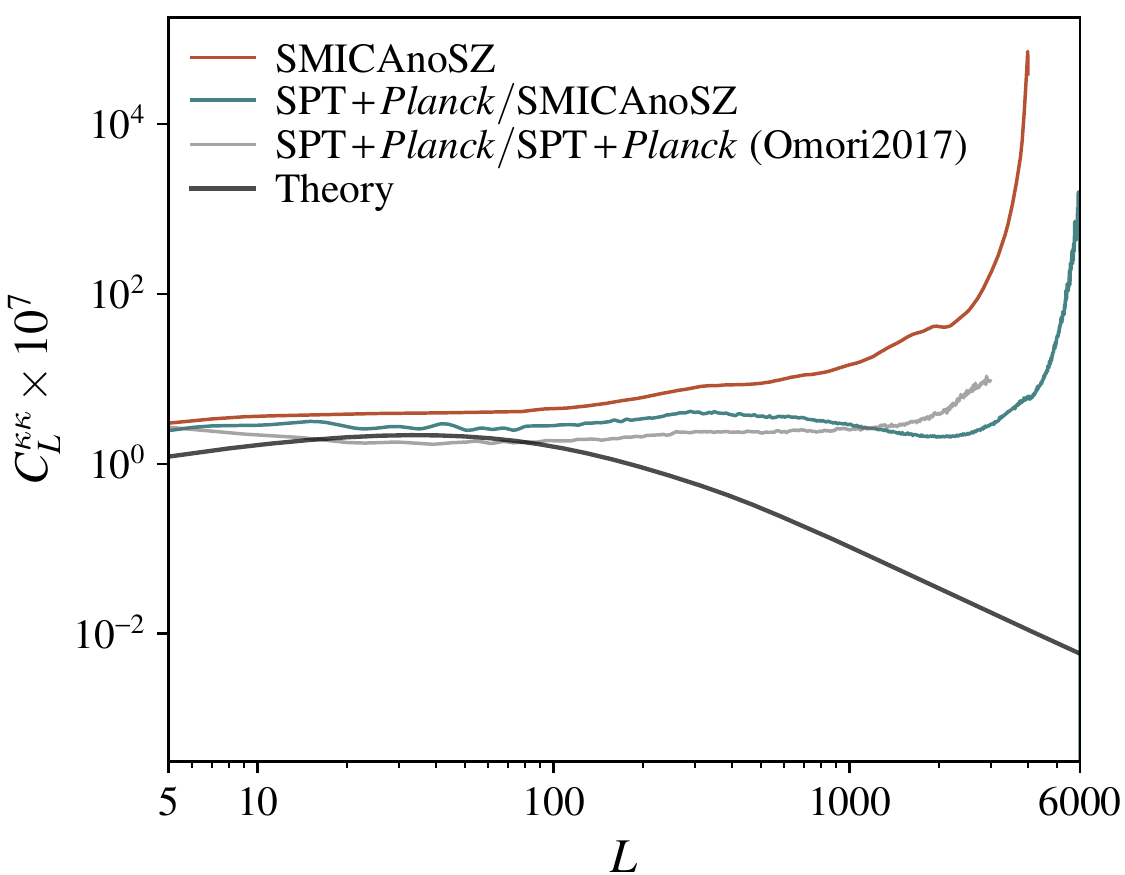}
\caption{Noise levels estimated from simulations for  SPT+{\it Planck}/SMICAnoSZ (teal) and {\it Planck} (orange) over the patch of sky that will be used to measure the cross-correlations. Also shown are the noise levels from \cite{omori2017} (light gray) and an analytical prediction for the convergence signal (black).  The procedures described in Section~\ref{sec:maps} eliminate tSZ contamination from the lensing maps at the cost of a small increase in the map noise (teal vs. light gray).
}
\label{fig:noiselevels}
\end{center}
\end{figure} 

Several approaches to obtaining the lensing response function have been proposed. Here we largely follow \cite{omori2017} in that we use the cross-spectrum with the input simulation:  
\begin{equation}
    \mathcal{R}_{L}^{\phi}=\frac{\langle C_{L}^{\bar{\phi}\phi^{*}}\rangle}{\langle C_{L}^{\phi\phi^{*}}\rangle},
\end{equation}
where $\bar\phi$ is the output reconstructed lensing map, the unbarred $\phi$ are the simulation input lensing potential maps, and the average is taken over the 300 simulation realizations. Our final reconstructed CMB lensing map is shown in Figure~\ref{fig:kcmb_map_ZEA}.   The calculated noise power spectrum of the lensing map is shown in Figure \ref{fig:noiselevels}.

\subsubsection{Validation of the CMB lensing map} \label{sec:kmap_validation}

\begin{figure}
\begin{center}
\includegraphics[width=1.0\linewidth]{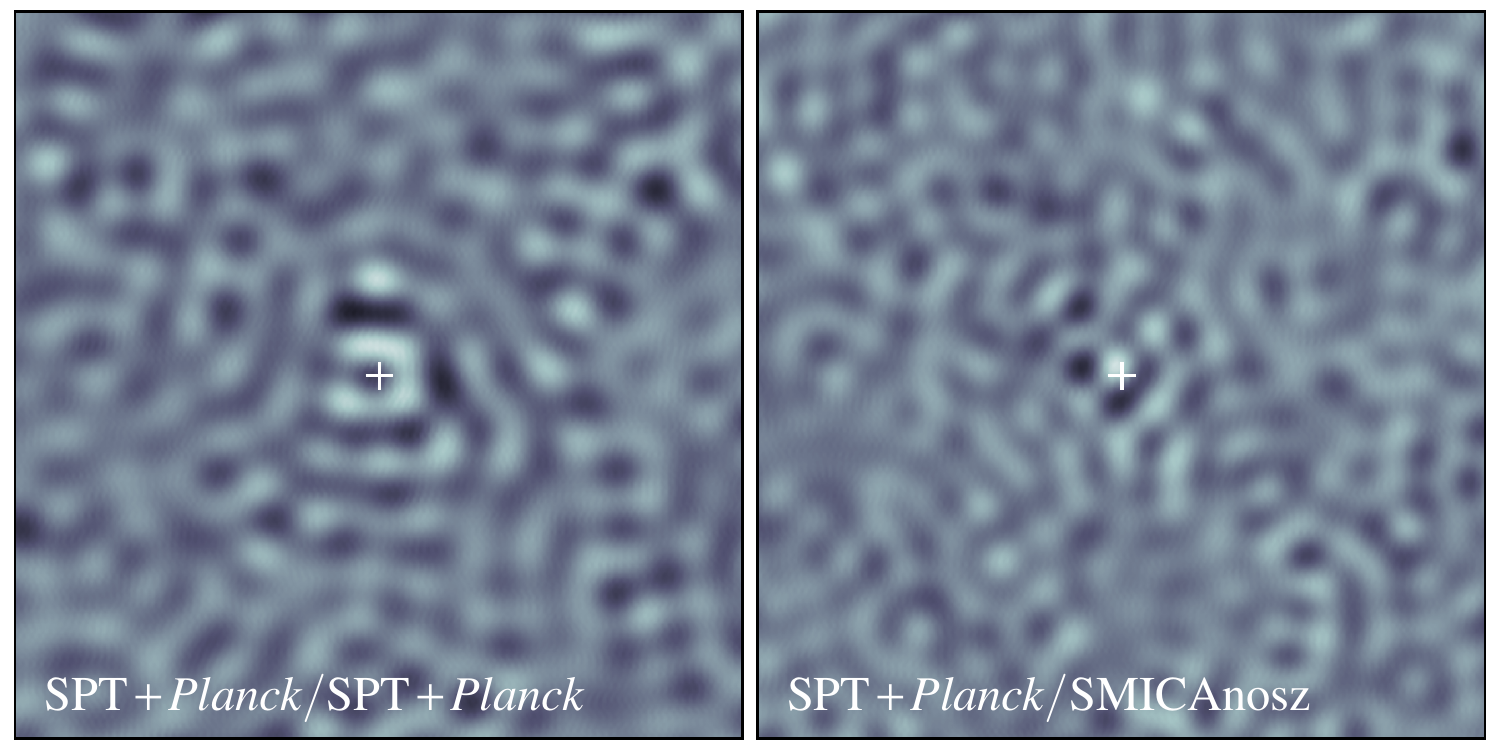}
\caption{Stacks of CMB lensing maps at the locations of clusters from \cite{bleem2015} with signal-to-noise in the range $5<S/N<10$.  Without tSZ nulling (left panel), the stacked CMB lensing map shows a strong feature at the cluster center due to tSZ contamination of the lensing estimator.  With tSZ nulling (right panel), the stacked map shows no strong features at the cluster center, as expected since the cluster lensing signal is weak. 
}
\label{fig:stack_clusters}
\end{center}
\end{figure}

As a test of the level of tSZ contamination in the new CMB lensing maps, we show stacks of the lensing maps at the locations of tSZ-selected clusters from \cite{bleem2015} in Figure~\ref{fig:stack_clusters}. The CMB cluster lensing signal is expected to be very small in SPT-SZ data \cite{Baxter:2015}, so we do not expect to see a significant signal at the cluster location.  However, as a result of tSZ bias, a significant artefact at the cluster location does appear for the map constructed using the SPT+{\it Planck} temperature maps for both legs of the quadratic estimator (left panel).  In contrast, when using the SMICAnosz map for one leg of the estimator, no significant artefact appears at the cluster location.  This suggests that the maps produced in this analysis have reduced the level of tSZ bias. Note that there is also some difference in the noise levels of the two maps, as seen also in Figure~\ref{fig:noiselevels}.

\begin{figure}
\begin{center}
\includegraphics[width=1.00\linewidth]{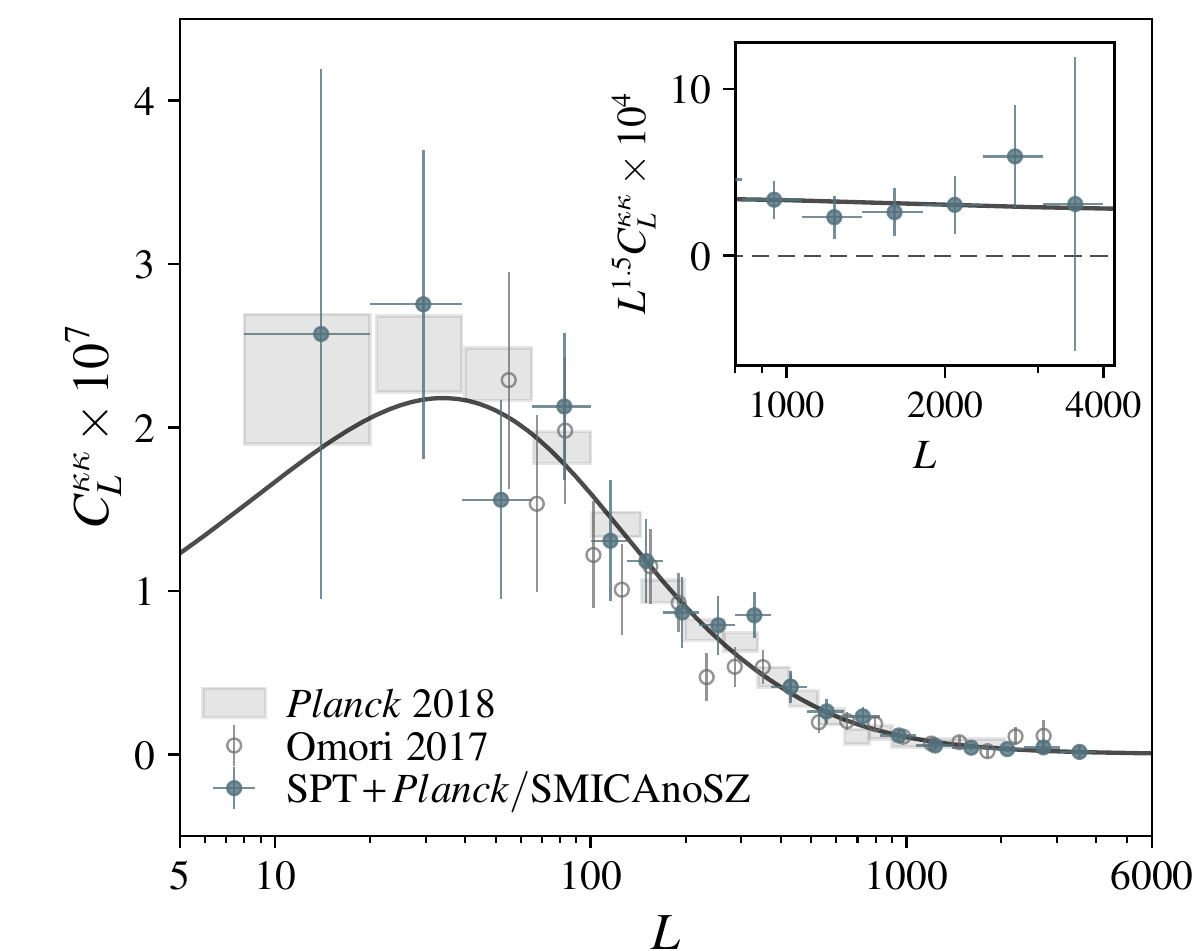}
\caption{The power spectrum of the convergence map constructed from the combination of SPT and {\it Planck} (blue points). Shown for reference are the points from  \cite{planck18-8} (gray squares) and points from \cite{omori2017} (gray open circles), and the analytical convergence power spectrum calculated using the fiducial cosmology assumed in our analysis (black solid line). The inset shows the power spectrum in the high-$L$ range, where possible contamination from the tSZ would show most strongly.} 
\label{fig:clkk}
\end{center}
\end{figure}

We next measure the CMB lensing auto-spectrum and check that it is consistent with that from other studies and theoretical predictions. The formulation of the auto-spectrum calculation is described in Appendix \ref{sec:cmbauto}, and the results are shown in Figure \ref{fig:clkk}. We find that our spectrum is highly consistent with other measurements, and we find no apparent signatures of foreground contamination at small angular scales. We additionally note that due to the inpainting procedure that we carry out prior to the lensing reconstruction, the mask becomes less complex, and the mean-field becomes better characterized, which allows us to reach lower $L$ modes than in \cite{omori2017}.

The procedure of nulling the tSZ in one of the input temperature maps to the quadratic estimator could amplify the CIB in that map (unless the CIB is explicitly nulled, which would result in an additional noise penalty).  This could in turn increase the level of CIB bias in the resultant CMB lensing map.  To test whether CIB contamination is significantly impacting our CMB lensing map, we cross-correlate the map with the \textit{Planck} map at 545~GHz, which is dominated by the CIB.  Since the CIB traces large-scale structure, we expect to detect a non-zero correlation (see also \cite{holder2013,planck13-18,vanengelen2015,maniyar2018,lenz2019,cao2020}).  We therefore compare our measured $\kappa$-CIB correlation with other measurements and predictions from simulations that are known to be uncontaminated by CIB.  The rationale behind this test is that any residual CIB contamination of our new lensing maps will correlate strongly with the CIB, causing the cross-correlation measurement to depart strongly from the predictions of the simulations and previous measurements.  To this end, we compare our measurements with (i) cross-correlation between CIB and the minimum-variance lensing map from SMICA (which has a lower input $\ell_{\rm max}$ cut of $\ell<2048$ in the lensing reconstruction and is therefore less affected by the CIB bias), (ii) cross-correlation between CIB and CMB lensing map of \cite{wu2019} based on the polarization data from SPTpol (since the polarization of CIB is known to be negligible, the bias is expected to be small), and finally (iii) cross-correlation between CIB and pure CMB lensing in simulations \cite{mdpl2synsky}.  

\begin{figure}
\begin{center}
\includegraphics[width=1.0\linewidth]{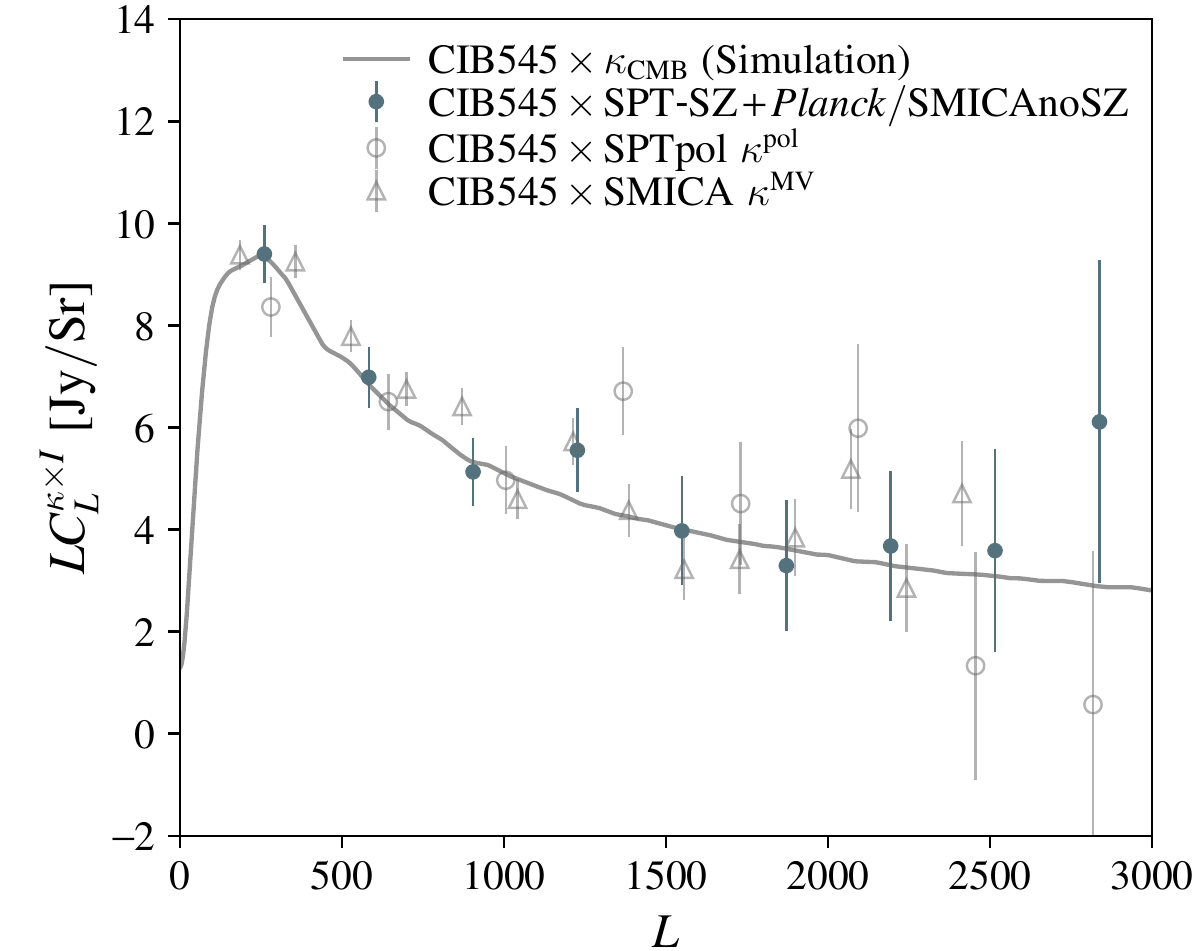}
\caption{Cross-correlation between various CMB lensing maps and the 545 GHz map from \cite{lenz2019}, which is dominated by CIB.  The correlation with the SPT+{\it Planck}/SMICAnoSZ lensing map (blue) produced in this analysis is consistent with other measurements and with an external simulation (gray curve), demonstrating that this lensing map is not significantly contaminated by CIB.}
\label{fig:kcmb-cib}
\end{center}
\end{figure}

The results of the CIB cross-correlation test are shown in Figure~\ref{fig:kcmb-cib}, where it can be seen that our cross-correlation measurement is consistent with all the external measurements.  This suggests that CIB contamination is not significantly biasing our lensing reconstruction.

\subsection{SMICAnoSZ lensing map}
Since the SPT-SZ data only reaches up to Dec=$-40^{\circ}$, we cover the remaining DES Y3 footprint using the {\it Planck} lensing map generated from the SMICA-noSZ temperature map,\footnote{Publicly available at \url{https://pla.esac.esa.int/}. } as shown in Figure \ref{fig:QE_flowchart}.   To simplify the nomenclature of the CMB lensing maps used in this analysis, we  refer to the SPT-SZ+\planck{}/SMICAnoSZ map as the ``SPT+\planck{} lensing map", and the SMICAnoSZ lensing map as the ``\planck{} lensing map" hereafter.

\section{Modeling the CMB lensing cross-correlation functions}
\label{sec:modeling}

The previous section described the construction of a CMB lensing map optimized for cross-correlation with DES data. In this section, we describe our model for the correlations between DES galaxies, galaxy shears and CMB lensing. As mentioned in Section~\ref{sec:intro}, our modeling framework is largely based on the DES Y1 analysis described in \cite{baxter2019}, but with several updates to match the analysis choices of the DES Y3 cosmology analysis \cite{y3-generalmethods}. We therefore only outline the essential modeling components here and refer the readers to the two papers above for details.  

For the remainder of the paper, we use $\delta_{\rm g}$, $\gamma$ and $\kappa_{\rm CMB}$ to refer to the three large-scale structure tracers of interest in this work: galaxy position, galaxy weak lensing (or shear), and CMB lensing convergence, respectively. We will also refer to the galaxies that are used for the galaxy density tracers as {\it lens} galaxies, and the galaxies that have weak lensing shear measurements as the {\it source} galaxies. Ultimately, we will consider the full set of six two-point correlation functions between these three fields. Modeling of correlations between $\delta_{\rm g}$ and $\gamma$ for DES Y3 data is described in detail in \cite{y3-generalmethods}, and we refer readers to that work for more details.  We refer the readers to \cite{planck18-8} for details of the modeling of the {\it Planck} CMB lensing auto-spectrum.

\begin{figure}
\begin{center}
\includegraphics[width=1.0\linewidth]{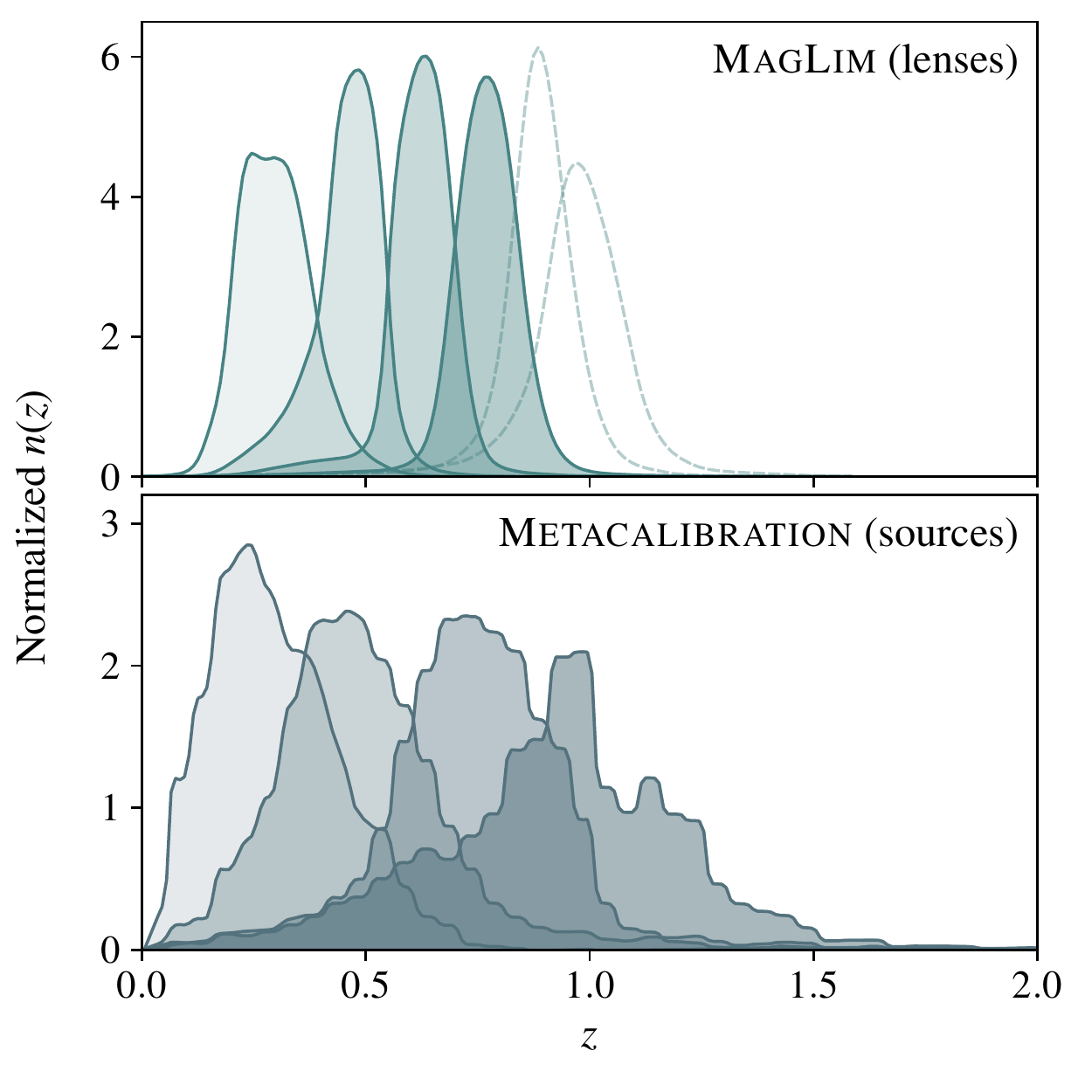}
\caption{Redshift distribution for the \maglim{} lens galaxy sample (upper) and  \metacal{} source galaxy sample (lower). The highest two redshift bins of the lens sample (in dashed lines) are not be used for the forecasting in this work.}
\label{fig:nofz}
\end{center}
\end{figure}

\subsection{Overview of DES galaxy samples}
\label{sec:galaxies}

Unlike analyses with DES Y1 data, the cosmological analyses of DES Y3 data use two different lens galaxy samples: a magnitude-limited sample (\maglim{} \cite{y3-2x2maglimforecast}) and a luminous red galaxy sample (\redmagic{} \cite{rykoff2014, rykoff2016}).  The tomographic bins of the \maglim{} lens sample are shown in Figure~\ref{fig:nofz}, while the number density of objects are listed in Table~\ref{tab:samples}. 

There are known trade-offs for each sample. The \redmagic{} sample was found to give internally inconsistent results: the galaxy bias preferred by galaxy-galaxy lensing was in conflict with that preferred by galaxy clustering \cite{y3-2x2ptbiasmodelling}. The \maglim{} sample, on the other hand, were shown to give poor fits to the baseline model, when the highest two lens galaxy redshift bins were included.  Given these considerations, the baseline DES Y3 cosmology results presented in \cite{y3-3x2ptkp} used only the first four bins of the \maglim{} sample, and we will adopt that approach here for our forecasts. Nevertheless, the methodology developed in this paper is general and can in principle be applied to alternative choices for the lens samples, including the full (i.e.~six tomographic bin) \maglim{} and the \redmagic{} galaxy samples. We will explore these possibilities in our forthcoming data analysis.  

The source galaxy sample used in this work is based on the \metacal{} shape catalog described in \cite{y3-shapecatalog}. The galaxies are divided into four tomographic bins and their redshift distributions are inferred via the \textsc{SOMPZ} method \cite{y3-sompz}; the corresponding distributions are shown in Figure~\ref{fig:nofz}. The number density of galaxies and shape noise estimate for each bin are listed in Table~\ref{tab:samples2}.

\begin{table}
\begin{tabular}{cc*{2}{c}}

\multicolumn{2}{c}{Lens sample}  \\
\midrule 
  Redshift bin& \hspace{0.2cm}$n_{\rm gal}$ (arcmin$^{-2}$) \hspace{0.2cm}   \\ 
\midrule
1 & 0.150 \\ 
2 & 0.107 \\ 
3 & 0.109 \\ 
4 & 0.146 \\ 
\textcolor{gray}{5} & \textcolor{gray}{0.106} \\ 
\textcolor{gray}{6}  & \textcolor{gray}{0.100} \\
 \bottomrule
\end{tabular}
\caption{Effective number density of galaxies in each redshift bin for the \maglim{} lens samples as calculated in \cite{y3-3x2ptkp}. These numbers are used to generate the covariance matrix. The highest two redshift bins will not be used for the forecasting in this work.}
\label{tab:samples}
\end{table}

\begin{table}
\begin{tabular}{ccc}

 & \hspace{-1cm}Source sample  & \\
\midrule 
Redshift bin & \hspace{0.2cm} $n_{\rm gal}$ (arcmin$^{-2}$)\hspace{0.2cm} & $\sigma_\epsilon$ \hspace{0.15cm}  \\ \midrule
1 & 1.672   & 0.247\hspace{0.2cm}  \\ 
2 & 1.695  & 0.266\hspace{0.2cm} \\ 
3 & 1.669  & 0.263\hspace{0.2cm} \\  
4 & 1.682  & 0.314\hspace{0.2cm} \\  
 \bottomrule
\end{tabular}
\caption{Effective number density of galaxies and shape noise for each source redshift bin as calculated in \cite{y3-3x2ptkp}.}
\label{tab:samples2}
\end{table}

\subsection{Galaxy-CMB lensing cross spectra}

We measure two-point functions between the galaxy position, galaxy shape, and CMB lensing observables as a function of angular separation between the points being correlated.  
To model these correlation, we begin by computing the harmonic-space cross-spectra between CMB lensing and galaxy density/shear using the Limber approximation \cite{Limber1953}:
\begin{equation}
C^{\kappa_{\rm CMB} X^{i}}(\ell) =\int d\chi \frac{q_{\kappa_{\rm CMB}}(\chi)q^i_{X} (\chi)}{\chi^2} P_{\rm NL} \left( \frac{\ell+1/2}{\chi}, z(\chi)\right),
\label{eq:Cl_basic}
\end{equation}
where $X \in \{\delta_g, \gamma \}$, $i$ labels the redshift bin, $P_{\rm NL}(k,z)$ is the non-linear matter power spectrum computed using \texttt{CAMB} and \texttt{Halofit} \cite{camb,takahashi2012}, and $\chi$ is the comoving distance to redshift $z$. The window functions, $q_{X}(\chi)$, are given by
\begin{equation}
q^i_{\delta_g}(\chi) = b^i(k,z(\chi)) n_{\delta_{g}}^i(z(\chi)) \frac{dz}{d\chi}
\label{eq:q_deltag}
\end{equation}
\begin{equation}
q^{i}_{\gamma}(\chi) = \frac{3H_0^2 \Omega_{\rm m}}{2c^2} \frac{\chi}{a(\chi)} \int_{\chi}^{\chi_h} d\chi' n_{\gamma}^i(z(\chi')) \frac{dz}{d\chi'}\frac{\chi' -\chi}{\chi'}, 
\label{eq:q_gamma}
\end{equation}
where $H_0$ and $\Omega_{\rm m}$ are the Hubble constant and matter density parameters respectively, $a(\chi)$ is the scale factor corresponding to comoving distance $\chi$, $b(k,z)$ is galaxy bias as a function of scale $k$ and redshift $z$, and $n^i_{{\delta_{\rm g}}/\gamma}(z)$ are the normalized redshift distributions of the lens/source galaxies.

\begin{table*}
\centering 
\begin{tabular}{ccc}
\hline
Parameter & Prior & Fiducial  \\ \hline
$\Omega_{\rm{m}} $ & $\mathcal{U}[0.1, 0.9]$ & 0.3 \\
$A_{\rm{s}}\times 10^{-9}$ & $\mathcal{U}[0.5, 5.0]$ & $2.19$\\
 
$\Omega_{\rm{b}}$ & $\mathcal{U}[0.03, 0.07]$ & 0.048 \\

$n_{\rm{s}}$ & $\mathcal{U}[0.87, 1.07]$  & 0.97\\

$h$ & $\mathcal{U}[0.55, 0.91]$ &  0.69\\

$\Omega_{\nu}h^2 \times 10^{-4} $& $\mathcal{U}[6.0, 64.4]$ & 8.3 \\  
$w$ & $\mathcal{U}[-2, -0.33]$ &-1.0\\ 
\hline
 $a_1$ & $\mathcal{U}[-5.0, 5.0]$ & 0.7\\
 $a_2$ & $\mathcal{U}[-5.0, 5.0]$ & -1.36\\
 $\eta_1$ & $\mathcal{U}[-5.0, 5.0]$ & -1.7\\
 $\eta_2$ & $\mathcal{U}[-5.0, 5.0]$ & -2.5\\
 $b_{\rm{ta}}$ & $\mathcal{U}[0.0, 2.0]$ & 1.0\\ 
\hline
 \maglim{} & & \\
$b^{1\cdots 6}$  & $\mathcal{U}[0.8, 3.0]$ & 1.5, 1.8, 1.8, 1.9, \textcolor{gray}{2.3, 2.3}\\
$b_{1}^{1\cdots 6}$  & $\mathcal{U}[0.66, 2.48]$ & 1.24, 1.49, 1.49, 1.60, \textcolor{gray}{1.90, 1.90}\\
$b_{2}^{1\cdots 6}$  & $\mathcal{U}[-3.41, 3.41]$ & 0.09, 0.23, 0.23, 0.28, \textcolor{gray}{0.48, 0.48}\\
 $C_{g}^{1\cdots 6}$ & fixed & 1.21, 1.15, 1.88, 1.97, \textcolor{gray}{1.78, 2.48} \\
$\Delta_z^{1...6} \times 10^{-2}$ & $\mathcal{N}[0.0, 0.7]$, $\mathcal{N}[0.0, 1.1]$, $\mathcal{N}[0.0, 0.6]$,  & 0.0, 0.0, 0.0, 0.0, \textcolor{gray}{0.0, 0.0} \\
 & $\mathcal{N}[0.0, 0.6]$, $\mathcal{N}[0.0, 0.7]$, $\mathcal{N}[0.0, 0.8]$ & \\
$\sigma_{z}^{1...6}$ & $\mathcal{N}[1.0, 0.062]$, $\mathcal{N}[1.0, 0.093]$, $\mathcal{N}[1.0, 0.054]$ & 1.0, 1.0, 1.0, 1.0, \textcolor{gray}{1.0, 1.0}\\
 &  $\mathcal{N}[1.0,  0.051]$, $\mathcal{N}[1.0,  0.067]$, $\mathcal{N}[1.0,  0.073]$ &  \\
 \hline
 \textsc{Metacalibration} & & \\
 $m^{1...4} \times 10^{-3}$ &  $\mathcal{N}[0.0, 9.1]$, $\mathcal{N}[0.0, 7.8]$, $\mathcal{N}[0.0, 7.6]$, $\mathcal{N}[0.0, 7.6]$ & 0.0, 0.0, 0.0, 0.0 \\
  $\Delta_z^{1...4} \times 10^{-2}$ & $\mathcal{N}[0.0, 1.8]$, $\mathcal{N}[0.0, 1.5]$, $\mathcal{N}[0.0, 1.1]$, $\mathcal{N}[0.0, 1.7]$ & 0.0, 0.0, 0.0, 0.0  \\
 \hline
\end{tabular}

\caption{Fiducial and prior values for cosmological and nuisance parameters included in our model. For the priors, $\mathcal{U}[a,b]$ indicates a uniform prior between $a$ and $b$, while $\mathcal{N}[a,b]$ indicates a Gaussian prior with mean $a$ and standard deviation $b$. The light faded entries are the values corresponding to the last two bins of the \maglim{} sample, not used in the fiducial analysis.}
\label{tab:params_all}
\end{table*}

The angular-space correlation functions are then computed via
\begin{align}
&w^{\delta_g^i \kappa_{\rm CMB}}(\theta) = \sum_{\ell} \frac{2\ell + 1}{4\pi} F(\ell) P_{\ell} (\cos(\theta)) C^{\delta_g^i \kappa_{\rm CMB}}(\ell), \\
&w^{\gamma_{t}^{i} \kappa_{\rm CMB}}(\theta)
=\sum_{\ell}\frac{2\ell+1}{4\pi\ell(\ell+1)}P_{\ell}^{2}(\cos\theta)F(\ell)C^{\kappa^{i}_{\gamma}\kappa_{\rm CMB}}(\ell)
\end{align}
where $P_{\ell}$ and $P^2_{\ell}$ are the $\ell$th order Legendre polynomial and associated Legendre polynomial, respectively, and $F(\ell)$ describes filtering applied to the $\kappa_{\rm CMB}$ maps. For correlations with the $\kappa_{\rm CMB}$ maps, we set $F(\ell)= B(\ell) \Theta(\ell - \ell_{\rm min}) \Theta(\ell_{\rm max} - \ell)$, where $\Theta(\ell)$ is a step function and $B(\ell) = \exp (-0.5\ell(\ell + 1)\sigma^2)$ with $\sigma \equiv \theta_{\rm FWHM}/\sqrt{8 \ln 2}$. The filtering choices ($\Theta_{\rm FWHM}$, $\ell_{\rm min}$ and $\ell_{\rm max}$) for the two $\kappa_{\rm CMB}$ maps are discussed in more detail in Section~\ref{sec:smoothing}.

We calculate the correlation functions within an angular bin $[\theta_{\rm min},\theta_{\rm max}]$ by averaging over the angular bin, i.e., replacing $P_\ell(\cos\theta)$ with their bin-averaged versions $\overline{P_\ell}$ defined by
 \begin{align}
     &\overline{P_\ell}\left(\theta_{\rm min},\theta_{\rm max}\right) \equiv \frac{\int_{\cos\theta_{\rm min}}^{\cos\theta_{\rm max}}dx\,P_\ell(x)}{\cos\theta_{\rm max}-\cos\theta_{\rm min}}\nonumber\\
     &\;\;\;\;\;\;\;\;\;\;\;\;\;\;\;\;\;\;\;\;\;\;\;\;\;\;=\frac{[P_{\ell+1}(x)-P_{\ell-1}(x)]_{\cos\theta_{\rm min}}^{\cos\theta_{\rm max}}}{(2\ell+1)(\cos\theta_{\rm max}-\cos\theta_{\rm min})}~.
\label{eq:bin_average}
 \end{align}

In the following subsections, we describe individual elements in the modeling framework beyond the basic formalism of Equation~\ref{eq:Cl_basic}. 

\subsection{Galaxy bias}
\label{sec:bias}

The \fivetwo{} analysis with DES Y1 data presented in \cite{5x2y1} relied on a linear bias model, where $b(k,z)$ is a constant that is different for each lens galaxy redshift bin.  That model was shown to yield unbiased cosmological constraints for the data analyzed therein.
For the analysis with DES Y3 data, we will use  both a linear galaxy bias model and a nonlinear galaxy bias model.  As we will show, the nonlinear galaxy bias analysis can be applied down to smaller scales than the linear bias analysis, resulting in tighter cosmological constraints. 

Briefly, the two models for the galaxy bias, $b(k,z)$, are: 

\begin{itemize}
    \item {\bf Linear galaxy bias:} We assume that the galaxy bias is independent of scale $b^{i}(k,z) = b^{i}$ and assume one effective bias value $b^{i}$ for each redshift bin. This is our fiducial analysis.
    \item {\bf Nonlinear galaxy bias:} Linear bias is known to break down on small scales, motivating the development of a nonlinear bias model that will allow us to access information on smaller scales. We follow the implementation of nonlinear bias presented in \cite{y3-2x2ptbiasmodelling}, using an effective 1-loop model with renormalized nonlinear bias parameters \cite{McDonaldRoy,Saito_bnl}: $b_1$ (linear bias),  $b_2$ (local quadratic bias), $ b_{s^2} $ (tidal quadratic bias) and $ b_{\rm 3nl} $ (third-order non-local bias). This effect impacts any correlation measured using the galaxy density field (i.e. \nn{}, \ngc{}, \nk{}). Effectively it replaces the galaxy-cross-matter power spectrum ($b P_{\rm NL}$) in Equation~\ref{eq:Cl_basic} with
    \begin{align}
    \label{eq:nlbias}
        P_{g\mathrm{m}}(k)=&\; b_{1} P_{\mathrm{mm}}(k) +  \frac{1}{2} b_{2} P_{ b_1 b_2}(k) \\ \notag 
        &+ \frac{1}{2} b_{s^2}P_{\rm b_1 s^2}(k) + \frac{1}{2} b_{\rm 3nl}P_{ b_1 b_{\rm 3nl} }(k).
    \end{align}
    Expressions for the power spectrum kernels $P_{b_1 b_2},$ etc., are given in \cite{Saito_bnl,fang2017}. 
\end{itemize}
The priors and ranges for the values $b^{i}$,$b_{1}^{i}$ and $b_{2}^{i}$ used in this analysis are summarized in Table \ref{tab:params_all}.

\subsection{Lensing magnification}

In addition to distorting or shearing shapes of galaxies, weak lensing also changes the observed flux, size and number density of the galaxies --- effects referred to as magnification \citep[see e.g.][]{Bartelmann2001}.  Magnification was ignored in the \fivetwo{} analysis with DES Y1 data presented in \cite{5x2y1}.   
Here, we ignore the impact of magnification on the shear-CMB lensing correlation, as the impact of source galaxy magnification is expected to be very small compared to our statistical precision \citep{y3-generalmethods}.  We do, however, incorporate the impact of magnification on the galaxy density-CMB lensing correlations.  Following \cite{y3-generalmethods}, we consider the change in projected number density due to geometric dilution as well as magnification effects on galaxy flux \citep{VernerV1995, Moessner1998} and size \citep{sizebias}, which modulate the selection function.

The effect of magnification can be modeled by modifying Equation~\ref{eq:q_deltag} to include the change in selection and geometric dilution quantified by the lensing bias coefficients $C_{\rm  g}^i$.
\begin{equation}
q^i_{\delta_{\rm g,mag}}(\chi) = q^i_{\delta_{\rm g}}(\chi) (1+C^{i}_{\rm g}\kappa^{i}_{\rm g }),
\label{eq:magnification}
\end{equation}
where 
\begin{equation}
C_{\rm g}^i = 5 \frac{\partial \ln n_{\rm g}^i}{\partial m}\bigg\rvert_{m_\mathrm{lim},r_\mathrm{lim}} +\frac{\partial \ln n_{\rm g}^i}{\partial \ln r}\bigg\rvert_{m_\mathrm{lim},r_\mathrm{lim}} -2,
\end{equation}
(here $m$ and $r$ represents the observed magnitude and radius respectively) and $\kappa_{\rm g}^i$ is the tomographic convergence field, as described in \cite{y3-generalmethods}.
The logarithmic derivatives are the slope of the luminosity and size distribution at the sample selection limit. The values of these lensing bias coefficients are estimated in \citep{y3-2x2ptmagnification} and fixed to the values listed in Table~\ref{tab:params_all}. 

\subsection{Intrinsic alignments}

The \fivetwo{} analysis with DES Y1 data considered the nonlinear alignment model \citep[NLA,][]{Hirata2004,Bridle2007} for galaxy intrinsic alignments (IA).   For the present analysis, we adopt the more flexible tidal alignment tidal torquing model (TATT) of \cite{blazek2019} to describe IA; more details of this model and its implementation in the context of DES Y3 cosmology analyses can be found in \cite{y3-generalmethods}.  
In this model, the intrinsic galaxy shape $\tilde{\gamma}_{\alpha,\mathrm{IA}}$, measured at the location of source galaxies, can be written as an expansion in the density $\delta_{\rm m}$ and tidal tensor $s_{ab}$, which can be decomposed into components $s_{\alpha}$:
\begin{equation}
\tilde{\gamma}_{\alpha,\mathrm{IA}} = A_1 s_{\alpha} +A_{1\delta} \delta_\mathrm{m} s_{\alpha} + A_2 \left(s\times s\right)_{\alpha} + \cdots\,.
\label{eq:TATT}
\end{equation}
The coefficients for the three terms in Equation~\ref{eq:TATT} can be expressed as follows:
\begin{align}
A_1(z) =&\ -a_1\bar{C}_1\frac{\rho_\mathrm{crit}\Omega_{\rm m}}{D(z)}\left(\frac{1+z}{1+z_0}\right)^{\eta_1}\\
A_{1 \delta} (z) =&\ b_{\rm ta} A_1 (z)  \\
A_2(z) =&\ 5a_2\bar{C}_1\frac{\rho_\mathrm{crit}\Omega_{\rm m}}{D(z)^2}\left(\frac{1+z}{1+z_0}\right)^{\eta_2},
\end{align}
where $\rho_{\rm crit}=H^{2}/8\pi G$ is the critical density of the universe, $z_{0}$ is a pivot scale fixed by convention, $\bar{C}_1$ is a normalization constant, which is fixed to $\bar{C}_1 = 5\times10^{-14} M_\odot h^{-2}\mathrm{Mpc}^2$, and $D(z)$ is the linear growth factor.

We use a total of five free parameters to describe IA: $a_1$, $\eta_1$, $a_2$, $\eta_2$, and $b_{\rm ta}$ and use flat priors as summarized in Table \ref{tab:params_all} .

\subsection{Smoothing of the CMB $\kappa$ map}
\label{sec:smoothing}
The noise power spectrum of the CMB lensing maps increases in amplitude at small scales.  Large-amplitude small-scale noise significantly impacts the covariance of the angular-space correlation function measurements that we consider in this analysis, making covariance computation difficult.  To reduce the effect of small-scale noise, we apply Gaussian smoothing and low-pass filtering to the CMB lensing maps.  This changes the expectation values of the correlation functions, but should not bias our analysis because we include the impact of filtering in our model.  The impact of the Gaussian smoothing amounts to a transformation of the cross spectra:
\begin{equation}
C_{\ell}^{\kappa_{\rm CMB}X} \rightarrow C_{\ell}^{\kappa_{\rm CMB}X} B_{\ell},
\end{equation}
where  $B_{\ell} = {\exp}(-\ell(\ell+1)\sigma^2)$ is the smoothing function and $\sigma=\theta_{\rm FHWM}/\sqrt{8{\ln 2}}$. For SPT+\planck{} and $\planck$ we use $\theta_{\rm FWHM}$ of $6'$ and $8'$ respectively. 
We additionally apply low-pass filtering to the maps, with $\ell_{\rm max} = 5000$ for the SPT+{\it Planck} lensing map and $\ell_{\rm max} = 3800$ for the {\it Planck}-only map.  The combination of the filtering and the smoothing ensures that the noise power spectrum of the filtered maps approaches zero at $\ell_{\rm max}$.

\subsection{Uncertainty in shear calibration and redshift distributions}

We model shear calibration and redshift biases for the DES galaxies as described in \cite{y3-generalmethods}.  We model shear calibration biases with a multiplicative factor such that the observed $C^{\kappa_{\rm CMB}\gamma}$ is modified by
\begin{equation}\label{eq:shear_bias}
C^{\kappa_{\rm CMB}\gamma^{i}}(\ell) \rightarrow (1+m^{i})C^{\kappa_{\rm CMB}\gamma^{i}}(\ell),\nonumber
\end{equation}
where $m^{i}$ is the shear calibration bias for source bin $i$.

Following \cite{y3-3x2ptkp}, our fiducial analysis models the uncertainty in the source galaxy redshift distributions with shift parameters, $\Delta_z^i$, where $i$ labels the redshift bin.  This parameter modifies the $n(z)$ as
\begin{equation}
\label{eq:nz}
n^{i}(z) \rightarrow n^{i}(z-\Delta^{i}_{z}).
\end{equation}

For the lens sample, we additionally introduce a stretch parameter ($\sigma_{z}$) in the redshift distribution such that (combining with the effect above):
\begin{equation}
n^{i}(z) \rightarrow \sigma_{z}^{i}n^{i}(\sigma_{z}^{i}[z-\langle z \rangle]+\langle z \rangle-\Delta_{z}^{i}).
\end{equation}
The fiducial values and priors used for $\sigma_{z}^{i}$ and $\delta_{z}^{i}$ are summarized in Table \ref{tab:params_all}.

We also consider an alternative method for parameterizing uncertainty in the redshift distributions known as \textsc{Hypperrank} \cite{y3-hyperrank}, which efficiently marginalizes over possible realizations of the redshift distributions. For the 3$\times$2pt analysis presented in \cite{y3-3x2ptkp}, \textsc{Hypperrank} was shown to give similar results as the simpler model shown in Equation~\ref{eq:nz}.  We verify that this is also the case for 5$\times$2pt in Appendix~\ref{sec:hyperrank}.

\section{Model Fitting}
\label{sec:model_fitting}

We adopt a Gaussian likelihood, $\mathcal{L}(\vec{d}| \vec{\theta})$, for analyzing the data:
\begin{eqnarray}
\label{eq:likelihood}
\ln \mathcal{L}(\vec{d}| \vec{\theta}) = -\frac{1}{2} \left[ \vec{d} - \vec{m}(\vec{\theta})\right]^T \mathcal{C}^{-1} \left[ \vec{d} - \vec{m}(\vec{\theta}) \right],
\end{eqnarray}
where $\vec{d}$ is the vector of observed correlation function measurements, $\vec{m}(\theta)$ is the vector of model predictions at parameter values $\vec{\theta}$, and $\mathcal{C}$ is the covariance matrix of the data.  The posterior on the model parameters is then given by 
\begin{eqnarray}
\mathcal{P}(\vec{\theta} | \vec{d}) = \mathcal{L}(\vec{d}| \vec{\theta}) \mathcal{P}(\vec{\theta}),
\end{eqnarray}
where $\mathcal{P}(\vec{\theta})$ are the priors on model parameters.  We summarize the priors on model parameters in Table~\ref{tab:params_all}. All values are consistent with those used in \cite{y3-3x2ptkp}.

\subsection{Covariance}
\label{sec:covariance_model}

Computing the likelihood in Equation~\ref{eq:likelihood} requires an estimate of the data covariance matrix.  For the block of this matrix consisting of DES-only cross-correlations (i.e. 3$\times$2pt), we use the halo model covariance described in \cite{y3-covariances}.  For the blocks involving cross-correlations with CMB lensing, we adopt a lognormal covariance model based on \cite{y3-covariances}.  We briefly describe the lognormal covariance model below.

In the lognormal model, the galaxy overdensity, galaxy lensing, and CMB lensing fields are modeled as shifted lognormal random fields \cite{hilbert2011}.  These are specified by 
\begin{eqnarray}
X = \lambda (e^{n+\mu}-1),
\end{eqnarray}
where $n$ is a Gaussian random field with mean zero, and $\lambda$ is the so-called shift parameter.  The power spectrum of $n$ can be chosen so that the power spectrum of $X$ matches that of the desired field (computed from our theory model), and $\mu$ can be chosen such that $\langle X \rangle  = 0$, leaving $\lambda$ to be specified.  

\cite{friedrich2018} and \cite{y3-covariances} describe a procedure for determining $\lambda$, and we follow a similar procedure here.  In particular, we choose the value of $\lambda$ so that the re-scaled cumulant of the log-normal field,
\begin{equation}
S_{3}(\vartheta)\equiv\frac{\langle X(\vartheta)^{3} \rangle}{ \langle X(\vartheta)^{2} \rangle^{2} },
\end{equation}
matches that predicted by leading order perturbation theory, where $\vartheta$ is a choice of smoothing scale.  Here we set $\vartheta = 10'$, and $\lambda$ is chosen separately for each field  ($\lambda=1.089,1.106,1.046,1.252,1.177,1.177$ for the  6 \maglim{} lens redshift bins, $\lambda=0.866,1.956,1.075,1.1486$ for the 5 \redmagic{} lens redshift bins, $\lambda=0.033,0.085,0.021,0.033$ for the 4 sources redshift bins and $\lambda=2.7$ for CMB lensing field).

The covariance of lognormal weak lensing fields can be written as the sum of a Gaussian contribution and higher-order covariance terms \cite{hilbert2011}.  \cite{y3-covariances} took these results and generalized them to describe the covariance of arbitrary fields:\footnote{This is an approximation retaining only the first order term after the Gaussian covariance term.}
\begin{align}
&\mathcal{C}_{\rm LN} \sim\mathcal{C}_{\rm G}[\xi_{ X_{\rm a}X_{\rm b} }, \xi_{X_{\rm c} X_{\rm d} } ] + \frac{\xi_{ X_{\rm a}X_{\rm b} }(\theta_{1})\xi_{ X_{\rm c}X_{\rm d} }(\theta_{2})}{A_{\rm survey}}\times\nonumber\\
&\bigg\{
\frac{\mathcal{C}_{S}(X_{\rm  a},X_{\rm  c}) }{\lambda_{\rm a}\lambda_{\rm c}} + 
\frac{\mathcal{C}_{S}(X_{\rm  a},X_{\rm  d}) }{\lambda_{\rm a}\lambda_{\rm d}} +
\frac{\mathcal{C}_{S}(X_{\rm  b},X_{\rm  c}) }{\lambda_{\rm b}\lambda_{\rm c}} + 
\frac{\mathcal{C}_{S}(X_{\rm  b},X_{\rm  d}) }{\lambda_{\rm b}\lambda_{\rm d}}
\bigg\},
\end{align}
where $A_{\rm survey}$ is the survey area (in particular we use the effective overlapping area between the galaxy and CMB surveys), and $\lambda$ are the shift parameters for the fields ${\rm a,b,c,d}$, and $\mathcal{C}_{S}$ denotes the covariance between two fields  after the two fields have been averaged over the entire survey footprint.

\begin{figure}
\begin{center}
\includegraphics[width=1.0\linewidth]{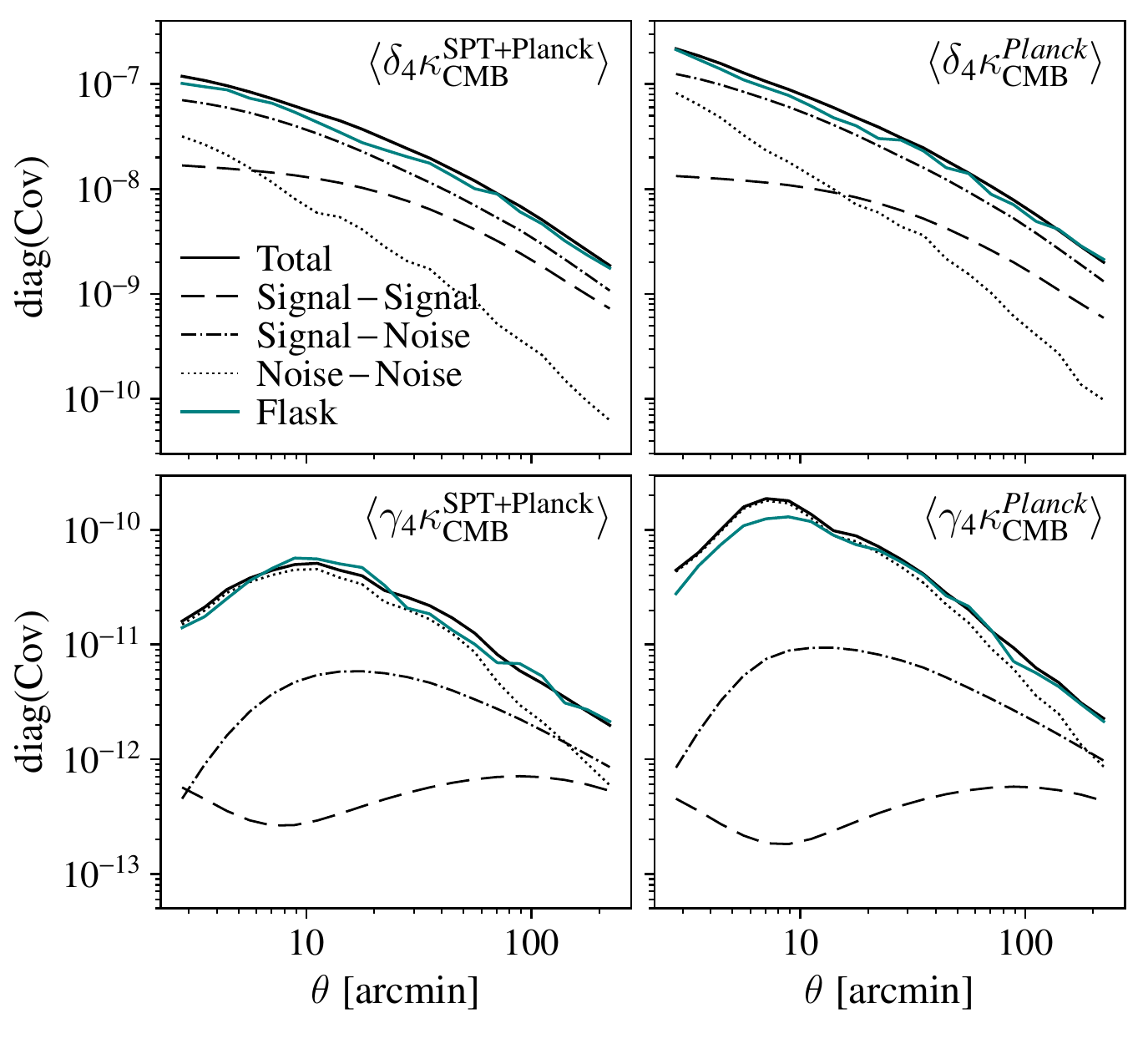}
\caption{Decomposition of the diagonal of the covariance matrix into the various terms in Eq.~\ref{eq:cov_breakdown}.  Results are shown for an arbitrary bin (bin four for both lens and source), but appear similar in other bins.  We also overlay the total covariance measured from the \textsc{Flask} simulations, described in Section~\ref{sec:nkgk_cov}.}
\label{fig:breakdown}
\end{center}
\end{figure}

Unlike the shot noise and shape noise that impact $\delta_g$ and $\gamma$, respectively, the CMB lensing noise varies strongly as a function of multipole.  For this reason, we adopt a special procedure to improve our estimate of noise contributions to the covariance matrix.   We note that without this treatment, the covariance validation tests described in Section~\ref{sec:nkgk_cov} do not pass. We decompose the total covariance into contributions from signal and noise:
\begin{equation}\label{eq:cov_breakdown}
\mathcal{C}_{\rm total}= \mathcal{C}_{\rm signal-signal} + \mathcal{C}_{\rm noise-noise} +  \mathcal{C}_{\rm signal-noise}. 
\end{equation}
The first two terms can be isolated by setting either the signal or noise power to zero; $\mathcal{C}_{\rm signal-noise}$ can be obtained by subtracting the signal-signal and noise-noise terms from the total covariance. 

Owing to the non-white power spectrum of the CMB lensing noise and the complexities of the DES mask, we compute the noise-noise term in Equation \ref{eq:cov_breakdown} using many noise simulations.   This approach takes into account the impact of the survey geometry.  Furthermore, in the case of the CMB lensing map, since the noise realizations are generated using the real data, this approach captures possible inhomogeneity in the noise over the sky area.   For the lens galaxies, we generate noise catalogs by drawing from the random point catalogs used to characterize the survey selection function.  We draw the same number of random points in the survey footprint as the number of galaxies in the data catalog.  For the galaxy weak lensing field, we take the data shear catalog and apply a random rotation such that \cite{mandelbaum2012}: \begin{align}
 e^{\rm rot}_{1}&=e_{1}'\cos(2\varphi)+e'_{2}\sin(2\varphi)\label{eq:random_rot_shear1},\\
 e^{\rm rot}_{2}&=-e_{1}'\sin(2\varphi)+e'_{2}\cos(2\varphi)\label{eq:random_rot_shear2},
 \end{align}
where $e'_{1},e'_{2}$ are the measured ellipticity components, and $\varphi$ is some random angle between  0 and $2\pi$.  We treat these rotated ellipticities as the noise.  For CMB lensing, our estimate of noise realizations is formed from the difference between reconstructed lensing maps from simulations (which include noise) and the noiseless input lensing maps that were used to lens the simulated temperature maps.   We use 300 noise realizations, since this is the number of noise realizations provided for the  {\it Planck} lensing maps.

The \nk{} and \gk{} cross-correlations are then measured for each of the 300 noise realizations and the covariance matrix across these realizations is computed.
The relative amplitudes of the covariance contributions as a function of angular scale are shown in Figure~\ref{fig:breakdown}.  While we only show the decomposition for one redshift bin, similar behavior is found for the other redshift bins. For \nk{}, the dominant term at all scales is the signal-noise term (this results from the relative amplitudes of the signal/noise terms for $\delta_g$ and $\kappa_{\rm CMB}$), and the signal-signal term is larger than the noise-noise term at large scales. For \gk{}, most of the angular bins are dominated by the noise-noise term.

\begin{figure}
\begin{center}
\includegraphics[width=1.00\linewidth]{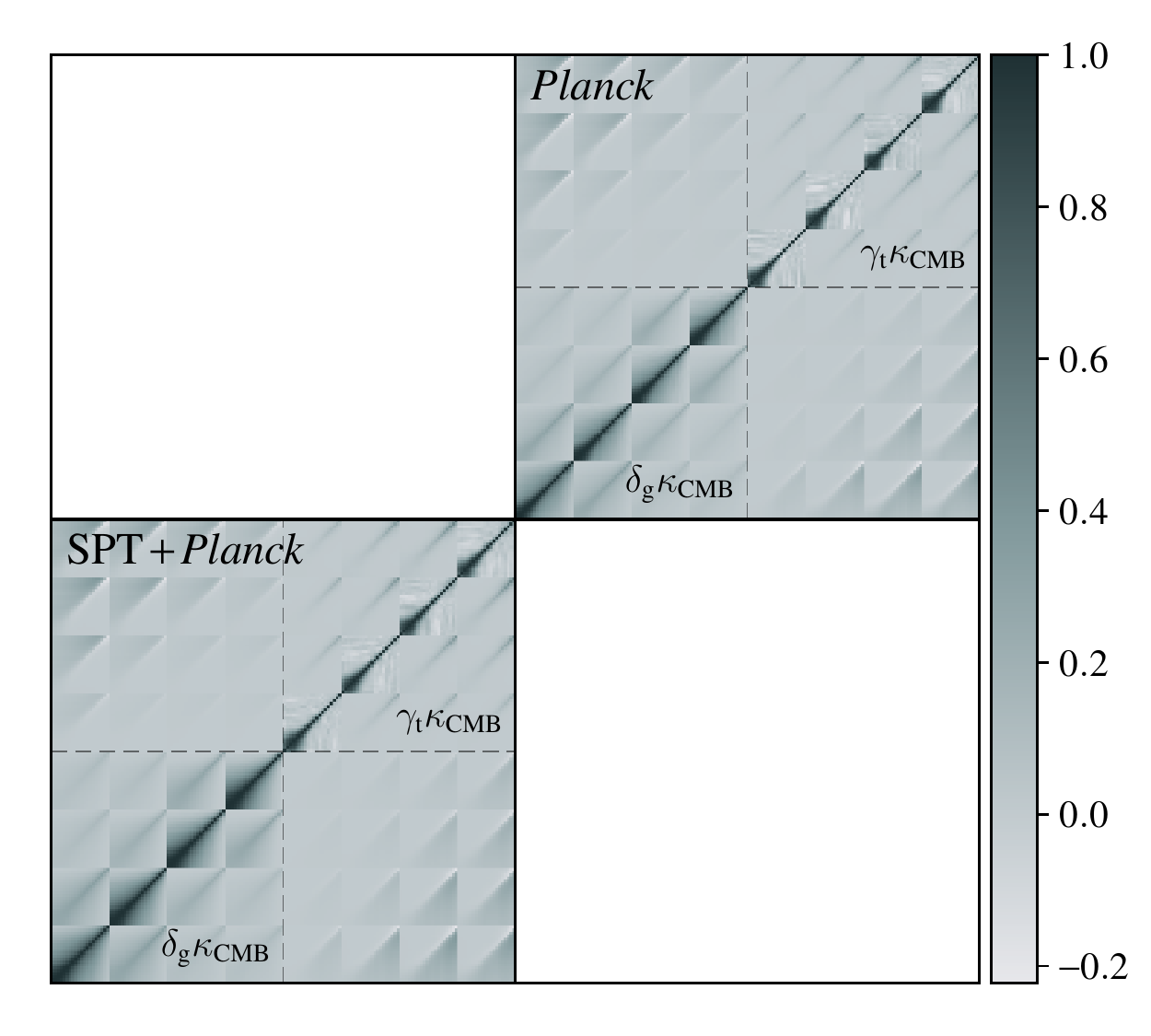}
\caption{Plot of the \nkgk{} correlation matrix. The off-diagonal cross (SPT+\planck{})-(\planck{}) blocks are set to zero as discussed in Section \ref{sec:spt_planck_independence}. Each \nk{} block has 100 elements (5 redshift bins with 20 angular bins each) and \gk{} has 80 elements (4 redshift bins with 20 angular bins each).}
\label{fig:cov_5x2pt}
\end{center}
\end{figure}

To complete our estimate of the covariance matrix, we must also determine the covariance between the SPT+\planck{} and \planck{} sky patches, and the covariance between \nk{} and \gk{} with the \threetwo{} correlations.  The covariance between the non-overlapping SPT+\planck{} and \planck{} sky patches is expected to be small, and we will take the approach of setting it to zero.  The validity of this approximation is tested in the next section.  To compute the covariance between \nk{} and \gk{} with the \threetwo{} data vector measured over the full DES patch, we rely on the log-normal covariance estimate.  We further make the approximation that each patch (SPT+\planck{} or \planck{}) only correlates with the \threetwo{} measurements over the overlapping fraction of sky, and that the measurement of the total \threetwo{} data vector can be expressed as a weighted combination of \threetwo{} measurements in the different patches.  The weights are assumed to be proportional to the corresponding sky areas. This approximation and a similar calculation is discussed in Appendix G of \cite{vanUitert:2018}.  We show the final correlation matrix for the \nkgk{} part in Figure~\ref{fig:cov_5x2pt}.

We note that the \threetwo{} analysis presented in \cite{y3-3x2ptkp} included a modification to the covariance matrix which accounts for possible variation in the galaxy-matter correlation at small scales \citep{y3-generalmethods}.  The galaxy-tangential shear correlation is a non-local quantity such that its value at a given angular scale depends on the galaxy-matter power spectrum down to arbitrarily small scales.  Using the technique developed in \cite{MacCrann:2020}, the analysis in \cite{y3-3x2ptkp} effectively marginalizes over a ``point mass'' contribution to the galaxy-tangential shear correlation at small scales by introducing a modification to the covariance matrix.  Our analysis of the galaxy-convergence correlation, on the other hand, need not account for a point mass contribution because convergence is a local quantity.  One caveat is that the application of smoothing to the convergence map introduces some non-locality.  However, because our angular scale cuts (see Section~\ref{sec:analysischoices}) remove angular scales comparable to the smoothing scale, this is not a worry for our analysis.  In principle, since the \gk{} correlation is also non-local, we could adjust its covariance to account for a point mass contribution.  However, since the signal-to-noise of the \gk{} correlation at small scales is low, we do not expect this to have a significant impact on our analysis.  Furthermore, as we demonstrate in Section~\ref{sec:analysischoices}, our analysis of \gk{} is robust to variations in the matter power spectrum caused by baryonic feedback.  We therefore do not include a point mass contribution to the covariance matrix for \gk{} in our analysis.

\subsection{Validation of the covariance matrix}
\label{sec:nkgk_cov}

\subsubsection{$\chi^2$ test}
\label{sec:chi2}

\begin{figure*}
\begin{center}
\includegraphics[width=0.85\linewidth]{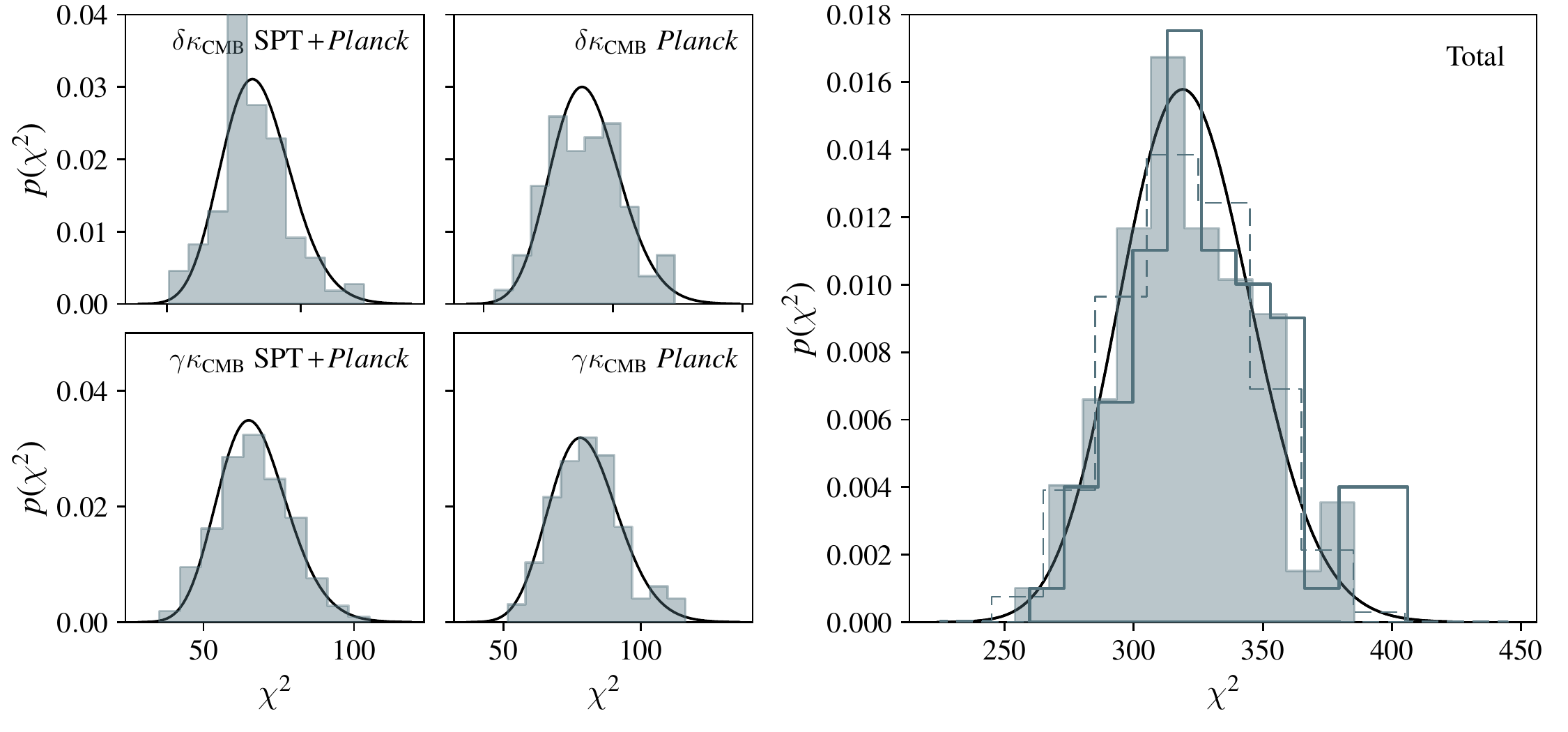}
\caption{\textit{Left:} Distribution of $\chi^2$ derived from \textsc{Flask} simulations and our covariance model for \nk{} and \gk{} data vectors in the SPT+{\it Planck} and {\it Planck} patches separately. The histograms are overlaid with a $\chi^{2}_{\nu}$ distribution (smooth black curve), and we only include data points after the scale cuts (see Section~\ref{sec:analysischoices}). {\it Right:} Same as the left panel but for the combined data vector of \nk{} and \gk{} in both patches (filled histogram). The open gray histogram represents the $\chi^2$ distribution prior to applying the 4\% correction and the dashed histogram corresponds to the $\chi^2$ distribution when combining different realizations for the SPT+{\it Planck} and {\it Planck} patches, effectively nulling the off-diagonal block of the covariance in the \textsc{Flask} realizations, as described in Section \ref{sec:spt_planck_independence} }
\label{fig:chi2}
\end{center}
\end{figure*}

As a test of the covariance matrix that we obtained in the previous section, we first show that using this covariance matrix recovers the correct $\chi^{2}$ distribution from a set of simulated data vectors.  To do this, we first generate simulated realizations of the galaxy position, galaxy weak lensing, and CMB lensing fields (see description of these simulations below).  For each simulation, $i$, we calculate the two-point correlation functions, $D_i$.  The $\chi^2$ is then computed via:
\begin{equation}
\label{eq:chi2_covtest}
    \chi^2_i = (D_i-M)^{T}\mathcal{C}^{-1}(D_i-M), 
\end{equation}
where $M$ is the true correlation function (which is known for the simulations), and $\mathcal{C}$ is the covariance matrix described in Section~\ref{sec:covariance_model}. If $\mathcal{C}$ is indeed a good estimation of the covariance matrix for $D$, we expect the distribution of $\chi_i^2$ to follow a $\chi^{2}_{\nu}$ distribution with $\nu$ equal to the dimensionality of $D$.  

This procedure tests several aspects of the covariance calculation.  First, it ensures that our approximation that the CMB lensing noise is uniform across the SPT+{\it Planck} and {\it Planck} patches is a good approximation (which is assumed for the signal-noise term in the covariance), since the simulated data vectors include non-uniformity in the noise. Second, this test validates our assumption that cross-covariance between observables computed from the SPT+{\it Planck} and {\it Planck} patches of the CMB lensing map can be ignored. Finally, it confirms that our treatment of survey geometry is sufficient to model the data covariance.  We note when computing the $\chi^2$ in these tests, we impose angular scale cuts that remove small-scale measurements. These cuts will be described in the next section.

The simulated data used for the $\chi^2$ covariance test are generated from log-normal realizations of the lens catalog (galaxy position), the source catalog (galaxy position and shape), and the CMB lensing map using the package \textsc{Flask} \cite{xavier2016}. We start with generating a set of noiseless maps of the galaxy density, galaxy lensing and CMB lensing fields given all the combinations of auto- and cross-correlation power spectrum $C_{\ell}$ as well as lognormal shift parameters associated with each field. 
The lens catalog is generated by Poisson sampling with expectation $N=\bar{n}(1+\delta)$, where $\bar{n}$ is the average galaxy density per pixel, and $\delta$ is the density field generated by \textsc{Flask} (which already includes the galaxy bias).  For the source catalog, we use the same random rotation approach described in Equations \ref{eq:random_rot_shear1} and \ref{eq:random_rot_shear2} on the DES Y3 galaxy shape catalog \citep{y3-shapecatalog}.  Shape noise obtained this way is added to the shear signal extracted from the \textsc{Flask} galaxy weak lensing maps evaluated at the locations of observed galaxies.\footnote{We note that this is a good approximation in the weak lensing regime. Formally, the galaxy ellipticity changes under an applied shear according to e.g. Equation 4.12 of \citep{Bartelmann2001}.}  For the CMB lensing map, we add the difference between the reconstructed lensing map and the input convergence map  to the noiseless \textsc{Flask} CMB lensing map, then apply the same filtering and smoothing to the maps as the data (described in Section \ref{sec:smoothing}).  We then compute the \nk{} and \gk{} data vectors from these simulations, and evaluate the $\chi^2$ with respect to the fiducial model as in Equation~\ref{eq:chi2_covtest}.  

Upon measuring the $\chi^{2}$ distribution from the flask realizations, we have found that the distribution is marginally skewed towards higher $\chi^{2}$ than we would expect.   To alleviate this, we have scaled up the \gk{} covariance by a small amount (4\%) such that the $\chi^2$ distribution matches with expectations, and we subsequently use this covariance in the analysis. The results of the covariance $\chi^2$ distribution test are shown in Figure~\ref{fig:chi2}. The four panels on the left show the $\chi^2$ distributions separately for the two patches of sky and for \nk{} and \gk{} (combining all redshift bins). We see that individually, all of them show good agreement with an analytical $\chi^2$ distribution.  The right panel shows the $\chi^2$ distribution for the combined data vector, which includes the cross-covariance between the two patches of the sky and between \nk{} and \gk.

\subsubsection{The independence of SPT+\planck{} and \planck{} patches}\label{sec:spt_planck_independence}
In the covariance we described in the previous section, we have assumed that the covariance between the patches is zero (i.e. the empty blocks in Figure \ref{fig:cov_5x2pt}). We further test this assumption using the \textsc{Flask} data vectors. The full \textsc{Flask} data vector includes the correlation between the patches since they were measured from catalogs generated from the same sky realization. We create a set of ``shuffled" data vectors, in which the the SPT+\planck{} patch data vectors from one sky realization are combined with the \planck{} patch data vectors from a different realization, and we compute $\chi^{2}$ or each of these sets of shuffled data vectors and original (correlated) data vectors. The comparison of the two $\chi^{2}$ distributions is shown in Figure~\ref{fig:chi2}. We see no significant differences in the two distributions, and we conclude that the ignoring the off-diagonal blocks is valid.

\subsubsection{The independence of \fivetwo{} and \planck{} full sky}\label{sec:kk_cov}

The end goal of this analysis is to perform a joint analysis of the \fivetwo{} data vector and the CMB lensing auto-spectrum as measured by {\it Planck}.  Since the sky area that DES observes lies within the sky area that was used for the {\it Planck} CMB lensing analysis, we expect the measurements to be correlated to some degree. In this section, we examine the degree of correlation.

There are several reasons to expect the covariance between the full-sky CMB lensing auto-spectrum from {\it Planck} and the \fivetwo{} data vector to be negligible. First, the CMB lensing auto-spectrum is most sensitive to redshift $z \sim 2$. The \fivetwo{} data vector, on the other hand, is most sensitive to structure at $z \lesssim 1$, because this is the regime probed by DES galaxy positions and shapes. Secondly, the bulk of information in the {\it Planck} CMB lensing auto-spectrum analysis is derived from outside the patch of sky over which we measure \fivetwo{} -- the overlap is approximately 15\% of the {\it Planck} lensing analysis area.  Finally, we note that over the SPT-SZ patch, the bulk of the lensing information comes from SPT-SZ data, which has instrumental noise that is uncorrelated with the \planck{} observations.

To determine whether the covariance between \fivetwo{} and the \planck~lensing auto-spectrum can be ignored, we proceed as follows. First, we compute the theoretical cross-covariance between the \fivetwo{} and full-sky CMB lensing angular-space auto-spectrum using the log-normal formulation  described in Section \ref{sec:covariance_model}.   We must account for the fact that \fivetwo{} is measured over a small patch of sky, while the CMB lensing auto-spectrum is measured over (nearly) the full-sky. To do this, we make the approximation that the full-sky CMB lensing measurements can be expressed as an inverse-variance weighted average of measurements inside the DES patch and outside of that patch, and that the covariance between \fivetwo{} and the outside-the-patch CMB lensing auto-spectrum measurements can be ignored.

Once the full \sixtwo{} covariance has been computed, we compute the likelihood of a \sixtwo{} datavector with and without setting the cross-covariance between \fivetwo{} and the CMB lensing auto-spectrum measurements to zero.  If the difference between these two likelihoods, $\Delta \ln \mathcal{L}$, is small, then we can ignore the cross-covariance.  For this purpose, we generate a \sixtwo{} datavector at the fiducial parameter values listed in Table~\ref{tab:params_all}.  We expect that as we consider parameter values farther away from this fiducial choice, the $\Delta \ln \mathcal{L}$ will increase.  However, since we are generally only interested in the parameter volume near the maximum likelihood, an increase in  $\Delta \ln \mathcal{L}$ at extreme parameter values is not problematic.  We find that for log-likelihoods within about 50 of the maximum likelihood, $\Delta \ln \mathcal{L} \lesssim 0.2$.  Such a small change in the likelihood will not significantly impact our parameter constraints.  We are therefore justified in ignoring cross-covariance between \fivetwo{} and the full-sky CMB lensing auto-spectrum.

\subsection{Shear ratio information}
\label{sec:SR}

As described in \cite{y3-3x2ptkp}, ratios of galaxy-lensing correlation functions that use the same lens sample, but different source galaxy samples can be used to constrain e.g. source galaxy redshifts and intrinsic alignment model parameters.  Since such ratios are essentially independent of the galaxy-matter power spectrum, these ratios can be used at much smaller scales than are employed in the standard \threetwo{} analysis \cite{y3-shearratio}.  We refer to these lensing ratios as shear ratios (SR).  The analysis presented in \cite{y3-3x2ptkp} treats the SR information as a separate likelihood that can be combined with the likelihood from the measured two-point functions.  

Our fiducial analysis of the \fivetwo{} observable will include SR information as a separate likelihood, as done in \cite{y3-3x2ptkp}.  A detailed description of the DES Y3 implementation of SR can be found in \cite{y3-shearratio}.

\section{Choice of angular scales}
\label{sec:analysischoices}

\begin{figure*}
    \centering
    \includegraphics[width=1.0\linewidth]{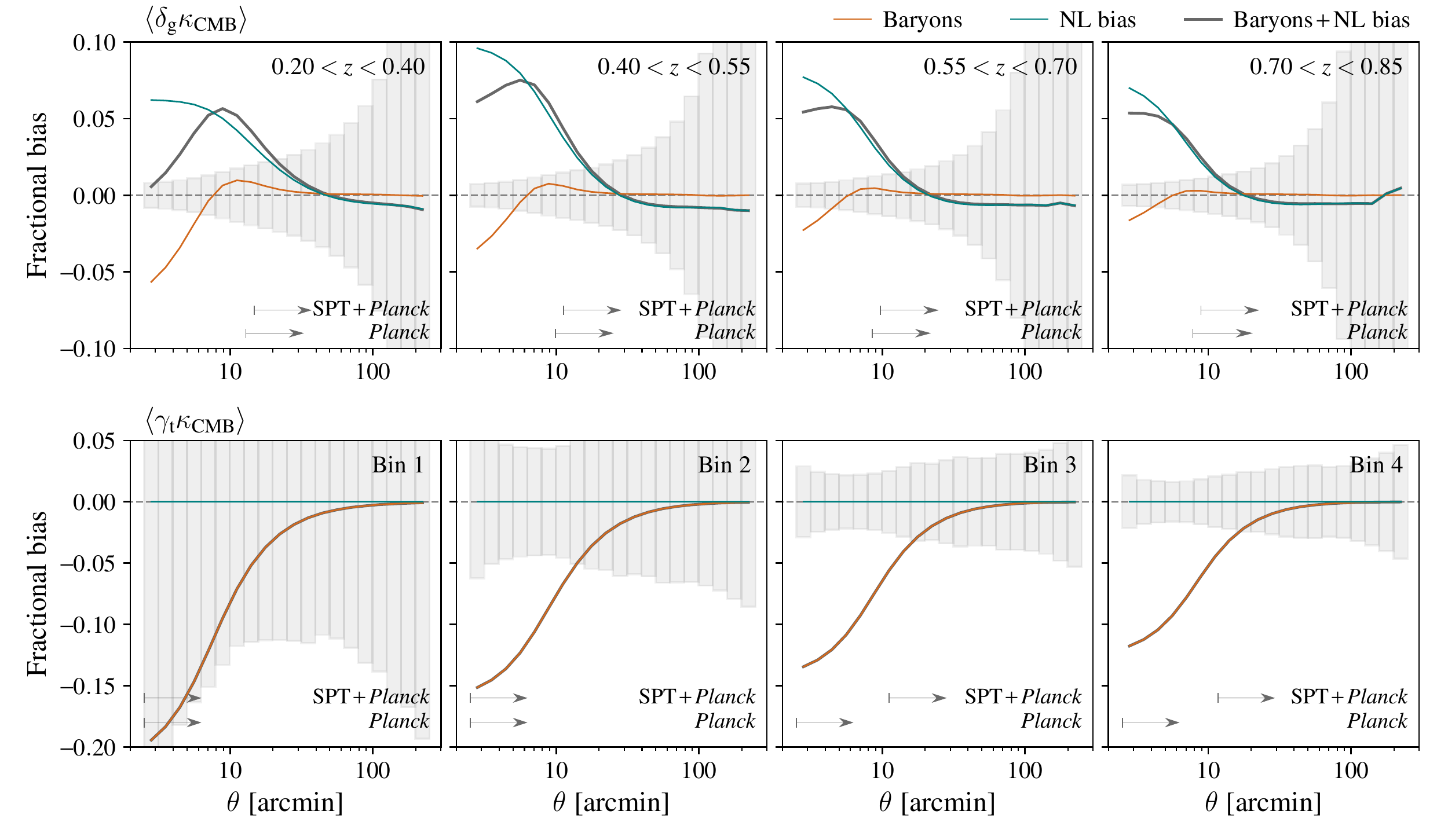}
    \caption{Fractional biases computed from the contaminated/uncontaminated data vectors with the effects of baryonic effects on the matter power spectrum (orange), non-linear galaxy bias (teal), and the sum of the two (dark gray). Also shown are the standard deviations of the  SPT+\planck{} data vectors scaled down by a factor of 10. The arrows indicate the angular scales used in the analysis. }
    \label{fig:biases}
\end{figure*}

The cross-correlations with CMB lensing that we consider in this analysis are impacted by several physical effects at small scales ($k\gtrsim0.2h{\rm Mpc}^{-1}$) that are challenging to model. For one, feedback from active galactic nuclei (AGN) impacts the distribution of baryons on small scales, leading to changes in the matter power spectrum that can reach the ten percent level \cite{Chisari2018, Huang2019}. Fully capturing feedback physics in an analytic model is very challenging given the complexity and large dynamic range of the problem. Since this astrophysical effect impacts the matter power spectrum, feedback will necessarily have an impact on  both \nk{} and \gk{}. Another small-scale effect that we must contend with is a breakdown in the linear bias model we use to describe the clustering of galaxies.  At small scales, galaxy bias becomes nonlinear \citep{Desjacques:2018}. Nonlinear galaxy bias will impact \nk{} (see discussion of a nonlinear bias model in Section~\ref{sec:bias}).

The impact of baryonic feedback and nonlinear bias on our analysis can be reduced by restricting the analysis to those physical scales that are least impacted.  In general, this corresponds to restricting the analysis to large physical scales.  The \threetwo{} analysis of \cite{y3-3x2ptkp} has taken this approach in their analysis of correlations of DES-only correlation functions, and we do the same here.  This approach is conservative in the sense that it is largely robust to detailed assumptions about feedback and nonlinear bias.  Of course, it also comes at the cost of reduced signal-to-noise.

\begin{figure*}
\begin{center}
\includegraphics[width=0.246\linewidth]{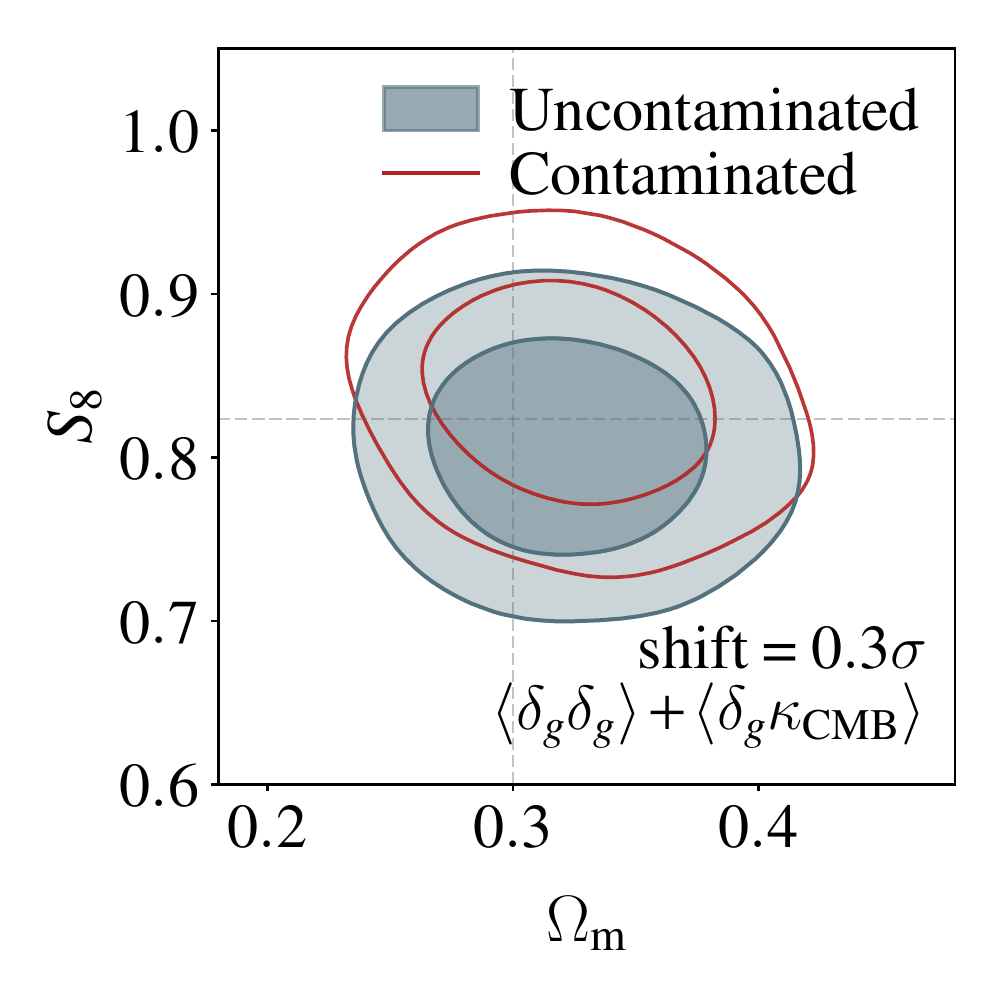}
\includegraphics[width=0.246\linewidth]{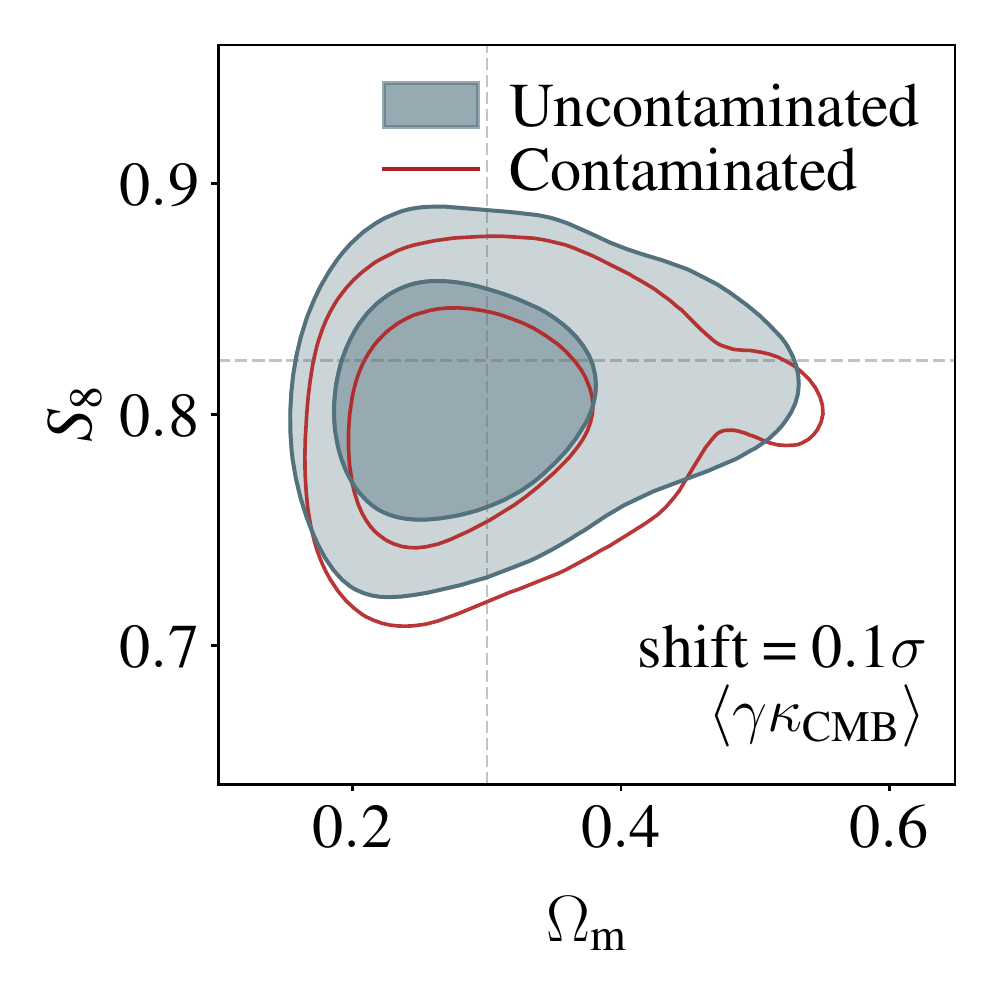}
\includegraphics[width=0.246\linewidth]{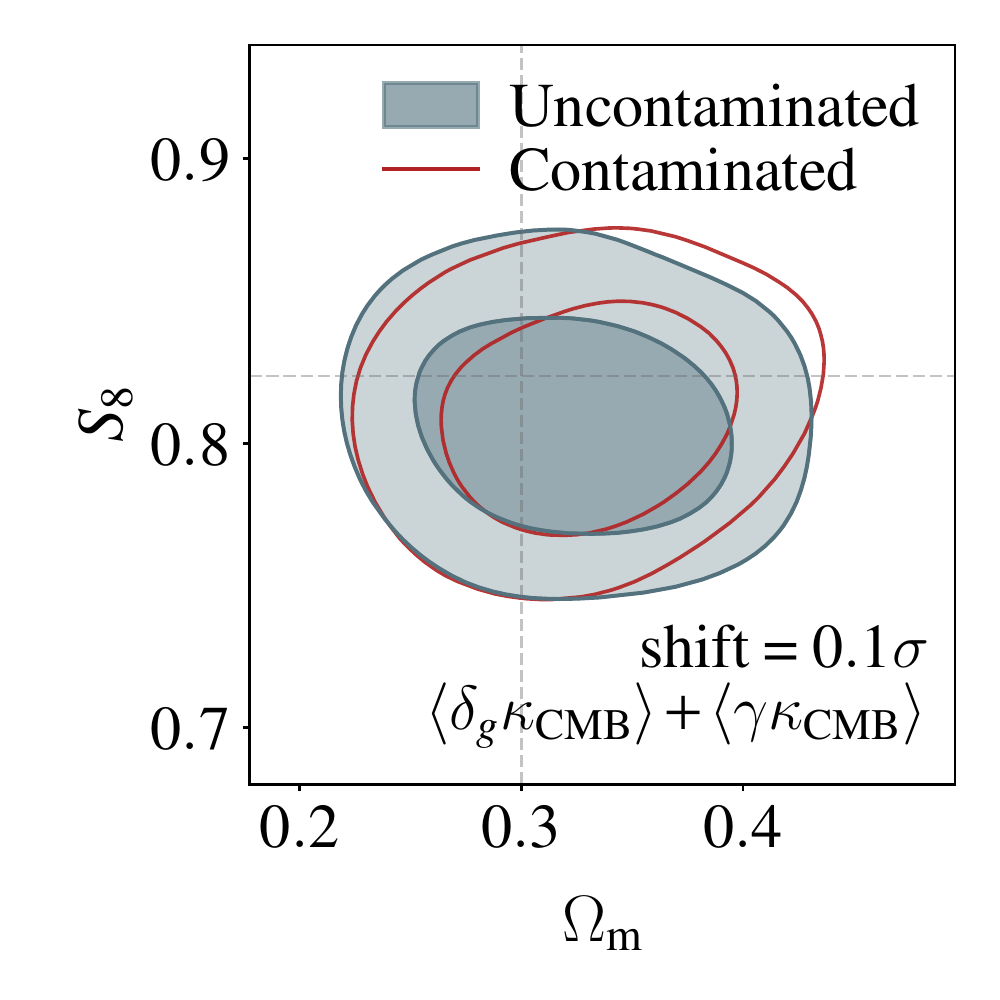}
\includegraphics[width=0.246\linewidth]{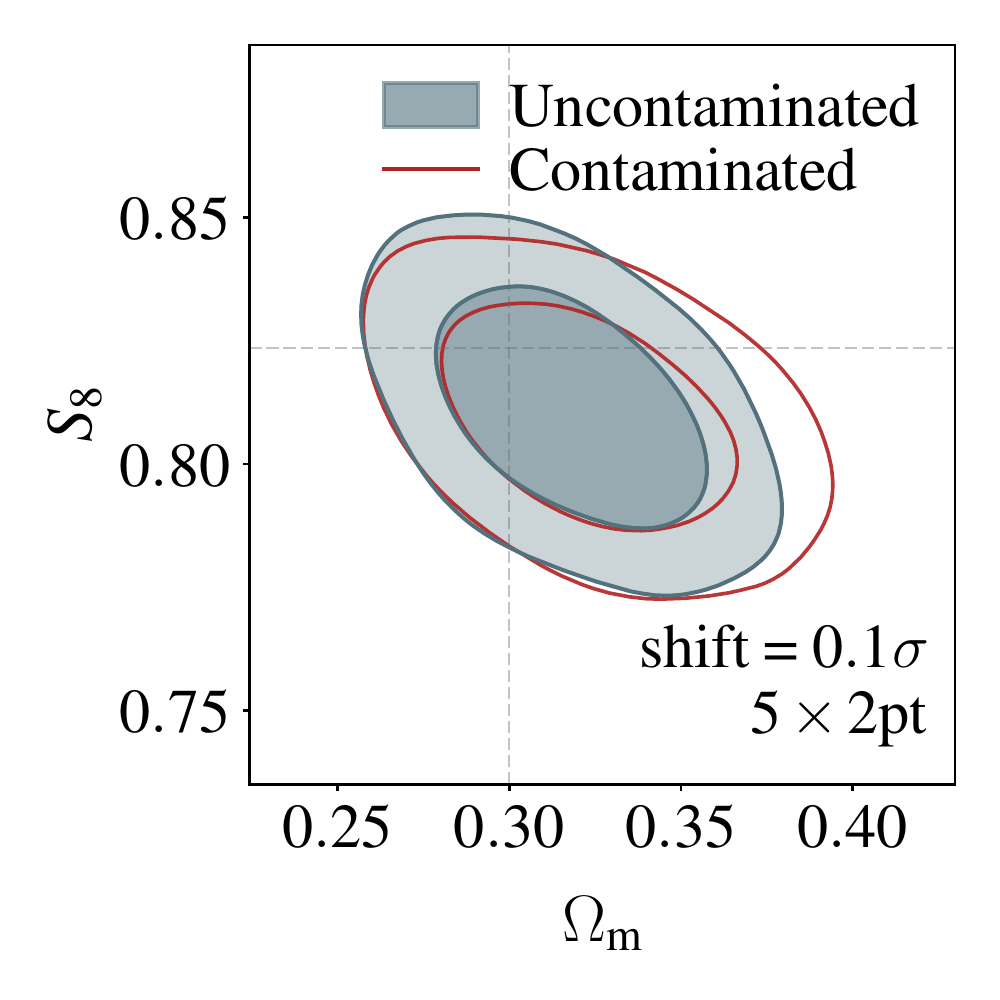}
\caption{Forecasted constraints on $\Omega_{\rm m}$ and $S_{8}$ using the fiducial data vector (blue) and a data vector contaminated with our model of nonlinear galaxy bias and baryonic effects on the small-scale matter power spectrum (red). The four panels show (from left to right) results for the combinations of \nn{}+\nk{}, \gk{}, \nk{}+\gk{} and \fivetwo{}. The shift in the two contours are shown in the bottom right of each panel. }
\label{fig:contam_contours}
\end{center}
\end{figure*}

\begin{table*}
\begin{center}
\begin{tabular}{cccccc*{3}{c}}
\toprule 
\multirow{2}{*}{Type} & \multirow{2}{*}{Redshift bin}& \multirow{2}{*}{\  }  & \multicolumn{2}{c}{$\theta_{\rm min}$} & \multicolumn{3}{c}{${\rm Forecasted\ S/N}$}   \\ \cmidrule{4-8} 
 &  &  & SPT+{\it Planck} & {\it Planck}  &            SPT+{\it Planck} & {\it Planck} & Combined  \\ \midrule 
\nk{} & 1  &    & 14.8$'$ (11.0$'$) & 12.9$'$ (11.1$'$)  & & \\
      & 2  &   & 11.3$'$ (8.5$'$) & 9.9$'$ (8.49$'$)    & & \\
      & 3  &   & 9.7$'$ (7.3$'$)  & 8.5$'$ (7.25$'$)    & & \\
      & 4  &   & 8.9$'$ (6.6$'$)  & 7.7$'$ (6.64$'$)    & &  \\ 
      & {All bins}& &                    &                   &   12.2 (14.9) & 11.6 (12.7) & 16.9 (19.6)\\
      
\midrule
\gk{}& 1 && 2.5$'$  & 2.5$'$ & & \\
      & 2 && 2.5$'$  & 2.5$'$ & & \\
      & 3 && 11.2$'$  & 2.5$'$ & & \\
      & 4 && 17.7$'$  & 2.5$'$ & & \\   
      & {All bins} &&   &  &  10.1  & 8.7  & 13.3 \\     
\midrule
\nkgk{}& All bins &&   &  & 13.9 (15.8) & 12.6 (13.5)  & 18.8 (20.8)\\ \bottomrule
\end{tabular}
\caption{Minimum angular scale cuts for \nk{} and \gk, for both the SPT+\planck{} and \planck{} patches. The maximum scale for all the data vectors is 250 arcmin. Numbers in parentheses correspond to the nonlinear galaxy bias analysis. } 
\end{center}
\label{tab:scalecuts}
\end{table*}

We now develop a choice of angular scales to include in our analysis of \nk{} and \gk.  Throughout this discussion, we refer to effects such as baryonic feedback and nonlinear bias which are not modelled in our analysis as ``unmodeled effects.''  The choice of angular scale cuts is motivated by two competing considerations.  First, biases to the analysis from unmodeled effects should be minimized, which requires excluding small angular scales from the analysis.  Second, we would like to maximize our constraining power, which motivates including more angular scales in the analysis.  To set a balance between these two considerations, our requirement is that the bias caused by unmodeled effects should be significantly smaller than our uncertainties.

In order to estimate the biases in our constraints caused by unmodeled effects and to make an appropriate choice of angular scales to include, we must have some (at least approximate) guess at the impact of these effects. Following \cite{y3-generalmethods}, for baryonic feedback, we adopt the OWLS AGN model \cite{schaye2010}; for nonlinear bias, we adopt the model described in Section~\ref{sec:bias}. We note that the OWLS AGN model is known to  over-predict the impact of baryonic feedback on the lensing signal, and therefore the scale cuts derived from this simulation tends to be conservative.  Once the bias has been estimated, our requirement is then that there is less than a $0.3\sigma$ shift in the  $S_{8}$-$\Omega_{\rm m}$ constraints relative to the constraints obtained using the uncontaminated data vector. This criterion is consistent with other DES Y3 analyses.

We note that the analysis of cross-correlations between DES Y1 data and SPT/{\it Planck} measurements of CMB lensing presented in \cite{5x2y1} also took the approach of removing small angular scale measurements in order to obtain unbiased cosmological constraints.  However, as noted previously, one of the main sources of bias in that analysis was from tSZ contamination of the CMB lensing maps.  This bias necessitated removal of a large fraction of the signal-to-noise.  In the present analysis, because we have endeavored to make a CMB lensing map that is free from tSZ bias, a larger fraction of the signal-to-noise can be retained.

The impact of baryonic feedback and nonlinear bias on the \nk{} and \gk{} data vectors is shown in Figure ~\ref{fig:biases}.  It is apparent that baryonic feedback suppresses the correlation functions at small scales, and has a larger impact on \gk{} than \nk{}.  Nonlinear bias, on the other hand, leads to an increase in \nk{} at small scales, and has no impact on \gk{} (since the latter does not use galaxies as tracers of the matter field).  The fact that \nk{} and \gk{} are most impacted by different biases, and that these two biases act in opposite directions presents a complication.  This ensures that the biases to cosmological parameters caused by unmodeled effects in \nk{} and \gk{} typically act in opposite directions, and to some extent will cancel each other in a joint analysis of both \nk{} and \gk{}.  In principle, this cancellation means that we could use very small angular scales in our analysis without sustaining a large bias to the cosmological constraints.  However, since the adopted models of nonlinear bias and baryonic effects also have associated uncertainties, we investigate the two biases separately. 

In determining the scale cuts, we first choose the scale cuts for \nk{} such that the inclusion of nonlinear bias in the joint analysis of \nk{} and \nn{}  results in an acceptably small bias to the cosmological posterior.  By considering \nk{} and \nn{} together, we maximize the impact of nonlinear bias (which would lead to a conservative scale cut), and also ensure that galaxy bias is well constrained. Our scale cuts for \nk{} are based on a physical scale evaluated at the mean redshift of the lens galaxies.  The minimum physical scale is then translated into angular scales for each of the lens galaxy bins. We consider different scale cuts for the correlations with the SPT+{\it Planck} and {\it Planck}-only CMB lensing maps, since these correlations have different signal-to-noise ratio. With the scale cuts applied, we run a simulated likelihood analysis with the \nn{} + \nk{} combination using the framework described in Section~\ref{sec:model_fitting}. As shown in the left panel of Figure~\ref{fig:contam_contours}, we find in the case of the linear bias analysis that a choice of 4 Mpc for SPT+\planck{} and 3.5 Mpc for \planck{}-only meets our acceptability criteria for the bias in cosmology, while maximizing signal-to-noise ratio.  Our definition of acceptable bias is that the maximum posterior point of the biased posterior should enclose at most $\mathrm{erf}(0.3/\sqrt{2})$ of the unbiased posterior mass in the $\Omega_{\rm m}$-$S_8$ plane (marginalizing over all other parameters). 

We next choose angular scales for \gk{} such that the joint analysis of \gk{} and \nk{} remains unbiased.  Since the \gk{} measurements at a single angular scale correspond to a wide range of physical scales, choosing a \gk{} scale cut based on a physical scale is less motivated than for \nk{}.  Instead, we remove angular scales in order of their contribution to the $\Delta \chi^2$ between the biased and unbiased data vectors.  This results in keeping most of the \gk{} data vector except for 6 (8) data points at the smallest scales for bin 3 (4) for the SPT+{\it Planck} patch. We show in the middle panel of Figure~\ref{fig:contam_contours} the resulting constraints on the $\Omega_{\rm m}-S_8$ plane using the \nkgk{} combination for the contaminated and uncontaminated data vectors.   
Lastly, we check that our choice of angular scales results in the \fivetwo{} data vector passing the same acceptable bias criteria as \nk{} for the combination of the nonlinear bias and baryonic feedback models. These results are shown in the right panel of Figure~\ref{fig:contam_contours}.

We adopt a slightly different procedure to that described above for determining an appropriate choice of angular scale cuts for the analysis that uses the nonlinear galaxy bias model described in Section~\ref{sec:bias}. Since in that case, nonlinear bias is not an unmodeled effect, we follow a procedure similar to \cite{y3-2x2ptbiasmodelling} to determine appropriate scale cuts. We determine the scale below which our nonlinear bias model fails to describe the 3D galaxy-matter correlation function in the MICE simulations \cite{Fosalba2015,Crocce2015}. We describe in detail our procedure in Appendix~\ref{sec:scalecut_nlb} -- we find that a scale cut of $3\ {\rm Mpc}$ meets our acceptability criteria for the bias in cosmology, while maximizing signal-to-noise.   Since nonlinear bias does not impact \gk{}, we adopt the same scale cuts as described above for analyzing \gk{}. 

The final choice of angular scale cuts to be applied to the analyses of \nk{} and \gk{} are summarized in Table~\ref{tab:scalecuts}, together with the resulting signal-to-noise ratios.  In the case of the linear bias analysis, for the \nk{} correlations, the minimum angular scales when correlating with the SPT/{\it Planck} CMB lensing map are $(14.8, 11.3, 9.7, 8.9)'$ for the four redshift bins.  These cuts are necessitated by possible breakdown in the linear galaxy bias model at small scales.  When using the nonlinear bias galaxy model, the corresponding minimum angular scales are $(11.0, 8.5, 7.3, 6.6)'$.  These cuts are in turn necessitated by uncertainty in the baryonic feedback model.  The minimum angular scale cuts for the correlations with the {\it Planck}-only lensing map are reduced compared to correlations with the SPT/{\it Planck} map because the signal-to-noise of the {\it Planck}-only lensing map is lower.  We can compare these angular scale cuts to those used in the DES Y1 analysis of \cite{5x2y1}, which were at $(15,25,25,15)'$ for redshift bins centered at approximately the same redshifts.  The more aggressive scale cuts in this analysis are made possible by the tSZ-cleaned CMB lensing map.

The increased range of angular scales afforded by the tSZ-cleaned CMB lensing map is even more significant for \gk{}.  In this case, the minimum angular scales are $(2.5, 2.5, 11.2, 17.7)'$ for the four redshift bins.  As can be seen in Figure~\ref{fig:biases}, the change in scale cuts across the different redshift bins is driven largely by the increase in signal-to-noise of the \gk{} measurements at high redshift.  These scale cuts can be compared to those imposed in the DES Y1 analysis of \cite{5x2y1}, where scale cuts at $(40,40,60,60)'$ were imposed for similar redshift bins.  Again, the significant reduction in minimum angular scales for the present analysis is enabled by the tSZ-cleaned CMB lensing map.  Because \gk{} is not impacted by nonlinear bias, but is strongly impacted by tSZ bias, tSZ cleaning has a more significant impact for this correlation than for \nk{}.

We can also compute the reduction in signal-to-noise caused by the angular scale cuts.  Relative to using a minimum scale of $2.5'$, the adopted scale cuts results in a signal-to-noise reduction for \nk{} of $45\%$ across all redshift bins for the linear bias analysis.  This reduction, which is still significant despite the tSZ-cleaned CMB lensing map, is necessitated by possible breakdown in the linear galaxy bias model at small scales.  When using the nonlinear bias galaxy model, the corresponding reduction in signal-to-noise is 36\%, necessitated by uncertainty in the baryonic feedback model.  For \gk{}, the reduction in signal-to-noise resulting from the scale cuts is 15\%.  These numbers highlight that future improvements in modeling of baryonic feedback can enable significant increases in the signal-to-noise that can be used for constraining cosmology with galaxy survey-CMB lensing cross-correlations.

The same procedure to determine the scale cuts is also performed for the \redmagic{} sample, and the results are presented in Appendix \ref{sec:redmagic}.

\section{Forecasts}
\label{sec:results}

We now use the methodology developed above to produce forecasts for cosmological constraints obtained from the CMB lensing cross-correlation functions. These forecasts will inform our forthcoming analysis with real data.

\begin{figure*}
\begin{center}
\includegraphics[width=1.00\linewidth]{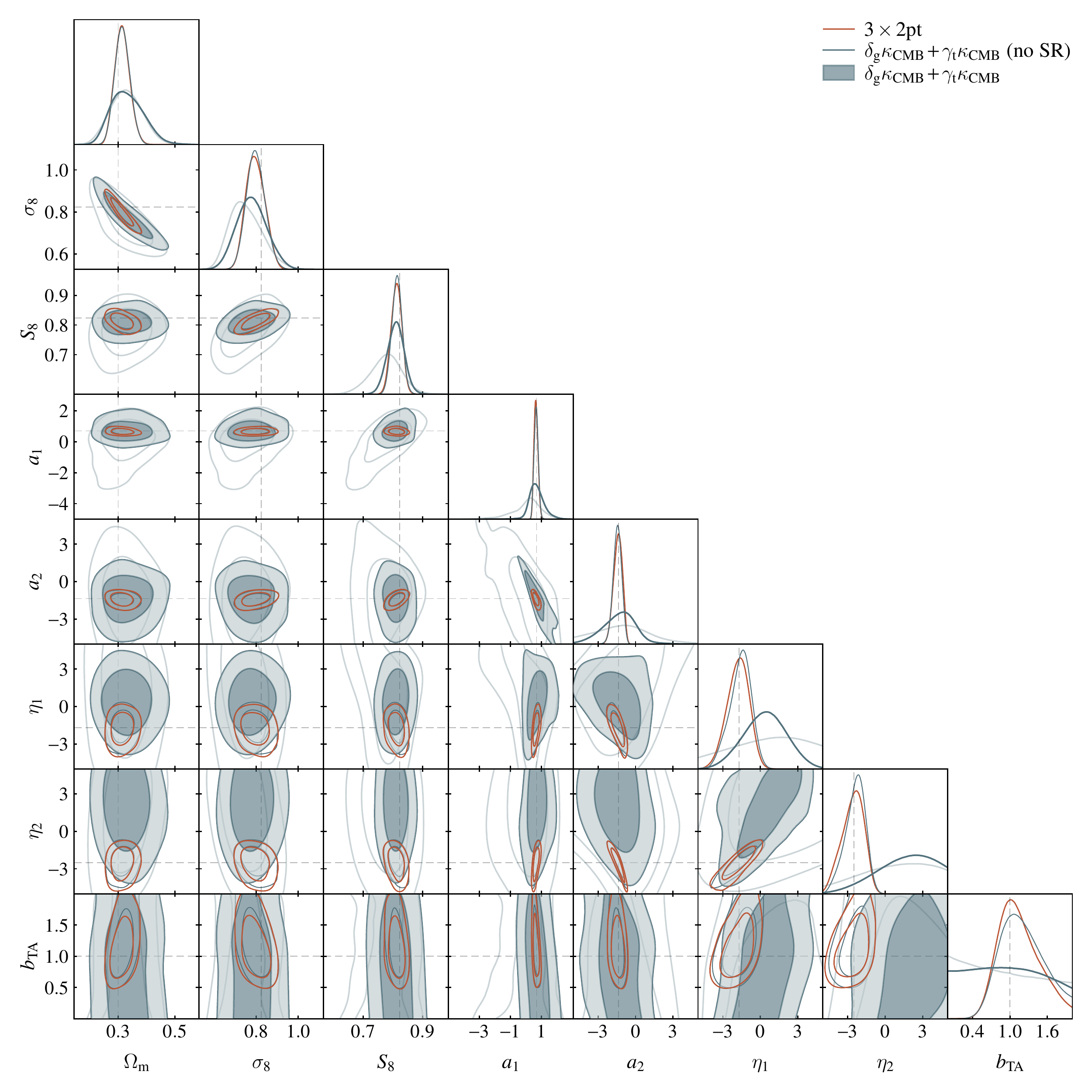}
\caption{Comparison of the  forecasted constraints on cosmological parameters $\Omega_{\rm m}$,$\sigma_{8}$,$S_{8}$ and intrinsic alignment parameters $a_{1}$, $a_{2}$, $\eta_{1}$, $\eta_{2}$, and $b_{\rm TA}$ using the combination of galaxy clustering and galaxy-CMB lensing correlation, with and without the addition of shear-ratio information, compared with the constraints from \threetwo{}. The dashed lines represent the input values for the individual parameters. }
\label{fig:nkgk_fid}
\end{center}
\end{figure*}

\subsection{\nkgk}

The forecasted cosmological constraints from the joint analysis of \nk{} and \gk{} are presented in Figure~\ref{fig:nkgk_fid}.  Constraints are presented with and without the inclusion of shear ratio (SR) likelihood described in Section~\ref{sec:SR}. We observe a significant improvement in the constraints when the SR likelihood is included. The improvement is particularly noticeable in the $S_8$ direction, which is roughly proportional to the amplitude of the lensing power spectrum. This improvement is not surprising 
since the SR likelihood can significantly improve constraints on IA parameters, as demonstrated in \cite{y3-shearratio,y3-cosmicshear1,y3-cosmicshear2}. We see in Figure~\ref{fig:nkgk_fid} the corresponding IA constraints and how $a_1$ is strongly degenerate with $S_8$. The SR constraints significantly reduce the IA parameter space allowed by the data, which in turn tightens the cosmological constraints.

For comparison, we also overlay constraints from the \threetwo{} data combination, analysed with the same analysis choices described in this paper. We see that when examining the $\Omega_{\rm m}$--$\sigma_8$ plane, our cross-correlation constraints are significantly larger than that of \threetwo{}. However, when projecting onto $S_8$, we expect our cross-correlation constraints to be only \results{1.4} times larger than \threetwo{}, with a \results{3\%} level constraint on $S_{8}$.  This suggests that the \nkgk{} combination could provide a powerful consistency check for the \threetwo{} data that is quite independent and robust to systematic effects that are only present in the galaxy surveys. 

\subsection{\fivetwo{}}

\begin{figure}
\begin{center}
\includegraphics[width=1.0\linewidth]{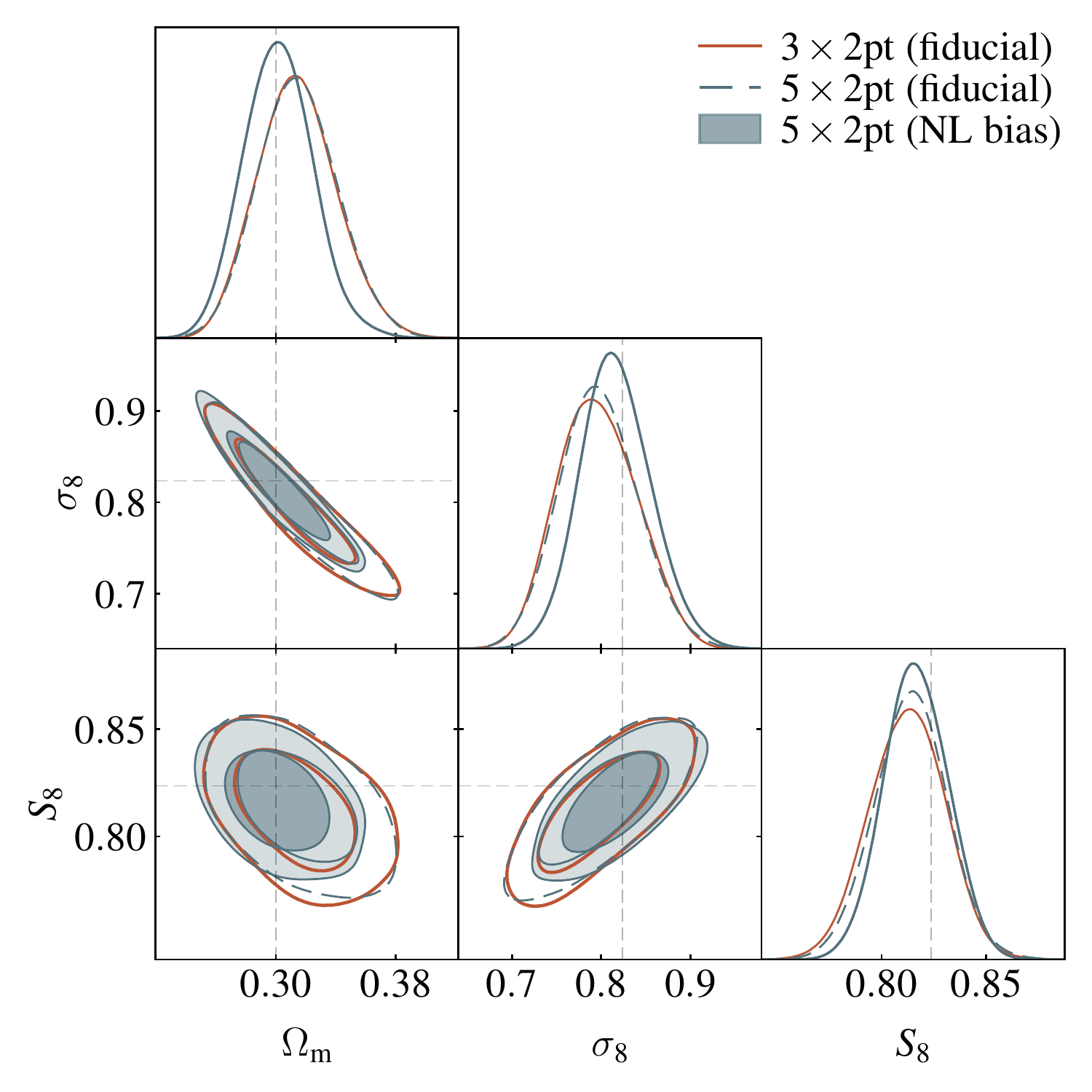}
\caption{Comparison of the forecast constraints on $\Omega_{\rm m},\sigma_{8}$ and $S_{8}$ from the \threetwo{} and \fivetwo{} probes using linear galaxy bias modeling (open orange and dashed blue contours) and non-linear galaxy modeling (filled blue contours). } 
\label{fig:5x2pt_fid}
\end{center}
\end{figure}

Next we combine \nkgk{} in the previous section with the \threetwo{} probes, including the SR likelihood, to arrive at Figure~\ref{fig:5x2pt_fid}. For reference, we also include the \threetwo{} constraints on the plot. We observe that although the overall improvement in constraining power over \threetwo{} is weak, \nkgk{} mildly breaks the degeneracy of the \threetwo{} constraints to give slightly tighter \fivetwo{} constraints.  We expect an improved  precision on $\Omega_{\rm m}$/$\sigma_8$/$S_8$ from \results{8.3/5.7/2.3\%} to \results{8.2/5.4/2.1\%}. It is worth emphasizing again that even though the added constraining power is not significant, the mere consistency (or inconsistency) between \nkgk{} and \threetwo{} could provide non-trivial tests for either systematics or new physics. This is because the cross-correlation probes include a dataset that is completely independent of all DES data processing pipelines, and therefore should not be sensitive to systematic effects that only exist in DES data (and vice versa for CMB datasets). In particular, given the somewhat puzzling inconsistencies between the galaxy-galaxy lensing and galaxy clustering signals using the \redmagic{} sample from the DES Y3 \threetwo{} analysis \cite{y3-3x2ptkp}, this consistency test will become extremely important for making progress in the future. 

In Figure~\ref{fig:5x2pt_fid} we also show the forecasted \fivetwo{} constraints assuming nonlinear galaxy bias. We find an overall gain in the constraining power compared to the linear galaxy bias mode. The gain in constraining power going from \threetwo{} to \fivetwo{} when using nonlinear galaxy bias is similar to that using linear galaxy bias, with a forecast constraint on $\Omega_{\rm m}$/$\sigma_8$/$S_8$ going from \results{7.9/5.2/2.0\%} to \results{7.7/4.7/1.9\%}.

\subsection{Constraints on shear bias parameters}

Cosmological constraints from galaxy surveys can be significantly degraded by systematic uncertainties impacting measurements of the lensing-induced shears, and the measurements of photometric redshift for the lensed galaxies.  Shear calibration systematics are especially pernicious, since a multiplicative bias in shear calibration is perfectly degenerate with the amplitude of the lensing correlation functions that we wish to constrain \cite{Hirata:2003}.  Typically, ancillary data is used to constrain these sources of systematic uncertainty.  In the case of multiplicative shear bias, one often relies on simulated galaxy images to constrain the bias parameters, $m$.  If the simulations do not accurately capture the properties of real galaxies, priors on $m$ may be untrustworthy. 

CMB lensing, on the other hand, provides a measure of the mass distribution that is independent of these sources of uncertainty.  As a result, cross-correlations of galaxy surveys with CMB lensing have different sensitivity to the nuisance parameters describing these effects than auto-correlations of galaxy survey observables.  By jointly analyzing the auto-correlations and the CMB lensing cross-correlations, one can obtain constraints on $m$ directly from the data \cite{Vallinotto:2012,Baxter:2016,Schaan:2017}.  The idea of using the data to obtain constraints on nuisance parameters is often referred to as {\it self-calibration}.

\begin{figure}
\begin{center}
\includegraphics[width=1.0\linewidth,bb=22 0 420 420]{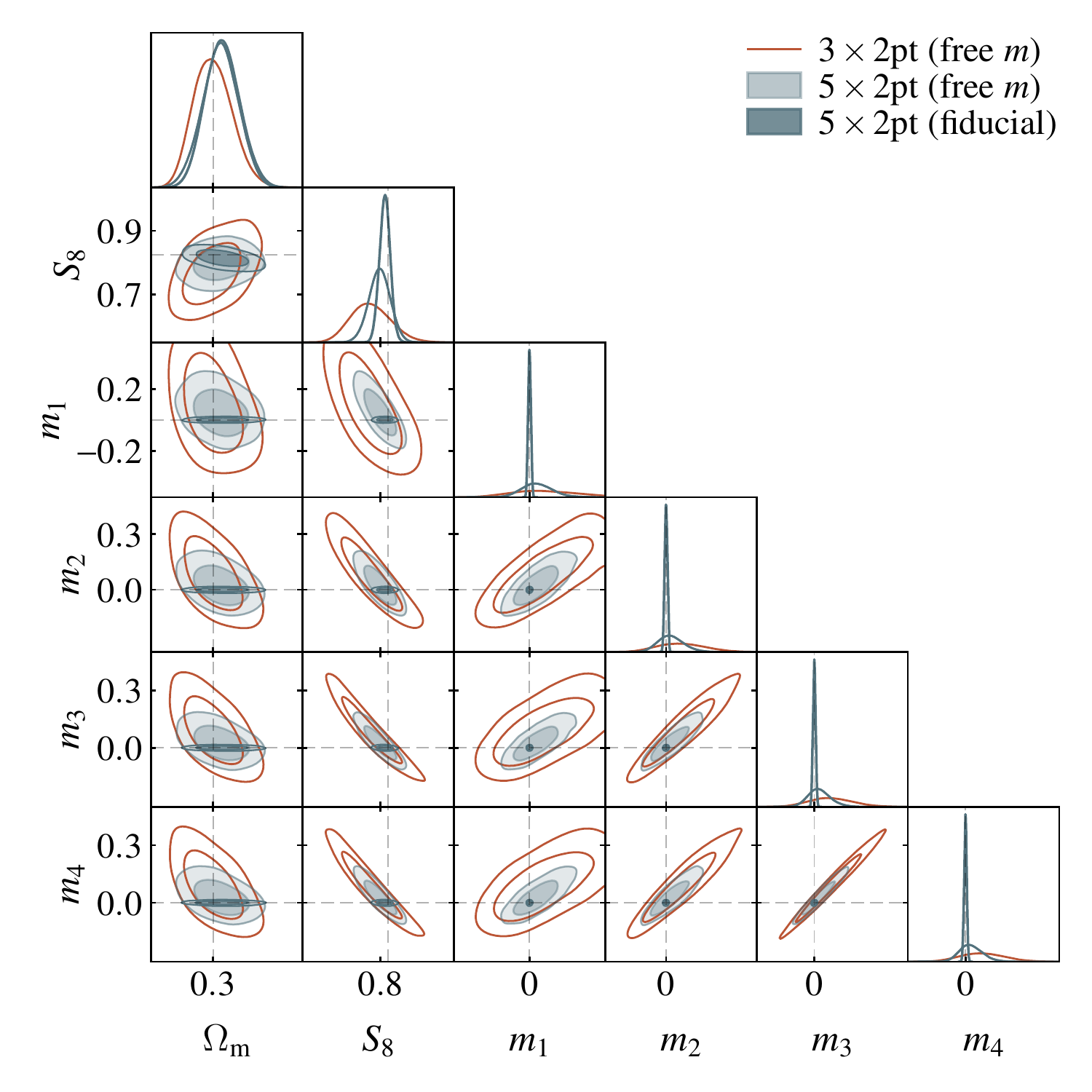}
\caption{Simulated constraints from \threetwo{} (red) and \fivetwo{} (blue) probes when the fiducial priors on the shear calibration parameters are replaced by very wide priors (free $m$). The results show the ability of the data to constrain these nuisance parameters with the \threetwo{} and \fivetwo{} probes respectively. Also overlaid are the fiducial \fivetwo{} constraints, where the $m_{i}$ parameters are informed by external priors.}
\label{fig:freem}
\end{center}
\end{figure}

Here we re-examine the case for self-calibrating $m$ using our new datasets and models. We perform our fiducial \threetwo{} and \fivetwo{} analyses removing the tight priors on the shear calibration parameters in all redshift bins, $m^{i}$, and replacing them with very wide flat priors.  We show in Figure~\ref{fig:freem} the constraints in the $\Omega_{\rm m}$--$S_8$ plane as well as the shear calibration parameters. We see that without any prior knowledge of the shear calibration parameter, both \threetwo{} and \fivetwo{} are able to place constraints on these parameters to some extent: \threetwo{} measures $S_8$ at the \results{8\%} level while \fivetwo{} is expected to significantly improve on that, and constrain $S_8$ at the \results{4\%} level.

These uncertainties on $m$ (\results{$\sim$0.1-0.2} for \threetwo{} and \results{$\sim$0.05-0.1} for \fivetwo{}) are still much larger than what we could achieve with other approaches using e.g. simulations, which are  currently below 0.01 \citep{y3-imagesims}. These findings are consistent with our results in \citep{5x2y1}.

\section{Summary}
\label{sec:discussion}

We have presented the key ingredients for our forthcoming analysis of cross-correlations between DES Y3 measurements of galaxy positions and galaxy shears, and measurements of CMB lensing from SPT and {\it Planck} data.  These include:
\begin{enumerate}
    \item A new CMB lensing map that is constructed to remove bias from the thermal SZ effect using a combination of SPT and {\it Planck} data in the SPT-SZ footprint. The removal of the tSZ bias will allow cosmological information to be extracted from the CMB lensing cross-correlations at much smaller angular scales than those used in DES Y1 analysis. This CMB lensing map will be useful for other cross-correlations analyses beyond those considered here.
    \item A modeling framework built on the DES Y3 \threetwo{} methods presented in \cite{y3-generalmethods}.  In particular, we describe our models for the galaxy and galaxy lensing cross-correlations with CMB lensing.
    \item A hybrid covariance matrix estimate for the \fivetwo{} data vector that combines three components: the \threetwo{} halo-model covariance matrix from \citep{y3-covariances}, an analytic log-normal covariance for the galaxy-CMB cross-covariance, and a model of the noise and mask contributions from realistic simulations.
    \item A choice of angular scales to use when analyzing the CMB lensing cross-correlations that ensures our cosmological constraints from data will be robust, even in the presence of baryonic feedback and nonlinear galaxy bias.  We describe two sets of angular scale choices, one set that is designed for the analysis that uses a linear galaxy bias model, and one designed for the analysis that uses a nonlinear galaxy bias model. 
\end{enumerate}

We use the methodological tools developed in this analysis to make forecasts for the cosmological constraints that will be obtained in our forthcoming analysis of actual data.  These forecasts make use of the true noise levels of the CMB lensing maps constructed here.  The main results from these forecasts are:
\begin{enumerate}
    \item We forecast that our cross-correlation data vector will have a total signal-to-noise of \results{18.8 (20.8)} when assuming linear (nonlinear) galaxy bias, which is about twice that obtained from past cross-correlation analyses between DES and SPT lensing using DES Y1 data \citep{5x2y1}.  
    \item When using the linear galaxy bias and the $\Lambda$CDM cosmology model, we expect to find a \results{3\%} constraint on $S_8$ using the cross-correlation data vectors \nkgk{} alone.\footnote{We note that our analysis of \nkgk{} includes the so-called shear ratio likelihood, which acts as a prior on e.g. intrinsic alignments and the source redshift distributions.} This constraint does not include any of the correlation functions that go into \threetwo{} data vector analyzed in \cite{y3-3x2ptkp} and therefore serves as a powerful consistency test. 
    \item We anticipate a \results{2\%} constraint on $S_8$ from the \fivetwo{} analysis. Similar constraints are obtained when the nonlinear galaxy bias model is used.   
    \item When we do not apply external priors on the shear calibration parameters, we find that both \threetwo{} and \fivetwo{} are able to calibrate the shear bias parameters, $m$, with \fivetwo{} roughly doubling the constraining power on these nuisance parameters. However, the resultant posteriors on the $m$ parameters are still significantly weaker than the current external priors used by DES, suggesting that self-calibration of shear biases from galaxy-CMB lensing cross-correlation is not likely to improve cosmological constraints in the near term.  However, we emphasize that \fivetwo{} offers significantly tighter constraints than \threetwo{} in the absence of external priors on shear calibration.
\end{enumerate}

Cross-correlations of measurements of large-scale structure from the Dark Energy Survey with measurements of CMB lensing from the South Pole Telescope and {\it Planck} offer tight cosmological constraints that are particularly robust against sources of systematic error.  Given the challenges of extracting unbiased cosmological constraints from increasingly precise measurements by galaxy surveys, we expect cross-correlations between galaxy surveys and CMB lensing to continue to play an important role in future cosmological analyses.

\acknowledgements
The South Pole Telescope program is supported by
the National Science Foundation (NSF) through
the grant OPP-1852617. Partial support is also
provided by the Kavli Institute of Cosmological Physics at the University of Chicago.
Argonne National Laboratory’s work was supported by
the U.S. Department of Energy, Office of Science, Office of High Energy Physics, under contract DE-AC02-
06CH11357. Work at Fermi National Accelerator Laboratory, a DOE-OS, HEP User Facility managed by the Fermi Research Alliance, LLC, was supported under Contract No. DE-AC02- 07CH11359. The Melbourne authors acknowledge support from the Australian Research Council’s Discovery Projects scheme (DP200101068). The McGill authors acknowledge funding from the Natural Sciences and Engineering Research
Council of Canada, Canadian Institute for Advanced research, and the Fonds de recherche du Qu\'{u}bec Nature et technologies. The CU Boulder group acknowledges support from NSF AST-0956135. The Munich group acknowledges the support by the ORIGINS Cluster (funded by the Deutsche Forschungsgemeinschaft (DFG, German Research Foundation) under Germany’s Excellence Strategy – EXC-2094 – 390783311), the MaxPlanck-Gesellschaft Faculty Fellowship Program, and
the Ludwig-Maximilians-Universit\"{a}t M\"{u}nchen. JV acknowledges support from the Sloan Foundation.

Funding for the DES Projects has been provided by the U.S. Department of Energy, the U.S. National Science Foundation, the Ministry of Science and Education of Spain, 
the Science and Technology Facilities Council of the United Kingdom, the Higher Education Funding Council for England, the National Center for Supercomputing 
Applications at the University of Illinois at Urbana-Champaign, the Kavli Institute of Cosmological Physics at the University of Chicago, 
the Center for Cosmology and Astro-Particle Physics at the Ohio State University,
the Mitchell Institute for Fundamental Physics and Astronomy at Texas A\&M University, Financiadora de Estudos e Projetos, 
Funda{\c c}{\~a}o Carlos Chagas Filho de Amparo {\`a} Pesquisa do Estado do Rio de Janeiro, Conselho Nacional de Desenvolvimento Cient{\'i}fico e Tecnol{\'o}gico and 
the Minist{\'e}rio da Ci{\^e}ncia, Tecnologia e Inova{\c c}{\~a}o, the Deutsche Forschungsgemeinschaft and the Collaborating Institutions in the Dark Energy Survey. 

The Collaborating Institutions are Argonne National Laboratory, the University of California at Santa Cruz, the University of Cambridge, Centro de Investigaciones Energ{\'e}ticas, 
Medioambientales y Tecnol{\'o}gicas-Madrid, the University of Chicago, University College London, the DES-Brazil Consortium, the University of Edinburgh, 
the Eidgen{\"o}ssische Technische Hochschule (ETH) Z{\"u}rich, 
Fermi National Accelerator Laboratory, the University of Illinois at Urbana-Champaign, the Institut de Ci{\`e}ncies de l'Espai (IEEC/CSIC), 
the Institut de F{\'i}sica d'Altes Energies, Lawrence Berkeley National Laboratory, the Ludwig-Maximilians Universit{\"a}t M{\"u}nchen and the associated Excellence Cluster Universe, 
the University of Michigan, NFS's NOIRLab, the University of Nottingham, The Ohio State University, the University of Pennsylvania, the University of Portsmouth, 
SLAC National Accelerator Laboratory, Stanford University, the University of Sussex, Texas A\&M University, and the OzDES Membership Consortium.

Based in part on observations at Cerro Tololo Inter-American Observatory at NSF's NOIRLab (NOIRLab Prop. ID 2012B-0001; PI: J. Frieman), which is managed by the Association of Universities for Research in Astronomy (AURA) under a cooperative agreement with the National Science Foundation.

The DES data management system is supported by the National Science Foundation under Grant Numbers AST-1138766 and AST-1536171.
The DES participants from Spanish institutions are partially supported by MICINN under grants ESP2017-89838, PGC2018-094773, PGC2018-102021, SEV-2016-0588, SEV-2016-0597, and MDM-2015-0509, some of which include ERDF funds from the European Union. IFAE is partially funded by the CERCA program of the Generalitat de Catalunya.
Research leading to these results has received funding from the European Research
Council under the European Union's Seventh Framework Program (FP7/2007-2013) including ERC grant agreements 240672, 291329, and 306478.
We  acknowledge support from the Brazilian Instituto Nacional de Ci\^encia
e Tecnologia (INCT) do e-Universo (CNPq grant 465376/2014-2).

This manuscript has been authored by Fermi Research Alliance, LLC under Contract No. DE-AC02-07CH11359 with the U.S. Department of Energy, Office of Science, Office of High Energy Physics.

\appendix

\section{CMB lensing auto-spectrum}
\label{sec:cmbauto}

As a validation of our CMB lensing map, we also measure its auto-power spectrum and compare to previous measurements. The raw CMB power spectrum contains noise bias terms which we must subtract off: 
\begin{equation}
\hat{C}_{L}^{\kappa\kappa}=C_{L}^{\hat{\kappa}\hat{\kappa}}-N_{L}^{(0)}-N_{L}^{(1)},
\end{equation}
where the $N_{L}^{(0)}$ and $N_{L}^{(1)}$ terms are the 
noise terms from the disconnected and connected 4-pt functions \cite{planck13-17}.
In practice, we replace the $N_{L}^{(0)}$ term with the ``realization dependent'' $N_{L}^{(0)}$ (RDN0) noise \cite{namikawa2012}, which uses a mixture of simulation realizations and the data map itself: 
\begin{align}
N_{L}^{\rm (0),RD}
=\biggl\langle &C_{L}^{\hat\kappa\hat\kappa}[\kappa(T^{x}_{\rm d}T^{\rm {SMICA}}_{ {\rm s}_{i},\phi_{i} })\kappa(T^{x}_{\rm d}T^{\rm {SMICA}}_{ {\rm s}_{i},\phi_{i} })]\nonumber\\
      + &C_{L}^{\hat\kappa\hat\kappa}[\kappa(T^{x}_{ {\rm s}_{i},\phi_{i} }T^{\rm {SMICA}}_{\rm d})\kappa(T^{x}_{\rm d}T^{\rm {SMICA}}_{ {\rm s}_{i},\phi_{i} })]\nonumber\\
      + &C_{L}^{\hat\kappa\hat\kappa}[\kappa(T^{x}_{\rm d}T^{\rm {SMICA}}_{ {\rm s}_{i},\phi_{i} })\kappa(T^{x}_{ {\rm s}_{i},\phi_{i} }T^{\rm {SMICA}}_{\rm d})]\nonumber\\
      + &C_{L}^{\hat\kappa\hat\kappa}[\kappa(T^{x}_{{{\rm s}_{i},\phi_{i}} }T^{\rm {SMICA}}_{\rm d})                  \kappa(T^{x}_{ {\rm s}_{i},\phi_{i} }T^{\rm {SMICA}}_{\rm d})]\nonumber\\
      - &C_{L}^{\hat\kappa\hat\kappa}[\kappa(T^{x}_{ {\rm s}_{i},\phi_{i} } T^{\rm {SMICA}}_{ {\rm s}_{j},\phi_{j} } )\kappa(T^{x}_{ {\rm s}_{i},\phi_{i} }T^{\rm {SMICA}}_{ {\rm s}_{j},\phi_{j} })]\nonumber\\
      - &C_{L}^{\hat\kappa\hat\kappa}[\kappa(T^{x}_{ {\rm s}_{i},\phi_{i} } T^{\rm {SMICA}}_{ {\rm s}_{j},\phi_{j} } )\kappa(T^{x}_{ {\rm s}_{j},\phi_{j} }T^{\rm {SMICA}}_{ {\rm s}_{i},\phi_{i} })]
\biggl\rangle_{i,j},
\end{align}
where the subscripts \{$d,s$\} refer to data and simulation realizations, $\phi_{i}$ represents the input lensing potential realization used to lens the CMB realization, and the superscript $x$/{\rm SMICA} denotes whether we are using the SPT+{\it Planck} or the SMICAnoSZ temperature maps. In this equation, we are representing the convergence maps used to compute the power spectrum inside the square brackets and the two temperature maps that were used to reconstruct the lensing map with the round brackets. 
The $N_{L}^{(1)}$ bias term can be computed using simulated maps with different CMB realizations lensed with using a common lensing field: 
\begin{align}
N_{L}^{(1)}
=\biggl\langle &C_{L}^{\hat\kappa\hat\kappa}[\kappa(T^{x}_{{\rm s}_{i},\phi_{i}}T^{\rm {SMICA}}_{{\rm s}_{j},\phi_{i}})\kappa(T^{x}_{{\rm s}_{i},\phi_{i}}T^{\rm {SMICA}}_{{\rm s}_{j},\phi_{i}})]\nonumber\\
              +&C_{L}^{\hat\kappa\hat\kappa}[\kappa(T^{x}_{{\rm s}_{i},\phi_{i}}T^{\rm {SMICA}}_{{\rm s}_{j},\phi_{i}})\kappa(T^{x}_{{\rm s}_{j},\phi_{i}}T^{\rm {SMICA}}_{{\rm s}_{i},\phi_{i}})]\nonumber\\
              -&C_{L}^{\hat\kappa\hat\kappa}[\kappa(T^{x}_{{\rm s}_{i},\phi_{i}}T^{\rm {SMICA}}_{{\rm s}_{j},\phi_{j}})\kappa(T^{x}_{{\rm s}_{i},\phi_{i}}T^{\rm {SMICA}}_{{\rm s}_{j},\phi_{j}})]\nonumber\\
              -&C_{L}^{\hat\kappa\hat\kappa}[\kappa(T^{x}_{{\rm s}_{i},\phi_{i}}T^{\rm {SMICA}}_{{\rm s}_{j},\phi_{j}})\kappa(T^{x}_{{\rm s}_{j},\phi_{j}}T^{\rm {SMICA}}_{{\rm s}_{i},\phi_{i}})]
\biggl\rangle_{i,j}.
\end{align}
where we highlight that the same CMB lensing potential is used to lens the CMB realizations $s_{i}$ and $s_{j}$.
The final debiased power spectrum is presented in Figure \ref{fig:clkk}. Compared to the results of \cite{omori2017}, we are able to extend our measurements to higher multipoles because of the nulling of the tSZ bias and improved treatment of point sources and clusters. 

\section{Validating the tSZ-nulling method}
\begin{figure}
\begin{center}
\includegraphics[width=1.0\linewidth]{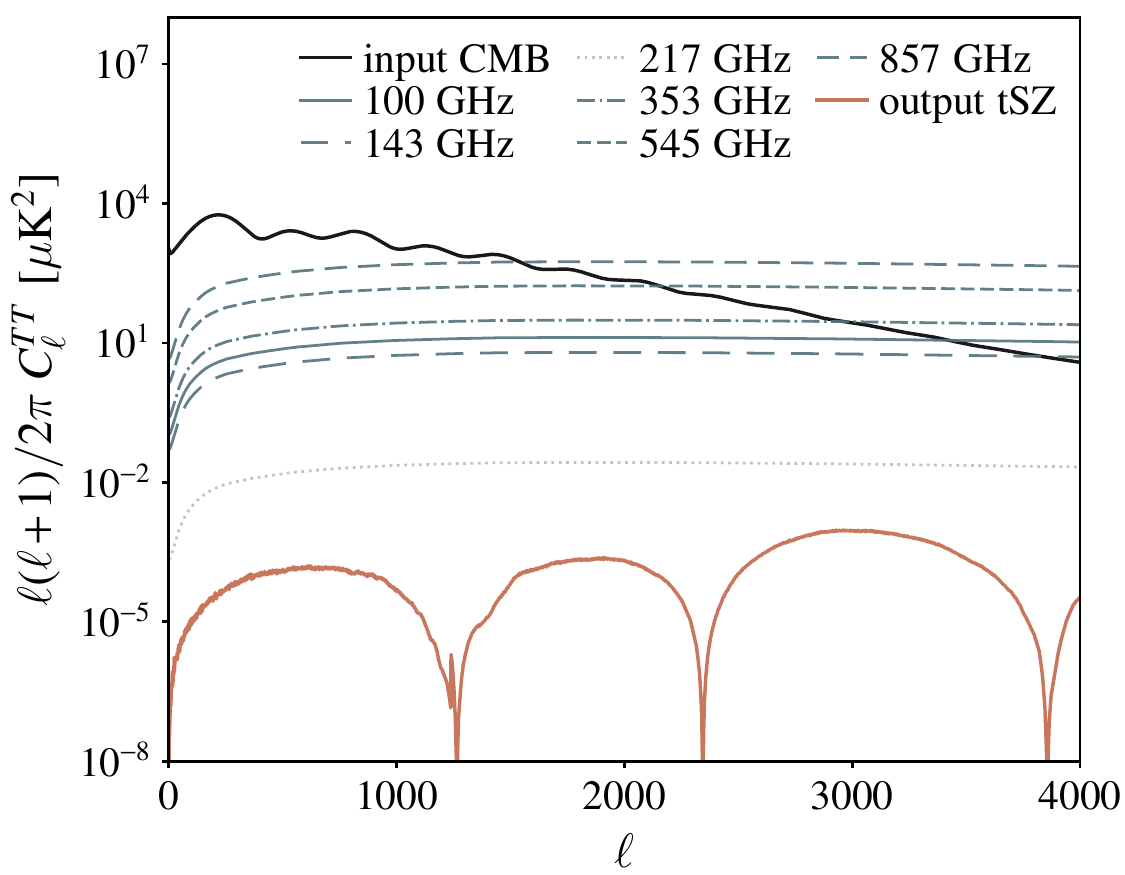}
\caption{tSZ power spectra at 100/143/217/353/545/857 GHz (various blue lines), as well as the tSZ residual power spectrum after passing the individual frequency maps through the SMICAnoSZ weights (orange).}
\label{fig:smicanosz_residual}
\end{center}
\end{figure}

In this section, we verify that the methodology described in Section \ref{sec:QE} results in a tSZ bias free CMB lensing map using a simplified two-component (CMB and tSZ) simulation. This is demonstrated in two steps:
\begin{enumerate}
\item We first show that SMICAnoSZ is free of the tSZ effect. 
\item We perform lensing reconstruction with one temperature map free of tSZ effect, and demonstrate that the reconstructed lensing map is free of tSZ bias.
\end{enumerate}
For the first step, we take a lensed CMB map and simulated tSZ maps at 100-857 GHz generated from an $N$-body simulation (Omori in prep.), and multiply each frequency channel with the weights given by the SMICA weight propagation code\footnote{\texttt{ COM\_Code\_SMICA\-weights\-propagation\_R3.00} available from Planck Legacy Archieve \url{https://pla.esac.esa.int/}. We specifically use the values from \texttt{weights\_T\_smica-nosz\_R3.00\_Xfull.txt}.}. The power spectra of the tSZ effect at 100/143/217/353/545/857 GHz channels and the resulting spectra after passing through the weights are shown in Figure \ref{fig:smicanosz_residual}. We find that the resulting tSZ amplitude is suppressed to negligible levels as expected.

\begin{figure}
\begin{center}
\includegraphics[width=1.0\linewidth]{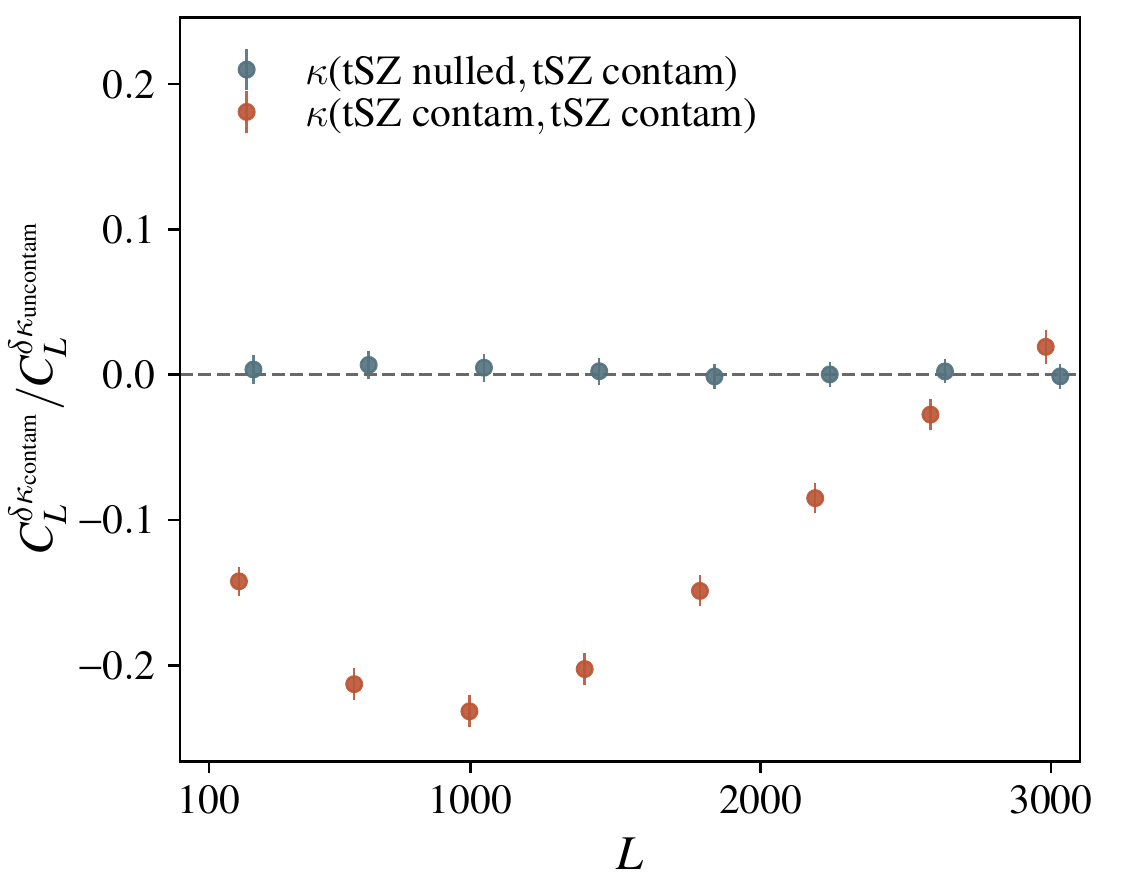}
\caption{Comparison of the density - CMB lensing correlation for two types of CMB lensing map reconstruction: one quadratic estimator leg contaminated with the tSZ effect and the other tSZ nulled (blue) and both temperature legs contaminated with the tSZ effect (orange).   }
\label{fig:tsznull_cross_density}
\end{center}
\end{figure}

Next, we construct a lensing map from the combination of two types of temperature maps
\begin{enumerate}
\item CMB only maps to mimic tSZ nulled CMB maps (i.e. {\it Planck} SMICAnoSZ map), and \item CMB + tSZ maps to mimic high resolution CMB maps (i.e. SPT+{\it Planck} map),
\end{enumerate}
which gives us three lensing maps (a) $T^{\rm CMB only} + T^{\rm CMB only}$, (b) $T^{\rm CMB\ only} + T^{\rm CMB+tSZ}$ and (c) $T^{\rm CMB+tSZ} + T^{\rm CMB+tSZ}$.
For the purpose of this demonstration, we assume $f_{\rm sky}=1$, and add noise that is reduced by a factor of 100 to reduce the computational cost of averaging over many realizations.
 We carry out the lensing reconstruction procedure, measure the cross-correlations between 
the reconstructed lensing maps and a mock galaxy density map, and compare the resulting cross-correlation amplitudes against the unbiased case (i.e. taking the ratios ((b)-(a))/(a) and ((c)-(a))/(a)). The results are shown in Figure \ref{fig:tsznull_cross_density}: we observe that the lensing map without any treatment of the tSZ effect is biased low, whereas the lensing map produced using the ``half-leg" method is compatible with the lensing map produced from ``CMB only" temperature maps.

\section{Hyperrank}
\label{sec:hyperrank}

In our fiducial analysis, we used the model described in Equation~\ref{eq:nz} to characterize the uncertainty in our knowledge of the redshift distribution. In \cite{y3-hyperrank}, however, the authors investigated a more generic way of sampling the uncertainties in the redshift distribution -- a framework referred to as \textsc{Hyperrank}. In principle, \textsc{Hyperrank} is more correct in marginalizing the uncertainty in photometric redshifts since it includes variation in the entire shape of the $n(z)$, but since the lensing kernel is typically broad, the approximation of only marginalizing the mean redshift is often a reasonable one. In \cite{y3-hyperrank} it is shown that the constraints on cosmic shear using \textsc{Hyperrank} are consistent with just marginalizing the mean redshift, which motivates the fiducial choice here and in \cite{y3-3x2ptkp}, which is computationally more efficient to sample. However, in \cite{y3-3x2ptkp} (Figure~23 in Appendix E), it is shown that when applied to data, using \textsc{Hyperrank} results in cosmological constraints that are shifted from the fiducial analysis by $\sim$0.5$\sigma$, with slightly tighter overall constraints. We compare in Figure~\ref{fig:hyperrank} our \fivetwo{} constraints using the fiducial approach in marginalizing the $n(z)$ with shift parameter, and \textsc{Hyperrank}. We find a slight improvement in the constraint -- the uncertainties on $\Omega_{\rm m}$/$\sigma_8$/$S_8$ went from \results{7.5/4.9/1.9} to \results{7.0/4.5/1.7\%}. 

\section{Deriving scale cuts for nonlinear galaxy bias model}
\label{sec:scalecut_nlb}

\begin{figure*}
\begin{center}
\includegraphics[width=0.95\linewidth]{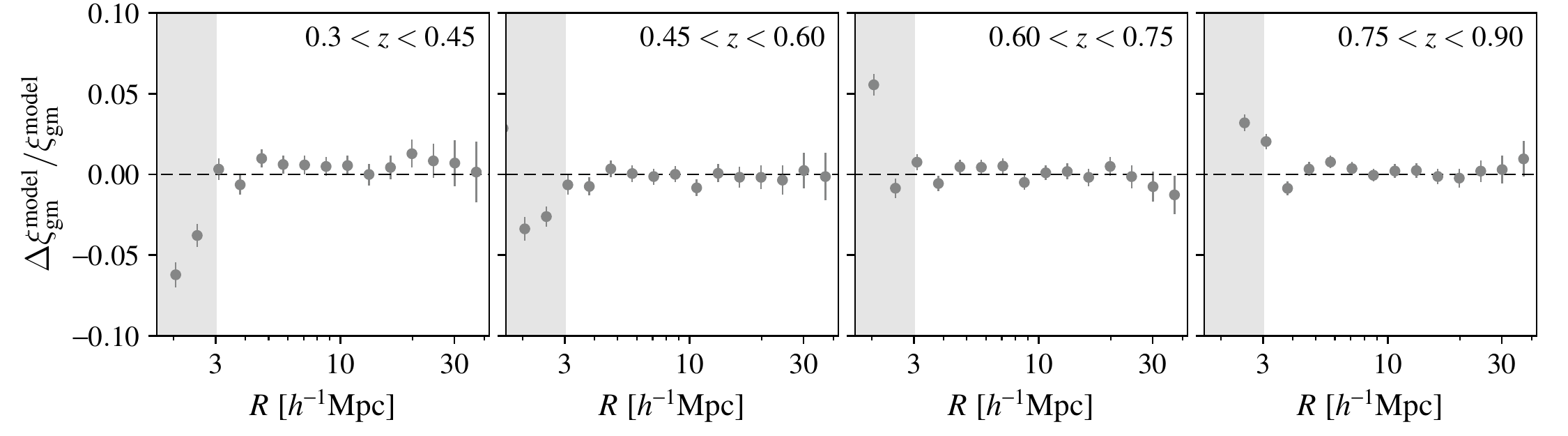}
\caption{Residuals of best-fit 3D galaxy-matter correlation function in the MICE simulation (with the MagLim galaxy sample) assuming a nonlinear galaxy bias model. The shaded region indicates bias exceeding 3\%, which we exclude in our analysis when we assume nonlinear galaxy bias.}
\label{fig:nlbias_mice}
\end{center}
\end{figure*}

As discussed in Section~\ref{sec:analysischoices}, when using the nonlinear galaxy bias model, we cannot apply the same framework of choosing scale cuts for \nk{} since the contaminated data vector that we use to perform the test is generated using our nonlinear bias model. Instead, we need an {\it a priori} criteria for where the PT-based nonlinear galaxy bias model fails to describe the galaxy-matter power spectrum. We take an approach similar to that used in \cite{pandey2020} where we measure the 3D galaxy-matter correlation function from a set of $N$-body simulations, namely the MICE simulations \cite{Crocce2015, Fosalba2015}. These simulations include mock galaxies that have similar selection functions as our lens galaxies (i.e. the \maglim{} and \redmagic{} samples). We fit the measurements using the nonlinear bias model described in Equation~\ref{eq:nlbias} and the input cosmological parameters to the simulations. Figure~\ref{fig:nlbias_mice} shows the relative residuals of the fit for the 4 tomographic lens bins for the \maglim{} sample. 

Based on Figure~\ref{fig:nlbias_mice}, we decide to include scales down to $\sim$3 Mpc/h. This gives at most 3\% difference between model and simulation data, compared to the statistical error bars in \nk{} at about 10\%. We note that out of the 50 or so data points, only 2 are above 1\%. In addition, in the real cosmological analysis, there will be many more degrees of freedom in the other nuisance parameters (IA, photo-z etc), which will further absorb this bias. These factors suggest that our scale cut choice is still relatively conservative.

\section{\redmagic{}}
\label{sec:redmagic}
In this section, we outline the parameter ranges used in the analysis, scale cut tables, forecasted signal-to-noise ratio as well as figures for the parameter contour shifts (equivalent to Table \ref{tab:params_all}, Table \ref{tab:scalecuts} and Figure \ref{fig:contam_contours} respectively) when using the \redmagic{} galaxy sample instead of our fiducial \maglim{} sample. We find that scale cuts similar to that of the \maglim{} sample allow us to pass our bias requirements, and we forecast that the signal-to-noise ratio will be marginally lower for the \redmagic{} sample.

\begin{figure*}
\begin{center}
\includegraphics[width=0.246\linewidth]{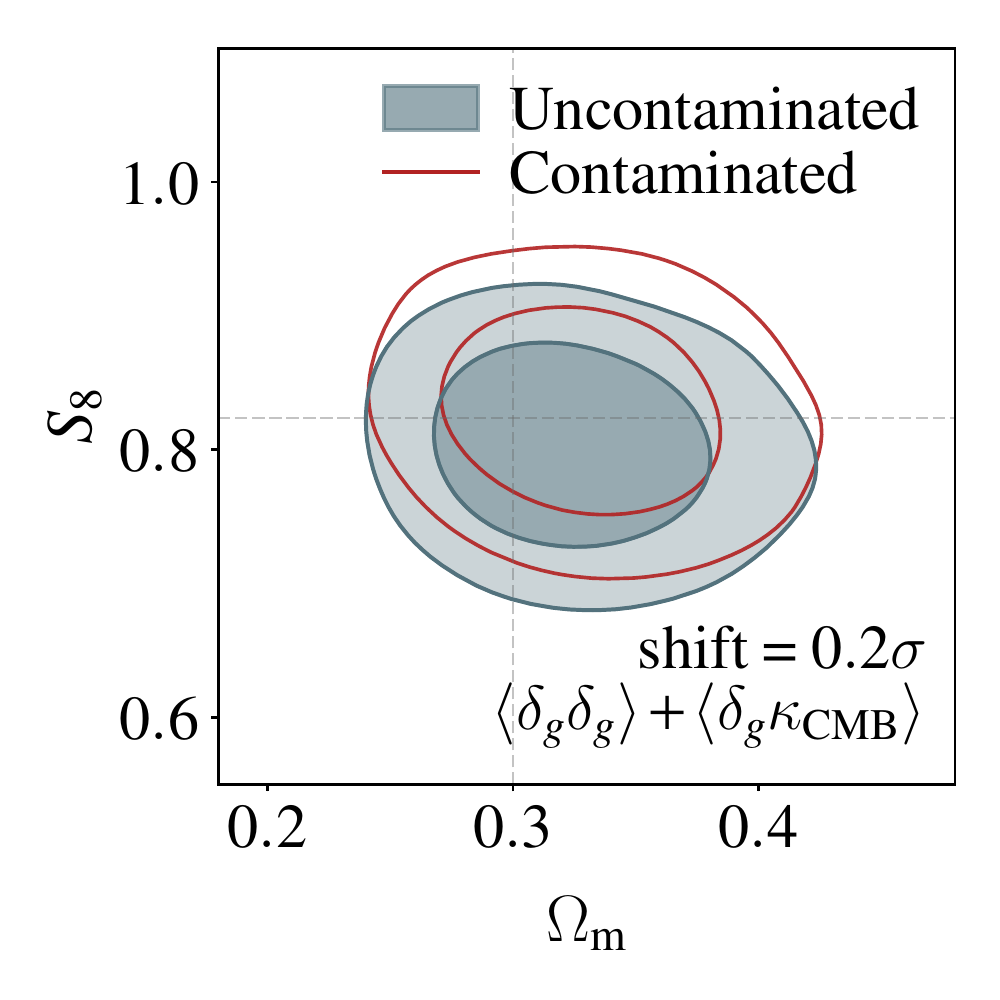}
\includegraphics[width=0.246\linewidth]{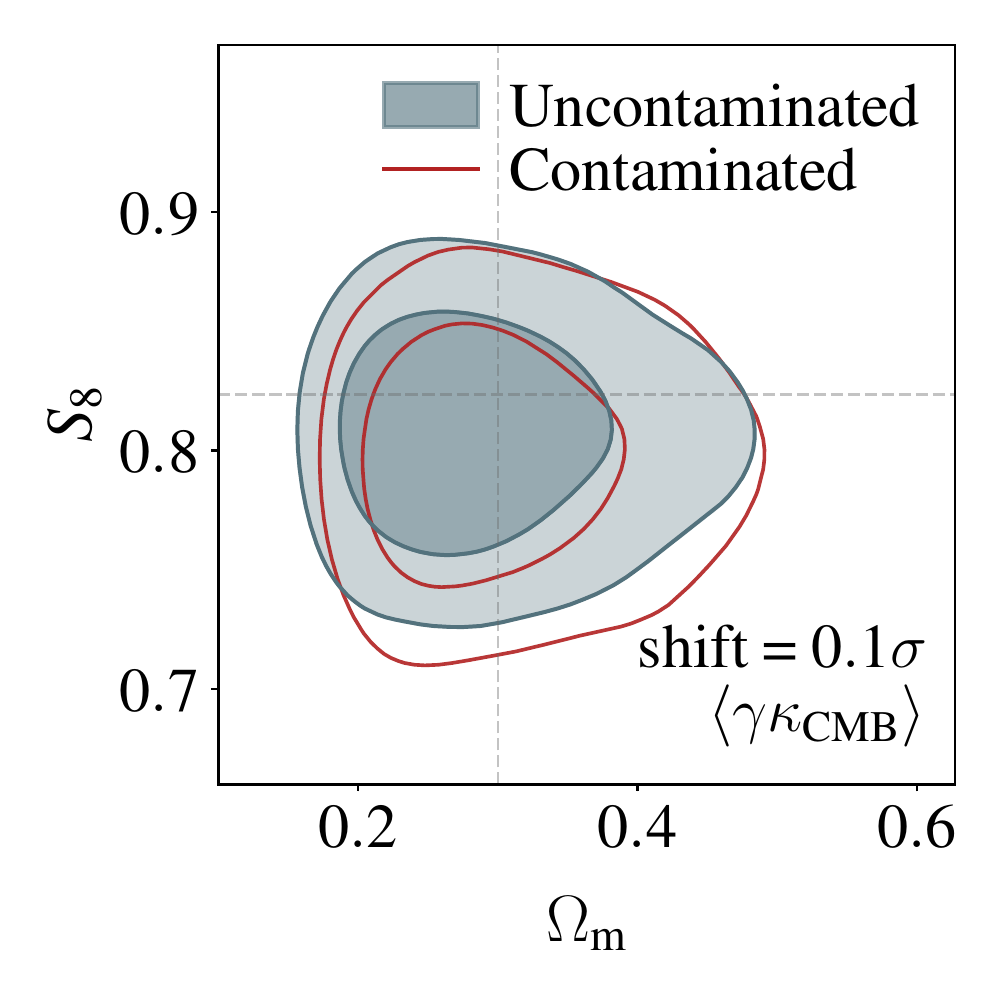}
\includegraphics[width=0.246\linewidth]{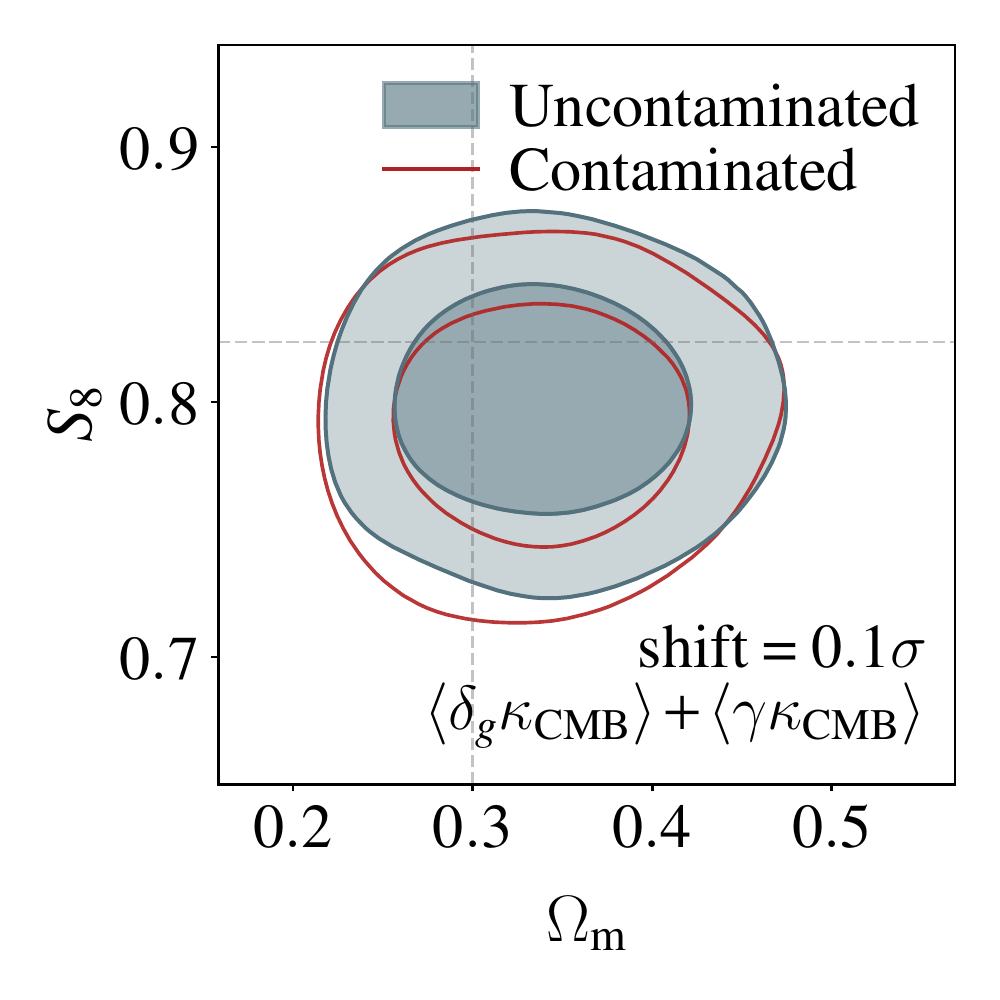}
\includegraphics[width=0.246\linewidth]{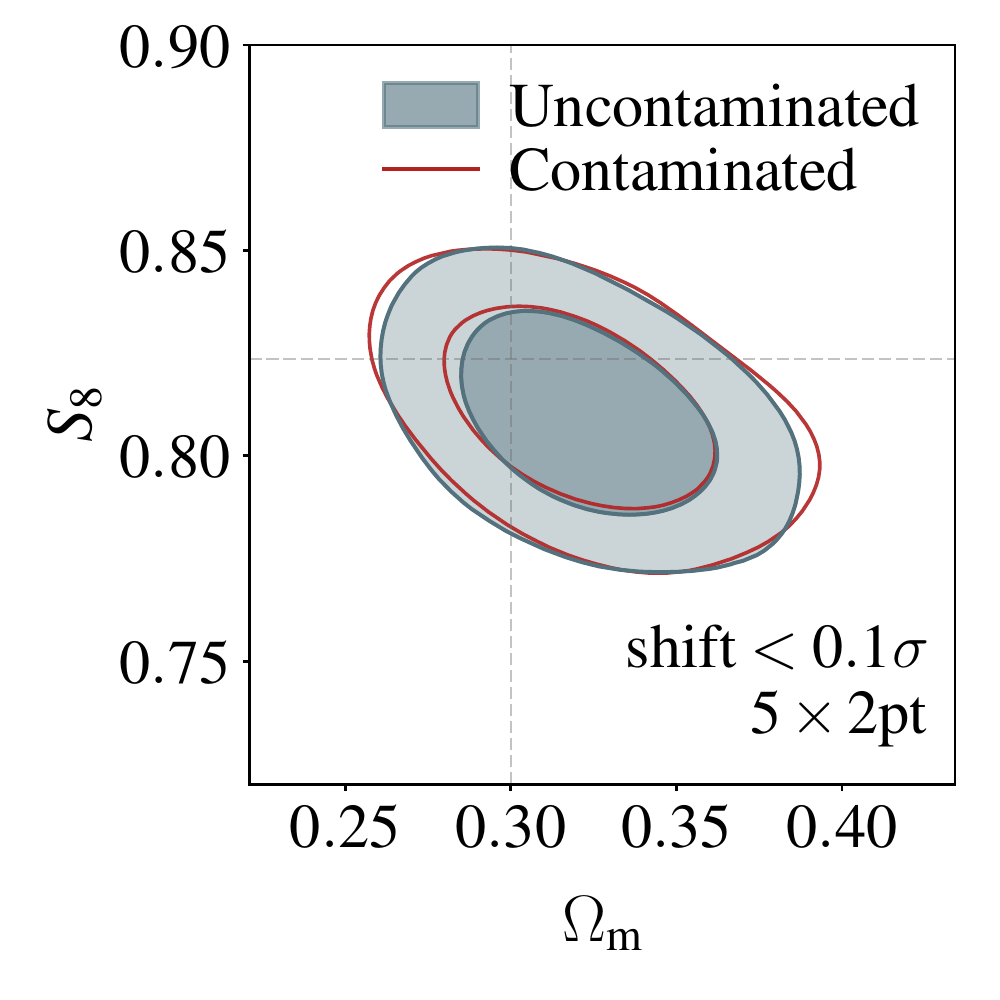}
\caption{Same as Figure \ref{fig:contam_contours} but for the \redmagic{} sample. }
\label{fig:contam_contours_redmagic}
\end{center}
\end{figure*}

\begin{table*}
\centering 
\begin{tabular}{ccc}
\hline
Parameter & Prior & Fiducial  \\ \hline
 \redmagic{} & & \\
$b^{1\cdots 5}$  & $\mathcal{U}[0.8, 3.0]$ & 1.7, 1.7, 1.7, 2.0, 2.0\\
$b_{1}^{1\cdots 5}$  & $\mathcal{U}[0.67, 3.00]$ & 1.40, 1.40, 1.40, 1.40,1.65,1.65\\
$b_{2}^{1\cdots 5}$  & $\mathcal{U}[-4.22, 4.22]$ & 0.16, 0.16, 0.16, 0.35,0.35, \\
 $C_{\rm l}^{1\cdots 5}$ & fixed & 1.31, -0.52, 0.34, 2.25, 1.97 \\
$\Delta_z^{1...5} \times 10^{-2}$ & $\mathcal{N}[0.0, 0.4]$, $\mathcal{N}[0.0, 0.3]$, $\mathcal{N}[0.0, 0.3]$,  & 0.0, 0.0, 0.0, 0.0, 0.0 \\
 & $\mathcal{N}[0.0, 0.5]$, $\mathcal{N}[0.0, 1.0]$ & \\
$\sigma_{z}^{1...5}$ & fixed, fixed, fixed, fixed  $\mathcal{N}[1.0, 0.054]$ & 1.0, 1.0, 1.0, 1.0, 1.0\\
 \hline
 \hline
\end{tabular}

\caption{Same as the lens galaxy section of Table \ref{tab:params_all} but for the \redmagic{} sample.}
\label{tab:params_all_redmagic}
\end{table*}

\begin{table*}
\begin{center}
\begin{tabular}{cccccc*{3}{c}}
\toprule 
\multirow{2}{*}{Type} & \multirow{2}{*}{Redshift bin}& \multirow{2}{*}{\  }  & \multicolumn{2}{c}{$\theta_{\rm min}$} & \multicolumn{3}{c}{${\rm Forecasted \ S/N}$}   \\ \cmidrule{4-8} 
 &  &  & SPT+{\it Planck} & {\it Planck}  &            SPT+{\it Planck} & {\it Planck} & Combined  \\ \midrule 
\nk{} & 1  &    & 15.8$'$ (11.8$'$) & 13.8$'$ (11.8$'$)  & & \\
      & 2  &   & 11.7$'$ (8.8$'$) & 10.2$'$ (8.8$'$)    & & \\
      & 3  &   & 10.0$'$ (7.5$'$)  & 8.7$'$ (7.5$'$)    & & \\
      & 4  &   & 9.0$'$ (6.8$'$)  & 7.9$'$ (6.8$'$)    & &  \\ 
      & 5  &   & 8.6$'$ (6.4$'$)  & 7.5$'$ (6.4$'$)    & &  \\ 
      
      & {All bins}& &                    &                   &   11.1 (13.0) & 10.9 (11.7) & 15.6 (17.5)\\

\midrule
\gk{}& 1 && 2.5$'$  & 2.5$'$ & & \\
      & 2 && 2.5$'$  & 2.5$'$ & & \\
      & 3 && 11.2$'$  & 2.5$'$ & & \\
      & 4 && 17.7$'$  & 2.5$'$ & & \\   
      & {All bins} &&   &  &  10.1  & 8.7 & 13.3 \\     
\midrule
\nkgk{}& All bins &&   &  & 13.2 (14.5) & 12.2 (12.8)  & 18.0 (19.4)\\ \bottomrule
\end{tabular}
\caption{Same as Table \ref{tab:scalecuts} but for the \redmagic{} sample.} 
\end{center}
\label{tab:scalecuts_redmagic}
\end{table*}

\begin{figure}
\begin{center}
\includegraphics[width=1.00\linewidth]{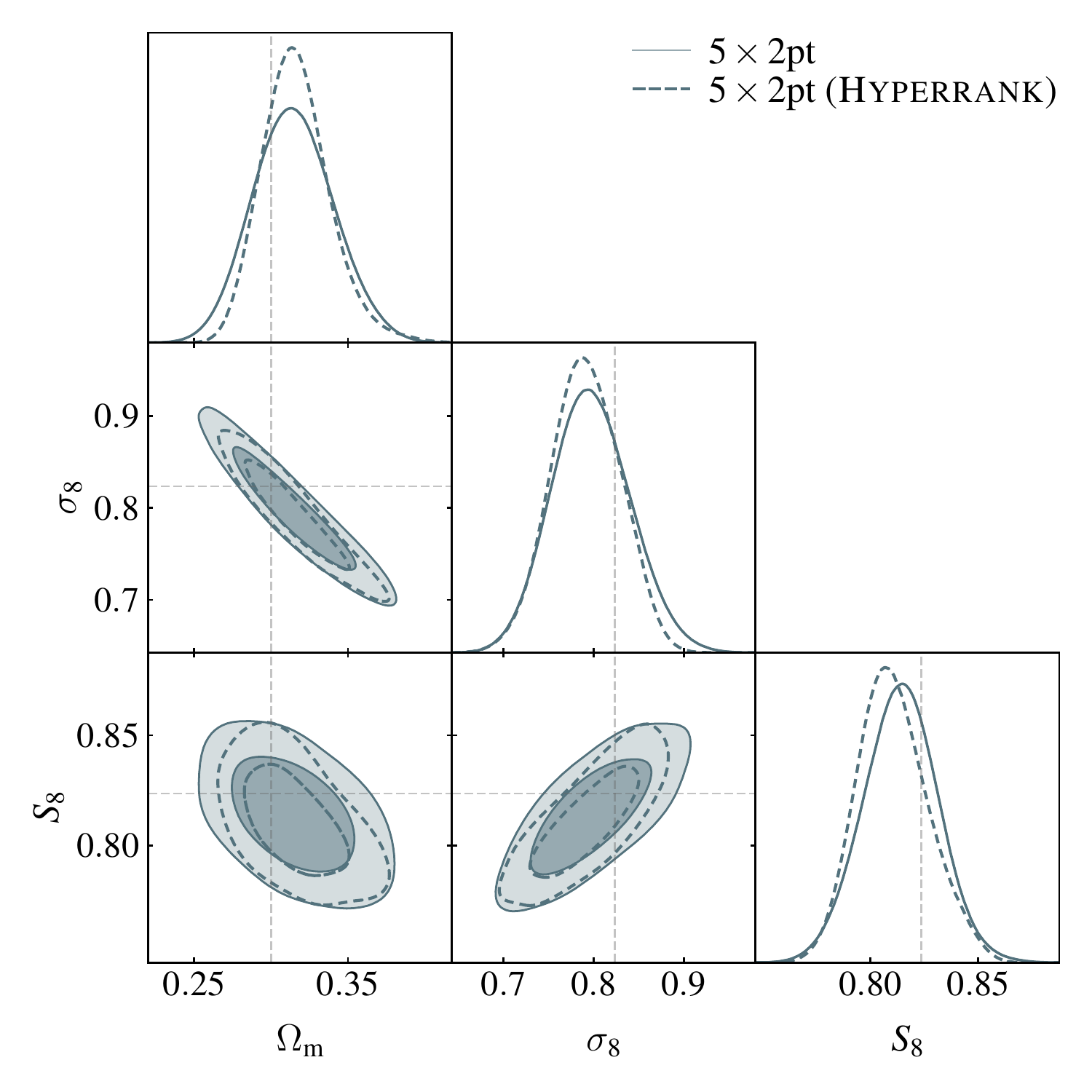}
\caption{Comparison of the forecasted constraints on $\Omega_{\rm m}-\sigma_{8}-S_{8}$ plane when using the fiducial model of assuming a shift in $n(z)$ and when drawing from possible realizations using \textsc{Hyperrank}.}
\label{fig:hyperrank}
\end{center}
\end{figure}

\bibliography{ref,y3kp}

\end{document}